\documentclass[16pt]{article}
\usepackage{amscd}
\usepackage{bbm}
\usepackage{mathrsfs}
\usepackage{amssymb}
\usepackage{graphics}
\usepackage{graphicx}
\usepackage{amsfonts}
\usepackage{amsmath}
\usepackage{array}
\usepackage{booktabs}
\usepackage{enumerate}
\newtheorem{theorem}{Theorem}

\newtheorem{lemma}{Lemma}

\usepackage[numbers,sort&compress]{natbib}
\begin{document}
\date{}
\title{Inverse scattering transform for the discrete nonlocal $PT$ symmetric nonlinear Schr\"{o}dinger equation with nonzero boundary conditions}

\author{Ya-Hui Liu, \ Rui Guo$\thanks{Corresponding author:
gr81@sina.com}$, \ Jian-Wen Zhang
\\
\\{\em
School of Mathematics, Taiyuan University  of} \\
{\em Technology, Taiyuan 030024, China}
} \maketitle

\begin{abstract}
In this paper, the inverse scattering transform for the integrable discrete nonlocal $PT$ symmetric nonlinear Schr\"{o}dinger equation with nonzero boundary conditions is presented. According to the two different signs of symmetry reduction and two different values of the phase difference between plus and minus infinity, we discuss four cases with significant differences about analytical regions, symmetry, asymptotic behavior and the presence or absence of discrete eigenvalues, namely, the existence or absence of soliton solutions. Therefore, in all cases, we study the direct scattering and inverse scattering problem, separately. The Riemann-Hilbert problem is constructed and solved as well as the reconstruction formula of potential is derived, respectively. Finally, combining the time evolution, we provide the dark, bright, dark-bright soliton solutions on the nonzero background in difference cases under the reflectionless condition.

\vspace{5mm}\noindent\emph{Keywords}: Discrete nonlocal nonlinear Schr\"{o}dinger equation; $PT$ symmetry; Inverse scattering transform; Riemann-Hilbert problem; Soliton solutions.
\end{abstract}
\vspace{7mm}\noindent\textbf{1  Introduction}
\hspace*{\parindent}
\renewcommand{\theequation}{1.\arabic{equation}}\\

Nonlinear phenomena are widely present in nature~\cite{ck1} and have received sufficient attention from various fields such as nonlinear optics, hydrodynamics, fluids~\cite{ck2,ck3,ck4,ck5}. Integrable system, which can be used to study nonlinear phenomena, has been studied worldwide over the past few decades and has developed rapidly in mathematics and physics~\cite{ck6,ck7,ck8,ck9}. The nonlinear Schr\"{o}dinger (NLS) equation
\begin{equation}\label{1.0}
\begin{aligned}
i\,q_{t}(x,t)+q_{xx}(x,t)\pm2\,|q(x,t)|^{2}q(x,t)=0
\end{aligned}
\end{equation}
is a typical and significant nonlinear integrable model that has been applied in many fields such as fluid dynamics, plasma physics, optical fibers and the atomic realm of Bose-Einstein condensates~\cite{ck10,ck11,ck12}. Therefore,  the importance of Eq.~(\ref{1.0}) further allows ones to find classes of its exact solutions by Darboux transformation~\cite{ck13}, the inverse scattering transform (IST)~\cite{ck14}, Hirota bilinear method~\cite{ck15}.

Although there is sufficient emphasis on studying continuous systems, the discrete ones have also been found to be important and are used to characterize many phenomena in physics, biology, nonlinear optics~\cite{ck16,ck17,ck18,ck19}, since matter itself is discrete~\cite{ck17}. Therefore, there is significant interest in the study of nonlinear waves in media that are governed by nonlinear discrete evolution systems. In 1975-1976, Ablowitz and Ladik~\cite{ck16,ck20} introduced and discussed the integrable discrete nonlinear Schr\"{o}dinger (NLS) equation
\begin{equation}\label{1.1}
\begin{aligned}
i\,q_{n,t}(t)=q_{n+1}(t)-2\,q_{n}(t)+q_{n-1}(t)-\sigma\left|q_{n}(t)\right|^{2}\left(q_{n+1}(t)+q_{n-1}(t)\right),
\end{aligned}
\end{equation}
where the function $q_{n}(t)$ depends on the discrete space variable $n\in\mathbb{Z}$ and the continuous time variable $t\in\mathbb{R}$, $q_{n,t}(t)=\frac{d q_{n}(t)}{d t}$, $\sigma=\mp1$ represents the focusing/defocusing cases, respectively and it is the so-called Ablowitz-Ladik equation. Eq.~(\ref{1.1}) has infinite conserved quantities and is also one of the most important nonlinear lattice models because it possesses nonlinearity and lattice dispersion as well as it can be applied in many physical phenomena including the dynamics of the condensed matter, anharmonic lattices, Heisenberg spin chains~\cite{ck19,ck20,ck21,ck22,ck23}. In fact, Eq.~(\ref{1.1}) has the more general form~\cite{ck24}:
\begin{subequations}\label{1.2}
\begin{align}
i\,q_{n,t}(t)&=q_{n+1}(t)-2\,q_{n}(t)+q_{n-1}(t)-q_{n}(t)r_{n}(t)\left(q_{n+1}(t)+q_{n-1}(t)\right)\,,\\
-i\,r_{n,t}(t)&=r_{n+1}(t)-2\,r_{n}(t)+r_{n-1}(t)-q_{n}(t)r_{n}(t)\left(r_{n+1}(t)+r_{n-1}(t)\right)\,.
\end{align}
\end{subequations}
Under the symmetry reduction condition $r_{n}(t)=\sigma q_{n}^{*}(t)$ (The symbol $*$ stands for the complex conjugate), Eqs.~(\ref{1.2}) are reduced to Eq.~(\ref{1.1}).

One of the most important aspects in the research of nonlinear integrable systems is to obtain exact solutions, among which the IST is one of the most effective and complete method and in recent years, many initial-value problems for Eq.~(\ref{1.1}) have already been solved via this method. Eq.~(\ref{1.1}) with zero boundary condition (ZBC) was solved via the IST and its  one-order~\cite{ck10,ck16,ck20} and higher-order~\cite{ck25} soliton solutions entirely related to discrete spectrum were obtained. For Eq.~(\ref{1.1}) with $\sigma=1$, Ref.~\cite{ck26} confirmed the $l^{2}$-Sobolev space bijectivity of the IST; under the nonzero boundary conditions (NZBCs), Ref.~\cite{ck27} studied the IST under the condition of $|q_{n}(t)|<1$ and obtained multisoliton solutions; Ref.~\cite{ck28} further relaxed the implicit requirement in Ref.~\cite{ck27} that the eigenfunctions are entire functions of the scattering parameter and obtained dark-soliton solutions as well as discussed the small-amplitude and continuum limits of the problem; Ref.~\cite{ck29} developed its IST with arbitrarily a large background at space infinity and then presented the discrete analog of Kuznetsov-Ma, Akhmediev, Peregrine solutions. For the IST of focusing Eq.~(\ref{1.1}) under the NZBCs, Ref.~\cite{ck23} constructed the robust IST by which the general Darboux matrix can possess the Riemann-Hilbert (RH) representation; Ref.~\cite{ck30} developed it without introducing a uniformization coordinate and  derived the left Marchenko equations to solve the inverse problem, but in Ref.~\cite{ck18} and~\cite{ck31}, the inverse problem is posed as a RH problem.

Recently, integrable discrete nonlocal $PT$ symmetric NLS equation
\begin{equation}\label{1.3}
\begin{aligned}
i\,q_{n,t}(t)=q_{n+1}(t)-2\,q_{n}(t)+q_{n-1}(t)-\sigma q_{n}(t)q^{*}_{-n}(t)\left(q_{n+1}(t)+q_{n-1}(t)\right),
\end{aligned}
\end{equation}
which corresponds to Eqs.~(\ref{1.2}) with the symmetry reduction condition $r_{n}(t)=\sigma q_{-n}^{*}(t)$, has been introduced by Ablowitz and Musslimani~\cite{ck22}. Eq.~(\ref{1.3}) is an integrable Hamiltonian system and also possesses infinite conservation laws as well as has particularly rich mathematical structure and interesting physical behavior~\cite{ck22,ck32}. Regarding the IST for Eq.~(\ref{1.3}), Ablowitz and Musslimani~\cite{ck22} obtained the discrete one-soliton solution by a left-right RH formulation with the potentials rapidly vanishing as $n\rightarrow\pm\infty$; Grahovski, Mohammed and Susanto~\cite{ck33} developed the IST and derived one and two-soliton solutions while they also proved the completeness relation for the associated Jost solutions so that they derived the expansion formula over the complete set of Jost solutions; In 2020, Ablowitz, Luo and Musslimani~\cite{ck24} further analyzed symmetries of the eigenfunctions, scattering data and norming constants for all nonlocal cases of discrete NLS equation under not only PT symmetry but also reverse-space-time symmetry and reverse-time symmetry, finally, they provided soliton solutions for both local and three nonlocal cases, respectively. But so far, the complete theory of the IST for Eq.~(\ref{1.3}) with nonzero boundary conditions (NZBCs) still remains open, which we will discuss in turn.

For Eq.~(\ref{1.3}), which is Eqs.~(\ref{1.2}) with $r_{n}(t)=\sigma q_{-n}^{*}(t)$, we consider the potential satisfying the following NZBCs:
\begin{equation}\label{1.4}
{\lim_{n\to \pm \infty}}q_{n}(t)=q_{\pm}(t)=q_{0}e^{i \vartheta_{\pm}(t)}\,,\vartheta_{\pm}(t)=\theta_{\pm}+2\sigma \delta t\,,
\end{equation}
\hspace{-0.2cm}
where $q_{0}$, $\theta_{\pm}$ are real constant parameters and $\delta=q_{0}^{2}e^{i\Delta\theta}$ with $\Delta\theta=\theta_{+}-\theta_{-}$ should also be real so that $\Delta\theta=k\pi$, $k\in\mathbb{Z}$, then the boundary conditions are bounded, that is the only case we consider here. We choose $\Delta\theta=0$ or $\Delta\theta=\pi$, so in the following content, we focus on the four cases:
\begin{enumerate}[(i)]
  \item $\sigma=1$, $\Delta\theta=0$, $\theta_{+}=\theta_{-}=\theta$, $\delta=q_{0}^{2}$, ${\lim_{n\to \pm \infty}}q_{n}(t)=q_{\pm}(t)=q_{0}e^{i (\theta+2q_{0}^{2}t)}$, ${\lim_{n\to \pm\infty}}r_{n}(t)=q_{\mp}^{*}(t)=q_{0}e^{-i (\theta+2q_{0}^{2}t)}$\,;\label{c1}
  \item $\sigma=1$, $\Delta\theta=\pi$, $\theta_{+}=\pi+\theta_{-}$, $\delta=-q_{0}^{2}$, ${\lim_{n\to \pm \infty}}q_{n}(t)=q_{\pm}(t)=q_{0}e^{i (\theta_{\pm}-2q_{0}^{2}t)}$, ${\lim_{n\to \pm\infty}}r_{n}(t)=q_{\mp}^{*}(t)=q_{0}e^{-i (\theta_{\mp}-2q_{0}^{2}t)}$\,;\label{c2}
  \item $\sigma=-1$, $\Delta\theta=0$, $\theta_{+}=\theta_{-}=\theta$, $\delta=q_{0}^{2}$, ${\lim_{n\to \pm \infty}}q_{n}(t)=q_{\pm}(t)=q_{0}e^{i (\theta-2q_{0}^{2}t)}$, ${\lim_{n\to \pm\infty}}r_{n}(t)=-q_{\mp}^{*}(t)=-q_{0}e^{-i (\theta-2q_{0}^{2}t)}$\,;\label{c3}
  \item $\sigma=-1$, $\Delta\theta=\pi$, $\theta_{+}=\pi+\theta_{-}$, $\delta=-q_{0}^{2}$, ${\lim_{n\to \pm \infty}}q_{n}(t)=q_{\pm}(t)=q_{0}e^{i (\theta_{\pm}+2q_{0}^{2}t)}$, ${\lim_{n\to \pm\infty}}r_{n}(t)=-q_{\mp}^{*}(t)=-q_{0}e^{-i (\theta_{\mp}+2q_{0}^{2}t)}$\,.\label{c4}
\end{enumerate}

The corresponding discrete scattering problem and the time-dependence equation of Eq.~(\ref{1.3}) are
\begin{subequations}\label{1.5}
\begin{align}
\varphi_{n+1}&=\mathbf{U}_{n}\varphi_{n}=(\mathbf{Z}+\mathbf{Q_{n}})\varphi_{n},\\
\varphi_{n,t}&=\mathbf{V}_{n}\varphi_{n}=i\sigma_{3}\left(\mathbf{Q_{n}}\mathbf{Q_{n-1}}-\mathbf{Z}\mathbf{Q_{n}}+\mathbf{Q_{n-1}}\mathbf{Z}-\frac{(z-z^{-1})^{2}}{2}\mathbf{I}_{2}\right)\varphi_{n},
\end{align}
\end{subequations}
where $\varphi_{n}=(\varphi_{n}^{(1)}(t)\,,\varphi_{n}^{(2)}(t))^{T}$ ($T$ is the transposition), $\mathbf{I}_{2}$ is the $2\times2$ identity matrix, $z$ is the scattering parameter that is independent of $n$ and $t$,
\begin{align}
\mathbf{Z}=\left (
\begin{array}{cc}
z &0\\
0 &z^{-1}\\
\end{array}
\right)\,,\,\sigma_{3}=\left (
\begin{array}{cc}
1 &0\\
0 &-1\\
\end{array}
\right)\,,\,\mathbf{Q_{n}}=\left (
\begin{array}{cc}
0 &q_{n}(t)\\
r_{n}(t) &0\\
\end{array}
\right)\nonumber
\end{align}
in which we still denote $\sigma q_{-n}^{*}(t)$ as $r_{n}(t)$ for simplifying writing in the following content. Eq.~(\ref{1.3}) is equivalent to the discrete compatibility condition $\frac{d}{d t}\varphi_{n+1}(t)=\left(\frac{d}{d t}\varphi_{j}(t)\right)_{j=n+1}$.

The main purpose of our work is to develop rigorous theory of the IST for Eqs.~(\ref{1.2}) with $PT$ symmetry reduction $r_{n}(t)=\sigma q_{-n}^{*}(t)$ or Eq.~(\ref{1.3}) for the above four cases. In each case, we discuss the direct scattering problem including the analysis of analyticity, asymptotic behavior, symmetry for eigenfunctions and scattering coefficients, inverse scattering problem consisting of constructing and solving a RH problem, as well as analyzing the existence of soliton solutions in detail. Due to the nonlocal case, there are many novel properties and conclusions different from the local one. Specifically, the highlights of this article are the following:
(i) For the four cases, the constraints for constant background, analytic regions, continuous spectrum, asymptotic behavior of scattering coefficients and so on have great differences, which we show in the text;
(ii) The second symmetry about $\zeta\mapsto\bar{\zeta}^{*}=\frac{r\zeta^{*}-1}{\zeta^{*}-r}$ is deduced by introducing the ``backward" modified scattering problem and analysing modified eigenfunctions, which has not been analyzed in local systems with NZBCs. Furthermore, the second symmetry for the four cases is also distinct; (iii) The discrete eigenvalues here are only obtained by comparing the trace formula with asymptotic behavior of scattering coefficients and the number of discrete eigenvalues is really different in certain cases; (iv) Due to the strong limitations on discrete eigenvalues, not all cases exist soliton solutions for Eq.~(\ref{1.3}) with NZBCs~(\ref{1.4}). For example, in the Case (\ref{c1}), the single eigenvalue is not exist so that there is no first order soliton solution. The simplest reflectionless potential generates second order soliton solutions, which include dark-dark, dark-bright, bright-bright ones. In the Case (\ref{c2}): $\sigma=1$, $\Delta\theta=\pi$, we have showed there are no discrete eigenvalues so that the soliton solutions are not exist.

This article is organized as follows. In Section $2$, we mainly present the IST for the case~(\ref{c1}). The direct scattering problem is discussed in Section $2.1$, which includes the definitions of eigenfunctions and scattering matrix, as well as the analysis of analyticity, asymptotic behavior and two types of symmetry. In Section $2.2$, the inverse scattering problem is organized: We construct and solve a suitable RH problem and then we get the reconstructed potential. In addition, we analysis the trace formula and obtain suitable discrete eigenvalues composed of quartets according the asymptotic behavior of scattering coefficients. The time evolution of eigenfunctions, scattering coefficients are studied in Section $2.3$ and under the reflectionless condition, the second-order soliton solutions containing dark-dark, dark-bright, bright-bright ones are displayed in Section $2.4$. Even in the defocusing ($\sigma=1$) and no phase difference ($\Delta\theta=0$) case, many analyses are different from Ref.~\cite{ck28}. Therefore, we will still provide detailed analysis except for deleting the proof of the same parts and only retaining the conclusion. The method in other three cases follows along similar progress but the details are quite distinct so that the different discussion is also developed such as the analysis of continuous spectrum, analytic region and discrete eigenvalues/soliton solutions, and the similar ones are only providing results in Section $3$-$5$.

\vspace{7mm}\noindent\textbf{2 Case~(\ref{c1}): $\sigma=1$, $\Delta\theta=0$}
\hspace*{\parindent}
\renewcommand{\theequation}{2.\arabic{equation}}\setcounter{equation}{0}\\

In this section, we will investigate the IST for Eq.~(\ref{1.3}) with NZBCs~(\ref{1.4}) meeting the case~(\ref{c1}).

\vspace{5mm} \noindent\textbf{2.1 The direct scattering problem}
\\\hspace*{\parindent}

For convenience, we will temporarily omit the time evolution of solutions now so that we rewrite the NZBCs~(\ref{1.4}) and ${\lim_{n\to \pm \infty}}r_{n}(t)$ as
\begin{equation}\label{2.1}
{\lim_{n\to \pm \infty}}q_{n}=q_{\pm}=q_{0}e^{i \vartheta}\,,{\lim_{n\to \pm \infty}}r_{n}=r_{\pm}=q_{0}e^{-i \vartheta}\,,
\end{equation}
where $\vartheta=\theta+2q_{0}^{2}t$ is also real.

\vspace{5mm}\noindent\textbf{2.1.1 Eigenfunctions, scattering matrix, uniformization variable}
\\\hspace*{\parindent}

Under the boundary conditions~(\ref{2.1}) and symmetry reduction condition $r_{n}(t)=q_{-n}^{*}(t)$, it is easy to prove that Eq.~(\ref{1.5}a) has the solution matrices $\mathbf{\Phi}_{n}(z)$ and $\mathbf{\Psi}_{n}(z)$ satisfying
\begin{subequations}\label{2.2}
\begin{align}
\mathbf{\Phi}_{n}(z)&=\left(\phi_{n}(z), \bar{\phi}_{n}(z)\right)\sim r^{n}\left (
\begin{array}{cc}
q_{-} &\lambda r-z\\
\lambda r-z &-r_{-}\\
\end{array}
\right)\mathbf{\Lambda}^{n},\ n\to -\infty\,,\\
\mathbf{\Psi}_{n}(z)&=\left(\bar{\psi}_{n}(z), \psi_{n}(z)\right)\sim r^{n}\left (
\begin{array}{cc}
q_{+} &\lambda r-z\\
\lambda r-z &-r_{+}\\
\end{array}
\right)\mathbf{\Lambda}^{n}\,,\,n\rightarrow+\infty\,,
\end{align}
\end{subequations}
where $\phi_{n}(z)$, $\bar{\phi}_{n}(z)$, $\bar{\psi}_{n}(z)$, $\psi_{n}(z)$ are eigenfunctions, $r=\sqrt{1-q_{0}^{2}}\in\mathbb{R}$ with $0<q_{0}<1$, $\lambda(z)=\xi\pm\sqrt{\xi^{2}-1}$ with $\xi(z)=\frac{z+\frac{1}{z}}{2r}$, $\mathbf{\Lambda}=diag(\lambda,\frac{1}{\lambda})$. $\lambda$ and $z$ satisfy $r(\lambda+\frac{1}{\lambda})=z+\frac{1}{z}$. It is easy to obtain that $\lambda(z)$ is a double-valued function of $z$ and has four branch points $\pm z_{0}$ and $\pm z^{*}_{0}$ meeting $\xi(z)^{2}=1$, where
\begin{equation}\label{2.3}
z_{0}=r+iq_{0}
\end{equation}
and all branch points are located on the unit circle $\{z\in\mathbb{C}:|z|=1\}$.
Furthermore, we can prove that the continuous spectrum of the scattering problem is any $z$ that meet $|\lambda(z)|=1$ and this corresponds to the set:
\begin{equation}\label{2.4}
\begin{aligned}
\{z\in\mathbb{C}:|z|=1\ and\ |Re z|<r\}\,,
\end{aligned}
\end{equation}
which is two arcs $(z_{0},\,-z^{*}_{0})$, $(z^{*}_{0},\,-z_{0})$ on the unit circle $\{z\in\mathbb{C}:|z|=1\}$.

Two solutions $\varphi_{n}$, $\bar{\varphi}_{n}$ of scattering problem obey the
recursion relation:
\begin{equation}\label{2.5}
\begin{aligned}
Det\left(\varphi_{n+1}, \bar{\varphi}_{n+1}\right)=(1-q_{n}r_{n})Det\left(\varphi_{n}, \bar{\varphi}_{n}\right)\,.
\end{aligned}
\end{equation}
Furthermore, one also has
\begin{equation}
\begin{aligned}
Det\left(r^{-n}\phi_{n}(z), r^{-n}\bar{\phi}_{n}(z)\right)\sim-(\lambda r-z)^{2}-q^{2}_{0}\,,n\to -\infty\,,\nonumber
\end{aligned}
\end{equation}
\begin{equation}
\begin{aligned}
Det\left(r^{-n}\bar{\psi}_{n}(z), r^{-n}\psi_{n}(z)\right)\sim-(\lambda r-z)^{2}-q^{2}_{0}\,,n\to +\infty\nonumber
\end{aligned}
\end{equation}
so that
\begin{subequations}\label{2.6}
\begin{align}
Det\left(\mathbf{\Phi}_{n}(z)\right)&=-\left((\lambda r-z)^{2}+q^{2}_{0}\right)r^{2n}\prod_{k=-\infty}^{n-1}\frac{1-q_{k}r_{k}}{1-q^{2}_{0}}\,,\\
Det\left(\mathbf{\Psi}_{n}(z)\right)&=-\left((\lambda r-z)^{2}+q^{2}_{0}\right)r^{2n}\prod_{k=n}^{+\infty}\frac{1-q^{2}_{0}}{1-q_{k}r_{k}}\,.
\end{align}
\end{subequations}
We can prove that $(\lambda r-z)^{2}+q^{2}_{0}=0$ if and only if $z=\pm z_{0}$ or $z=\pm z^{*}_{0}$, which are the four branch points of $\lambda(z)$. In addition, we assume $0<q_{0}<1$ and impose a small norm condition on potential functions $q_{n}$, $r_{n}$ for any $n\in\mathbb{Z}$, then Eqs.~(\ref{2.6}) don't equal zero so that $\phi_{n}(z)$ and $\bar{\phi}_{n}(z)$, $\bar{\psi}_{n}(z)$ and $\psi_{n}(z)$ are linearly independent for any $z\in\mathbb{C}$ except $\pm z_{0}$ and $\pm z^{*}_{0}$, respectively so that both $\mathbf{\Phi}_{n}(z)$ and $\mathbf{\Psi}_{n}(z)$ are the fundamental solutions of scattering problem for any $z\in\mathbb{C}$ except $\pm z_{0}$ and $\pm z^{*}_{0}$. As a consequence, we can introduce a $n$-independent and invertible scattering matrix $T(z)$ to relate these two solution matrices
\begin{equation}\label{2.7}
\begin{aligned}
\mathbf{\Phi}_{n}(z)=\mathbf{\Psi}_{n}(z)\mathbf{T}(z)\,,\,\mathbf{T}(z)=\left(
\begin{array}{cc}
t_{11}(z) &t_{12}(z)\\
t_{21}(z) &t_{22}(z)\\
\end{array}
\right)
\end{aligned}
\end{equation}
and $t_{ij}(z)$, $i,j=1,2$ are defined as the scattering coefficients. According to Eq.~(\ref{2.7}) and using Eqs.~(\ref{2.6}), one obtains
\begin{equation}\label{2.8}
\begin{aligned}
t_{11}(z)t_{22}(z)-t_{21}(z)t_{12}(z)=\Theta_{-\infty}\,,
\end{aligned}
\end{equation}
where
\begin{equation}\label{2.9}
\begin{aligned}
\Theta_{-\infty}=\lim_{n\to -\infty}\Theta_{n}\,,\Theta_{n}=\prod_{k=n}^{+\infty}\frac{1-q_{k}r_{k}}{1-q^{2}_{0}}
\end{aligned}
\end{equation}
and the scattering coefficients can be represented by determinant expressions of eigenfunctions as
\begin{subequations}\label{2.10}
\begin{align}
t_{11}(z)&=\frac{Det\left(\phi_{n}(z), \psi_{n}(z)\right)}{Det\left(\bar{\psi}_{n}(z), \psi_{n}(z)\right)}\,,
t_{22}(z)=\frac{Det\left(\bar{\psi}_{n}(z), \bar{\phi}_{n}(z)\right)}{Det\left(\bar{\psi}_{n}(z), \psi_{n}(z)\right)}\,,\\
t_{21}(z)&=\frac{Det\left(\bar{\psi}_{n}(z), \phi_{n}(z)\right)}{Det\left(\bar{\psi}_{n}(z), \psi_{n}(z)\right)}\,,
t_{12}(z)=\frac{Det\left(\bar{\phi}_{n}(z), \psi_{n}(z)\right)}{Det\left(\bar{\psi}_{n}(z), \psi_{n}(z)\right)}\,.
\end{align}
\end{subequations}

From Eqs.~(\ref{2.10}), we know that if $z_{j}$ is a zero of $t_{11}(z)$, we have $Det\left(\phi_{n}(z_{j}), \psi_{n}(z_{j})\right)=0$, namely, $\phi_{n}(z_{j})=b_{j}\psi_{n}(z_{j})$ with a complex parameter $b_{j}$. Likewise, if $\bar{z}_{j}$ is a zero of $t_{22}(z)$, then $Det\left(\bar{\psi}_{n}(\bar{z}_{j}), \bar{\phi}_{n}(\bar{z}_{j})\right)=0$, namely, $\bar{\psi}_{n}(\bar{z}_{j})=\bar{b}_{j}\bar{\phi}_{n}(\bar{z}_{j})$ with a complex parameter $\bar{b}_{j}$.

As mentioned above, $\lambda(z)$ is a double-valued function of $z$ so that we should define two-sheeted Riemann surface to ensure that eigenfunctions and the scattering coefficients are single-valued functions of $z$. This complexity can be addressed by defining the uniformization variable:
\begin{equation}\label{2.11}
\zeta(z)=\lambda(z)/z\,,
\end{equation}
so that the mappings are given by
\begin{equation}\label{2.12}
z^{2}=\frac{\zeta-r}{\zeta(\zeta r-1)}\,,\,\lambda^{2}=\zeta\frac{\zeta-r}{\zeta r-1}\,,\,z\lambda=\frac{\zeta-r}{\zeta r-1}
\end{equation}
and all even functions of $\lambda$ and $z$ can be expressed as a ratio of two rational functions of $\zeta$.

Next, we will discuss the mapping $z$, $\lambda\rightarrow \zeta$. We can directly obtain that the continuous spectrum, which refers to two arcs $(z_{0},\,-z^{*}_{0})$, $(z^{*}_{0},\,-z_{0})$ on the unit circle $\{z\in\mathbb{C}:|z|=1\}$, is mapped onto the unit circle $\{\zeta\in\mathbb{C}:|\zeta|=1\}$. The four branch points $\pm z_{0}$, $\pm z^{*}_{0}$ are mapped onto two points $\zeta_{0}=\zeta(\pm z^{*}_{0})=z_{0}$, $\zeta^{*}_{0}=\zeta(\pm z_{0})=z^{*}_{0}$ in the $\zeta-$plane. Furthermore, we can show that
\begin{equation}\label{2.13}
|\lambda|\leq 1 \Leftrightarrow |\zeta|\leq 1\,.
\end{equation}
Defining $\Sigma=\{\zeta\in\mathbb{C}\mid \left|\zeta\right|=1\}$ and its positive orientation is anticlockwise as shown in Fig.~$1$. Therefore, the regions
\begin{equation}\label{2.14}
D_{+}=\{\zeta\in\mathbb{C}\mid \left|\zeta\right|<1\}\,,\,D_{-}=\{\zeta\in\mathbb{C}\mid \left|\zeta\right|>1\}
\end{equation}
are the corresponding positive region for $|\lambda|< 1$ and negative region for $|\lambda|> 1$.

Constructing the modified eigenfunctions
\begin{subequations}\label{2.15}
\begin{align}
&(M_{n}(z),\bar{M}_{n}(z))=r^{-n}\mathbf{\Omega}\,\mathbf{\Phi}_{n}(z)\mathbf{\Omega}^{-1}\mathbf{\Lambda}^{-n}\,,\\
&(\bar{N}_{n}(z),N_{n}(z))=r^{-n}\mathbf{\Omega}\,\mathbf{\Psi}_{n}(z)\mathbf{\Omega}^{-1}\mathbf{\Lambda}^{-n}
\end{align}
\end{subequations}
with $\mathbf{\Omega}=diag(1,\lambda)$, then they can be represented by $\zeta$ and the modified scattering problems can all be written in terms of $\zeta$.

\begin{center}
\includegraphics[scale=0.6]{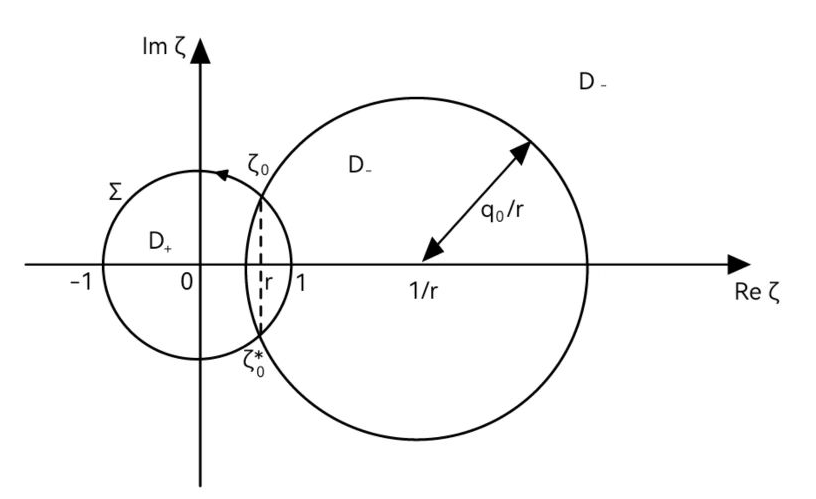}\hfill
\flushleft{\footnotesize
\textbf{Fig.~$1$.} The $\zeta$-plane:  $\Sigma=\{\zeta\in\mathbb{C}\mid \left|\zeta\right|=1\}$ is the continuous spectrum and it prescribed orientation identifies the regions $D_{+}$ and $D_{-}$ as in Eq.~(\ref{2.14}).}
\end{center}

$M_{n}(\zeta)$, $\bar{N}_{n}(\zeta)$ meet
\begin{equation}\label{2.16}
r \mathcal{W}_{n+1}(\zeta)=\left(
\begin{array}{cc}
\frac{1}{\zeta} &\frac{\zeta r-1}{\zeta(\zeta-r)}q_{n}\\
r_{n} &\frac{\zeta r-1}{\zeta-r}\\
\end{array}
\right)\mathcal{W}_{n}(\zeta)\,,
\end{equation}
$\bar{M}_{n}(\zeta)$, $N_{n}(\zeta)$ meet
\begin{equation}\label{2.17}
r \mathcal{\tilde{W}}_{n+1}(\zeta)=\left(
\begin{array}{cc}
\frac{\zeta-r}{\zeta r-1} &q_{n}\\
\frac{\zeta(\zeta-r)}{\zeta r-1}r_{n}  &\zeta\\
\end{array}
\right)\mathcal{\tilde{W}}_{n}(\zeta).
\end{equation}
In addition, their boundary conditions at $n\rightarrow\pm\infty$ are
\begin{subequations}\label{2.18}
\begin{align}
M_{n}(\zeta)&\sim\left (
\begin{array}{cc}
q_{-}\\
\zeta-r\\
\end{array}
\right),\ \bar{M}_{n}(\zeta)\sim \left (
\begin{array}{cc}
r-\frac{1}{\zeta}\\
-r_{-}\\
\end{array}
\right),\ n\to -\infty,\\
\bar{N}_{n}(\zeta)&\sim \left (
\begin{array}{cc}
q_{+}\\
\zeta-r\\
\end{array}
\right),\ N_{n}(\zeta)\sim \left (
\begin{array}{cc}
r-\frac{1}{\zeta}\\
-r_{+}\\
\end{array}
\right),\ n\to +\infty.
\end{align}
\end{subequations}

Applying the same transformation as Eqs.~(\ref{2.15}) to Eq.~(\ref{2.7}), we obtain
\begin{equation}\label{2.19}
(M_{n}(\zeta),\bar{M}_{n}(\zeta))=(\bar{N}_{n}(\zeta),N_{n}(\zeta))\mathbf{\Lambda}^{n}\tilde{\mathbf{T}}(\zeta)\mathbf{\Lambda}^{-n}\,,
\end{equation}
in which
\begin{equation}\label{2.20}
\tilde{\mathbf{T}}(\zeta)=\mathbf{\Omega}\,\mathbf{T}(z)\mathbf{\Omega}^{-1}=\left (
\begin{array}{cc}
\tilde{t}_{11}(\zeta) &\tilde{t}_{12}(\zeta)\\
\tilde{t}_{21}(\zeta) &\tilde{t}_{22}(\zeta)\\
\end{array}
\right)
\end{equation}
with $\tilde{t}_{11}(\zeta)=t_{11}(\zeta)\,,\,\tilde{t}_{22}(\zeta)=t_{22}(\zeta)\,,\,\tilde{t}_{12}(\zeta)=\frac{t_{12}(\zeta)}{\lambda}\,,\,
\tilde{t}_{21}(\zeta)=\lambda t_{21}(\zeta)$ is the modified scattering matrix and each columns can be written as
\begin{subequations}\label{2.21}
\begin{align}
M_{n}(\zeta)&=\tilde{t}_{11}(\zeta)\bar{N}_{n}(\zeta)+\lambda^{-2n}\tilde{t}_{21}(\zeta)N_{n}(\zeta)\,,\\
\bar{M}_{n}(\zeta)&=\tilde{t}_{22}(\zeta)N_{n}(\zeta)+\lambda^{2n}\tilde{t}_{12}(\zeta)\bar{N}_{n}(\zeta).
\end{align}
\end{subequations}

Now we construct another determinant expressions of the scattering coefficients. According to Eq.~(\ref{2.6}b) and Eq.~(\ref{2.15}b), we obtain
\begin{equation}\label{2.22}
Det(N_{n}(\zeta),\bar{N}_{n}(\zeta))=r\frac{\zeta+\frac{1}{\zeta}-2r}{\Theta_{n}}\,,
\end{equation}
where $\Theta_{n}$ is defined in Eq.~(\ref{2.9}). Combining with Eqs.~(\ref{2.10}), (\ref{2.15}), we obtain
\begin{subequations}\label{2.23}
\begin{align}
\tilde{t}_{11}(\zeta)&=-\Theta_{n}\frac{Det\left(M_{n}(\zeta), N_{n}(\zeta)\right)}{r(\zeta+\frac{1}{\zeta}-2r)}\,,\ \ \ \ \ \ \ \ \ \ \ \ \
\tilde{t}_{22}(\zeta)=\Theta_{n}\frac{Det\left(\bar{M}_{n}(\zeta), \bar{N}_{n}(\zeta)\right)}{r(\zeta+\frac{1}{\zeta}-2r)}\,,\\
\tilde{t}_{21}(\zeta)&=\Theta_{n}(\zeta\frac{\zeta-r}{\zeta r-1})^{n}\frac{Det\left(M_{n}(\zeta), \bar{N}_{n}(\zeta)\right)}{r(\zeta+\frac{1}{\zeta}-2r)}\,,\,
\tilde{t}_{12}(\zeta)=-\Theta_{n}(\zeta\frac{\zeta-r}{\zeta r-1})^{-n}\frac{Det\left(\bar{M}_{n}(\zeta), N_{n}(\zeta)\right)}{r(\zeta+\frac{1}{\zeta}-2r)}\,.
\end{align}
\end{subequations}

\vspace{5mm} \noindent\textbf{2.1.2 Analyticity of modified eigenfunctions and scattering coefficients}
\\\hspace*{\parindent}

To begin with, we introduce three modifications to define new eigenfunctions, whose corresponding Green's functions are relatively simple and have no singularities.

Introducing the first modification:
\begin{equation}\label{2.24}
\hat{\mathbf{Y}}_{n}=\mathbf{\Omega} \mathbf{Y}_{n}\mathbf{\Omega}^{-1}\,,
\end{equation}
where $\mathbf{Y}_{n}$ is the $2\times2$ matrix solution of Eq.~(\ref{1.5}a). $\hat{\mathbf{Y}}_{n}$ meets the following scattering problem
\begin{equation}\label{2.25}
\hat{\mathbf{Y}}_{n+1}=\hat{\mathbf{U}}_{n}\hat{\mathbf{Y}}_{n}\,,\hat{\mathbf{U}}_{n}=\mathbf{\Omega} \mathbf{U}_{n}\mathbf{\Omega}^{-1}\,,
\end{equation}
where $\hat{\mathbf{U}}_{n}$ has the limit
\begin{equation}\label{2.26}
\hat{\mathbf{U}}_{\pm}=\lim_{n\to \pm\infty}\hat{\mathbf{U}}_{n}=\left (
\begin{array}{cc}
z &\frac{q_{\pm}}{\lambda}\\
\lambda\,r_{\pm} &\frac{1}{z}\\
\end{array}
\right)
\end{equation}
and it is easy to derive that
\begin{equation}
\hat{\mathbf{E}}_{\pm}=\left (
\begin{array}{cc}
q_{\pm} &\frac{r\lambda-z}{\lambda}\\
\lambda(r\lambda-z) &-r_{\pm}\\
\end{array}
\right)
\nonumber
\end{equation}
is the eigenvectors of the matrix $\hat{\mathbf{U}}_{\pm}$ corresponding to eigenvalues $r\lambda$, $r\lambda^{-1}$ for each column. According to $\hat{\mathbf{E}}_{\pm}$ we can define
\begin{equation}
\hat{\mathbf{E}}_{n}=\left (
\begin{array}{cc}
\dot{q}_{n} &\frac{r\lambda-z}{\lambda}\\
\lambda(r\lambda-z) &-\dot{r}_{n}\\
\end{array}
\right),
\nonumber
\end{equation}
the modified potentials $\dot{q}_{n}$, $\dot{r}_{n}=\dot{q}_{-n}^{*}$ satisfy the boundary conditions ${\lim_{n\to \pm \infty}}\dot{q}_{n}=q_{\pm},\ {\lim_{n\to \pm \infty}}\dot{r}_{n}=r_{\pm}$
and the constraint $\dot{q}_{n}\dot{r}_{n}=q_{0}^{2}\,,\,n\in\mathbb{Z}$.

We define the second modification as
\begin{equation}\label{2.27}
\dot{\mathbf{Y}}_{n}=\hat{\mathbf{E}}_{n}^{-1} \hat{\mathbf{Y}}_{n}\,,
\end{equation}
where $\dot{\mathbf{Y}}_{n}$ is also a $2\times2$ matrix with the boundary condition $\dot{\mathbf{Y}}_{n}\sim r^{n}\mathbf{\Lambda}^{n}, n\to \pm \infty$ and meets the  following scattering problem
\begin{equation}\label{2.28}
\dot{\mathbf{Y}}_{n+1}=\dot{\mathbf{U}}_{n}\dot{\mathbf{Y}}_{n},
\dot{\mathbf{U}}_{n}=\hat{\mathbf{E}}_{n+1}^{-1}\hat{\mathbf{U}}_{n}\hat{\mathbf{E}}_{n}\,.
\end{equation}
In fact, $\dot{\mathbf{U}}_{n}$ can be decomposed into two parts:
\begin{equation}\label{2.29}
\dot{\mathbf{U}}_{n}=\left (
\begin{array}{cc}
r\lambda &0\\
0 &\frac{r}{\lambda}\\
\end{array}
\right)+\dot{\mathbf{U}}'_{n}\,,
\end{equation}
where
\begin{align}\label{2.30}
&\dot{\mathbf{U}}'_{n}=\frac{1}{q_{0}^{2}+(r\lambda-z)^{2}}\nonumber\\
&\left(
\begin{array}{cc}
(r\lambda-z)(\eta_{n}^{[1]}+\eta_{-n}^{[1]*})-(r\lambda q_{n}-z \eta_{n}^{[2]})\eta_{-(n+1)}^{[3]*} &\frac{(r\lambda-z)^{2}\eta_{-n}^{[2]*}-\dot{r}_{n}\eta_{-n}^{[1]*}-(z(r\lambda-z)-q_{n}\dot{r}_{n})\eta_{-(n+1)}^{[3]*}}{\lambda}\\
\lambda\left((r\lambda-z)^{2}\eta_{n}^{[2]}-\dot{q}_{n}\eta_{n}^{[1]}-(z^{-1}(r\lambda-z)+\dot{q}_{n}r_{n})\eta_{n}^{[3]}\right) &-(r\lambda-z)(\eta_{n}^{[1]}+\eta_{-n}^{[1]*})+(r\lambda^{-1} r_{n}-z^{-1}\eta_{-n}^{[2]*})\eta_{n}^{[3]}\\
\end{array}
\right)
\end{align}
with $\eta_{n}^{[1]}=\dot{q}_{n}r_{n}-q_{0}^{2}$, $\eta_{n}^{[2]}=q_{n}-\dot{q}_{n}$, $\eta_{n}^{[3]}=\dot{q}_{n+1}-\dot{q}_{n}$ all decaying as $n\rightarrow\pm\infty$ so that $\dot{\mathbf{U}}'_{n}$ decays as $n\rightarrow\pm\infty$.

We introduce the $2\times1$ new eigenfunction to remove the $n$ power coefficient in the boundary condition as the third modification:
\begin{equation}\label{2.31}
\dot{M}_{n}=r^{-n}\lambda^{-n}\dot{\mathbf{Y}}_{n,1}\,,
\end{equation}
where subscript `1' of $\dot{\mathbf{Y}}_{n,1}$ is the first column of the matrix so that we can get the boundary condition
\begin{equation}\label{2.32}
\dot{M}_{n}\sim\left (
\begin{array}{cc}
1\\
0\\
\end{array}
\right)\,,\,n\rightarrow -\infty\,.
\end{equation}
Under the above three modifications, $\dot{M}_{n}$ and $M_{n}$ have the following relationship:
\begin{equation}
M_{n}=\hat{\mathbf{E}}_{n}\,\dot{M}_{n}\,\nonumber
\end{equation}
so that $M_{n}$ has the same analytic region with $\dot{M}_{n}$.

According to Eqs.~(\ref{2.28}) and (\ref{2.31}), we obtain the difference equation
\begin{equation}\label{2.33}
\dot{M}_{n+1}(\zeta)=\left (
\begin{array}{cc}
1 &0\\
0 &\frac{r\zeta-1}{\zeta(\zeta-r)}\\
\end{array}
\right)\dot{M}_{n}(\zeta)+\frac{1}{r\lambda}\dot{\mathbf{U}}'_{n}(\zeta)\,\dot{M}_{n}(\zeta)\,,
\end{equation}
where $\frac{1}{r\lambda}\dot{\mathbf{U}}'_{n}(\zeta)\rightarrow0$ as $n\rightarrow\pm\infty$. Furthermore, for $\frac{1}{r\lambda}\dot{\mathbf{U}}'_{n}(\zeta)$, the coefficients of potentials in its entries can be denoted by $\zeta$ and all the entries are bounded for $\zeta$ except $\zeta_{0}$ and $\zeta_{0}^{\ast}$ so that we can set a matrix $\mathcal{U}_{n}$ which is $\zeta$-independent such that $\|\frac{1}{r\lambda}\dot{\mathbf{U}}'_{n}(\zeta)\|\leq\|\mathcal{U}_{n}\|$ for all $\zeta\neq\zeta_{0},\zeta_{0}^{\ast},\infty$.

We can set the solution of Eq.~(\ref{2.33}) meeting the following summation equation:
\begin{equation}\label{2.34}
\dot{M}_{n}(\zeta)=\left(
\begin{array}{cc}
1\\
0\\
\end{array}
\right)+\sum\limits _{k=-\infty}^{+\infty}\mathbf{\dot{G}}_{n-k}\,\frac{1}{r\lambda}\dot{\mathbf{U}}'_{k}(\zeta)\,\dot{M}_{k}(\zeta)\,,
\end{equation}
where the Green's function corresponding to $\dot{M}_{n}(\zeta)$ is
\begin{equation}\label{2.35}
\mathbf{\dot{G}}_{n}=\theta(n-1)\left (
\begin{array}{cc}
1 &0\\
0 &(\frac{r\zeta-1}{\zeta(\zeta-r)})^{n-1}\\
\end{array}
\right)
\end{equation}
so that $\dot{M}_{n}(\zeta)$ can be expressed as
\begin{equation}\label{2.36}
\dot{M}_{n}(\zeta)=\left (
\begin{array}{cc}
1\\
0\\
\end{array}
\right)+\sum\limits _{k=-\infty}^{n-1}\frac{1}{r\lambda}\left (
\begin{array}{cc}
1 &0\\
0 &(\frac{r\zeta-1}{\zeta(\zeta-r)})^{n-1-k}\\
\end{array}
\right)
\,\dot{\mathbf{U}}'_{k}(\zeta)\,\dot{M}_{k}(\zeta)\,.
\end{equation}
The solution of Eq.~(\ref{2.36}) can be written as a Neumann series
\begin{equation}\label{2.37}
\dot{M}_{n}(\zeta)=\sum\limits _{j=0}^{\infty}C_{n}^{j}(\zeta)\,,
\end{equation}
where
\begin{equation}
C_{n}^{0}(\zeta)=\left (
\begin{array}{cc}
1\\
0\\
\end{array}
\right)\,,\,C_{n}^{j+1}(\zeta)=\sum\limits _{k=-\infty}^{n-1}\frac{1}{r\lambda}\left (
\begin{array}{cc}
1 &0\\
0 &(\frac{r\zeta-1}{\zeta(\zeta-r)})^{n-1-k}\\
\end{array}
\right)
\,\dot{\mathbf{U}}'_{k}(\zeta)\,C_{k}^{j}(\zeta).\nonumber
\end{equation}

In fact, we have the analogous lemma:
\begin{lemma}\label{L1}
Under the condition the potentials $q_{n}, r_{n}, \dot{q}_{n}, \dot{r}_{n}\in \left\{ f_{n}: \sum\limits _{l=\pm\infty}^{n}|f_{l}-\lim_{k\to \pm \infty}f_{k}|<\infty, \forall\,l, n\in\mathbb{Z}\right\}$ and the fact $\eta_{n}^{[j]}$, $j=1,2,3$ are summable, $\dot{M}_{n}(\zeta)$ defined by the series~(\ref{2.37}) exists and is the unique solution of Eq.~(\ref{2.36}).
\end{lemma}

$\mathbf{Proof}$:
\noindent\textbf{A. The existence of series (\ref{2.37})}
\\\hspace*{\parindent}

We will first prove the series representation~(\ref{2.37}) converges absolutely and uniformly in $n$ and uniformly in $\zeta$ in the region $|\zeta|\geq1$ by induction on $j$:
\begin{equation}\label{2.38}
\|C_{n}^{j}(\zeta)\|\leq\frac{1}{j!}\left(\sum\limits _{k=-\infty}^{n-1}\|\frac{1}{r\lambda}\dot{\mathbf{U}}'_{k}(\zeta)\|\right)^{j}\leq\frac{1}{j!}
\left(\sum\limits _{k=-\infty}^{n-1}\|\mathcal{U}_{k}\|\right)^{j}\,.
\end{equation}
For $j=0$,
\begin{equation}
\|C_{n}^{0}(\zeta)\|\leq1\leq\frac{1}{0!}
\left(\sum\limits _{k=-\infty}^{n-1}\|\mathcal{U}_{k}\|\right)^{0}\,;\nonumber
\end{equation}
For $j=1$,
\begin{equation}
\|C_{n}^{1}(\zeta)\|\leq\sum\limits _{k=-\infty}^{n-1}\|\left(
\begin{array}{cc}
1 &0\\
0 &(\frac{r\zeta-1}{\zeta(\zeta-r)})^{n-1-k}\\
\end{array}
\right)\|\|\frac{1}{r\lambda}\dot{\mathbf{U}}'_{k}(\zeta)\|\|C_{n}^{0}(\zeta)\|
\leq\frac{1}{1!}
\left(\sum\limits _{k=-\infty}^{n-1}\|\mathcal{U}_{k}\|\right)^{1}\,;\nonumber
\end{equation}
Assuming that Eq.~(\ref{2.38}) holds for $j$, then for $j+1$, we have $|\zeta|\geq1$ if and only if $\left|\frac{\zeta(\zeta-r)}{r\zeta-1}\right|\geq1$ so that
\begin{align}
\|C_{n}^{j+1}(\zeta)\|&\leq\sum\limits _{k=-\infty}^{n-1}\|\left(
\begin{array}{cc}
1 &0\\
0 &(\frac{r\zeta-1}{\zeta(\zeta-r)})^{n-1-k}\\
\end{array}
\right)\|\|\frac{1}{r\lambda}\dot{\mathbf{U}}'_{k}(\zeta)\|\|C_{n}^{j}(\zeta)\|\nonumber\\
&\leq\sum\limits_{k=-\infty}^{n-1}\|\mathcal{U}_{k}\|
\frac{1}{j!}
\left(\sum\limits _{l=-\infty}^{n-1}\|\mathcal{U}_{l}\|\right)^{j}\nonumber\\
&\leq\frac{1}{(j+1)!}\left(\sum\limits _{k=-\infty}^{n-1}\|\mathcal{U}_{k}\|\right)^{j+1}\,.\nonumber
\end{align}
If the potentials $q_{n}, r_{n}, \dot{q}_{n}, \dot{r}_{n}\in \left\{ f_{n}: \sum\limits _{l=\pm\infty}^{n}|f_{l}-\lim_{k\to \pm \infty}f_{k}|<\infty, \forall n\in\mathbb{Z}\right\}$ in $\dot{U}'_{n}(\zeta)$, we can prove $\sum\limits_{k=-\infty}^{+\infty}|\eta_{k}^{[j]}|<+\infty$, $j=1,2,3$. For example,
\begin{align}
|\eta_{k}^{[3]}|=|\dot{q}_{k+1}-\dot{q}_{k}|\leq|\dot{q}_{k+1}-q_{\mp}|+|\dot{q}_{k}-q_{\mp}|=\{\begin{array}{cc}
|\dot{q}_{k+1}-q_{-}|+|\dot{q}_{k}-q_{-}|\,,as\ j\leq n\\
|\dot{q}_{k+1}-q_{+}|+|\dot{q}_{k}-q_{+}|\,,as\ j\geq n\\
\end{array}\,,
\nonumber
\end{align}
so that
\begin{align}
\sum\limits_{k=-\infty}^{+\infty}|\eta_{k}^{[3]}|\leq\sum\limits_{k=-\infty}^{n}(|\dot{q}_{k+1}-\dot{q}_{-}|+|\dot{q}_{k}-\dot{q}_{-}|)+\sum\limits_{k=n+1}^{+\infty}(|\dot{q}_{k+1}-q_{+}|+|\dot{q}_{k}-q_{+}|)<+\infty\,.\nonumber
\end{align}
The above means that $\sum\limits_{k=-\infty}^{+\infty}\|\mathcal{U}_{k}\|<+\infty$. Therefore, For $n\leq N\in\mathbb{Z}^{+}$, $\sum\limits_{k=-\infty}^{n-1}\|\mathcal{U}_{k}\|\leq\sum\limits_{k=-\infty}^{N-1}\|\mathcal{U}_{k}\|\doteq\sigma$, where $\sigma$ is a constant independent of $n$. Then $\|C_{n}^{j}(\zeta)\|\leq\frac{\sigma^{j}}{j!}$, we can obtain
\begin{equation}
\|\sum\limits _{j=0}^{\infty}C_{n}^{j}(\zeta)\|\leq\sum\limits _{j=0}^{\infty}\|C_{n}^{j}(\zeta)\|\leq\sum\limits _{j=0}^{\infty}\frac{\sigma^{j}}{j!}=e^{\sigma}\,,\nonumber
\end{equation}
namely, the series~(\ref{2.37}) converges absolutely and uniformly.

According to Eq.~(\ref{2.37}) and $C_{n}^{j}(\zeta)$, we have
\begin{align}
\sum\limits _{j=0}^{\infty}C_{n}^{j}(\zeta)&=C_{n}^{0}(\zeta)+\sum\limits _{j=1}^{\infty}C_{n}^{j}(\zeta)=C_{n}^{0}(\zeta)+\sum\limits _{j=1}^{\infty}\sum\limits _{k=-\infty}^{n-1}\frac{1}{r\lambda}\left (
\begin{array}{cc}
1 &0\\
0 &(\frac{r\zeta-1}{\zeta(\zeta-r)})^{n-1-k}\\
\end{array}
\right)
\,\dot{\mathbf{U}}'_{k}(\zeta)\,C_{k}^{j-1}(\zeta)\nonumber\\
&=C_{n}^{0}(\zeta)+\sum\limits _{k=-\infty}^{n-1}\frac{1}{r\lambda}\left (
\begin{array}{cc}
1 &0\\
0 &(\frac{r\zeta-1}{\zeta(\zeta-r)})^{n-1-k}\\
\end{array}
\right)
\,\dot{\mathbf{U}}'_{k}(\zeta)\,\sum\limits _{j=0}^{\infty}C_{k}^{j}(\zeta)\nonumber\\
&=\left (
\begin{array}{cc}
1\\
0\\
\end{array}
\right)+\sum\limits _{k=-\infty}^{n-1}\frac{1}{r\lambda}\left (
\begin{array}{cc}
1 &0\\
0 &(\frac{r\zeta-1}{\zeta(\zeta-r)})^{n-1-k}\\
\end{array}
\right)
\,\dot{\mathbf{U}}'_{k}(\zeta)\,\sum\limits _{j=0}^{\infty}C_{k}^{j}(\zeta)\,,\nonumber
\end{align}
therefore $\sum\limits _{j=0}^{\infty}C_{n}^{j}(\zeta)$ is the solution of Eq.~(\ref{2.36}).

\vspace{5mm} \noindent\textbf{B. The uniqueness of solution of Eq.~(\ref{2.36})}
\\\hspace*{\parindent}

Assuming $\sum\limits _{j=0}^{\infty}C_{n,1}^{j}(\zeta)$ and $\sum\limits _{j=0}^{\infty}C_{n,2}^{j}(\zeta)$ are both the solutions of Eq.~(\ref{2.36}) so that their difference $C_{n,-}(\zeta)=\sum\limits _{j=0}^{\infty}\left(C_{n,1}^{j}(\zeta)-C_{n,2}^{j}(\zeta)\right)$ is such that
\begin{align}
C_{n,-}(\zeta)=\sum\limits _{k=-\infty}^{n-1}
\frac{1}{r\lambda}\left (
\begin{array}{cc}
1 &0\\
0 &(\frac{r\zeta-1}{\zeta(\zeta-r)})^{n-1-k}\\
\end{array}
\right)
\,\dot{\mathbf{U}}'_{k}(\zeta)C_{k,-}(\zeta)\,,\nonumber
\end{align}
therefore in $D_{-}$,
\begin{align}
\|C_{n,-}(\zeta)\|\leq\sum\limits _{k=-\infty}^{n-1}
\|\frac{1}{r\lambda}\left (
\begin{array}{cc}
1 &0\\
0 &(\frac{r\zeta-1}{\zeta(\zeta-r)})^{n-1-k}\\
\end{array}
\right)
\,\dot{\mathbf{U}}'_{k}(\zeta)\|\|C_{k,-}(\zeta)\|\leq\sum\limits _{k=-\infty}^{n-1}\|\mathcal{U}_{k}\|\|C_{k,-}(\zeta)\|\,.\nonumber
\end{align}
If the potentials $q_{n}, r_{n}, \dot{q}_{n}, \dot{r}_{n}\in \left\{ f_{n}: \sum\limits _{l=\pm\infty}^{n}|f_{l}-\lim_{k\to \pm \infty}f_{k}|<\infty, \forall n\in\mathbb{Z}\right\}$ in $\dot{U}'_{n}(\zeta)$, we can set $\sum\limits_{k=-\infty}^{+\infty}\|\mathcal{U}_{k}\|<+\infty$. According to Dispersed Bellman Inequality, we obtain $C_{n,-}(\zeta)\equiv0$. So far, we have proved the uniqueness of solution of Eq.~(\ref{2.36}) and completed the proof of $\mathbf{Lemma~\ref{L1}}$.

Furthermore, because of the analyticity of $C_{n}^{0}(\zeta)$ and $\frac{1}{r\lambda}\,diag(1,(\frac{r\zeta-1}{\zeta(\zeta-r)})^{n-1-k})\,\dot{\mathbf{U}}'_{k}(\zeta)$ in $D_{-}$, $C_{n}^{j}(\zeta)$ are analytic in $D_{-}$ for all $j$, which indicates that $\dot{M}_{n}(\zeta)=\sum\limits_{j=0}^{\infty}C_{n}^{j}(\zeta)$ is analytic in $D_{-}$ so that $M_{n}$ is analytic in $D_{-}$ except for the only possible pole $\zeta=\infty$. Similarly, we can prove the analyticity for other modified eigenfunctions and scattering coefficients~(\ref{2.23}), then we conclude the following theorem:
\begin{theorem}\label{T1}
Under the condition of the potentials $q_{n}, r_{n}, \dot{q}_{n}, \dot{r}_{n}\in \left\{ f_{n}: \sum\limits _{l=\pm\infty}^{n}|f_{l}-\lim_{k\to \pm \infty}f_{k}|<\infty, \forall n\in\mathbb{Z}\right\}$, $M_{n}$, $N_{n}$, $t_{11}(\zeta)=\tilde{t}_{11}(\zeta)$ are analytic in $D_{-}$ (except for the only possible pole $\zeta=\infty$) and $\bar{M}_{n}$, $\bar{N}_{n}$, $t_{22}(\zeta)=\tilde{t}_{22}(\zeta)$ are analytic in $D_{+}$ (except for the only possible pole $\zeta=0$).
\end{theorem}

\vspace{5mm} \noindent\textbf{2.1.3 Asymptotic behavior of modified eigenfunctions and scattering coefficients}
\\\hspace*{\parindent}

According the definition of uniformization variable $\zeta$, when $z\rightarrow0$, $\lambda\rightarrow0$, we have $\zeta\rightarrow r$; When $z\rightarrow\infty$, $\lambda\rightarrow\infty$, we have $\zeta\rightarrow \frac{1}{r}$ so that we should also consider the asymptotic behavior when $\zeta\rightarrow r, \frac{1}{r}$. By direct calculations, the asymptotic behavior of $z^{2}, \lambda^{2}$ and $z\lambda$ can be obtained shown in Table~\ref{tab:1}.
\begin{table}
\begin{center}
\renewcommand\arraystretch{2.25}
\setlength{\tabcolsep}{10mm}
\caption{The asymptotic behavior of $z^{2}$, $\lambda^{2}$ and $z\lambda$}
\label{tab:1}
\vspace{3mm}
        \begin{tabular}{|c|c|c|c|c|}
        \hline
        \  &$\zeta\rightarrow\infty$  &$\zeta\rightarrow0$ & $\zeta\rightarrow\frac{1}{r}$ & $\zeta\rightarrow r$\\
        \hline
        $z^{2}=\frac{\zeta-r}{\zeta(\zeta r-1)}$ &$\frac{1}{r\zeta}$ &$\frac{r}{\zeta}$ &$\frac{q_{0}^{2}}{r\zeta-1}$ &$-\frac{\zeta-r}{rq_{0}^{2}}$  \\
        \hline
        $\lambda^{2}=\zeta\frac{\zeta-r}{\zeta r-1}$ &$\frac{\zeta}{r}$ &$r\zeta$ &$\frac{q_{0}^{2}}{r^{2}(r\zeta-1)}$ &$-\frac{r(\zeta-r)}{q_{0}^{2}}$ \\
        \hline
        $z\lambda=\frac{\zeta-r}{\zeta r-1}$ &$\frac{1}{r}$ &$r$ &$\frac{q_{0}^{2}}{r(r\zeta-1)}$ &$-\frac{\zeta-r}{q_{0}^{2}}$ \\
        \hline
        \end{tabular}
        \end{center}
            \end{table}

Then, we investigate the asymptotic behavior of modified eigenfunctions and scattering coefficients. It is easy to obtain the Laurent series expansion as $\zeta\rightarrow\infty$ of $\frac{1}{\lambda^{2}}, \frac{1}{z\lambda}$:
\begin{equation}\label{2.39}
\frac{1}{\lambda^{2}}=\frac{r}{\zeta}-q_{0}^{2}\sum\limits _{j=0}^{\infty}\frac{r^{j}}{\zeta^{j+2}}\,,\,\frac{1}{z\lambda}=r-q_{0}^{2}\sum\limits _{j=0}^{\infty}\frac{r^{j}}{\zeta^{j+1}}\,.
\end{equation}
Setting the Laurent series expansion as $\zeta\rightarrow\infty$ of the modified eigenfunction $M_{n}(\zeta)$:
\begin{equation}\label{2.40}
M^{(1)}_{n}=\sum\limits _{k=0}^{\infty}\frac{M^{(1),k}_{n}}{\zeta^{k}}\,,\,M^{(2)}_{n}=\sum\limits _{k=-1}^{\infty}\frac{M^{(2),k}_{n}}{\zeta^{k}}\,,\,
\end{equation}
where the constants $M_{n}^{(1),k}$, $M_{n}^{(2),k}$ are unknown and substituting it into Eq.~(\ref{2.16}), we have
\begin{subequations}\label{2.41}
\begin{align}
r\,\sum\limits _{k=0}^{\infty}\frac{M^{(1),k}_{n+1}}{\zeta^{k}}&=\sum\limits _{k=1}^{\infty}\frac{M^{(1),k-1}_{n}}{\zeta^{k}}+r\,q_{n}\sum\limits _{k=0}^{\infty}\frac{M^{(2),k-1}_{n}}{\zeta^{k}}-q_{0}^{2}\,q_{n}\sum\limits _{k=-1}^{\infty}M^{(2),k}_{n}\sum\limits _{j=0}^{\infty}\frac{r^{j}}{\zeta^{k+j+2}}\,,\\
r\,\sum\limits _{k=-1}^{\infty}\frac{M^{(2),k}_{n+1}}{\zeta^{k}}&=r_{n}\sum\limits _{k=0}^{\infty}\frac{M^{(1),k}_{n}}{\zeta^{k}}+r\sum\limits _{k=-1}^{\infty}\frac{M^{(2),k}_{n}}{\zeta^{k}}-q_{0}^{2}\sum\limits _{k=-1}^{\infty}M^{(2),k}_{n}\sum\limits _{j=0}^{\infty}\frac{r^{j}}{\zeta^{k+j+1}}.
\end{align}
\end{subequations}
Combining with the boundary condition of $M_{n}(\zeta)$ to calculate the coefficients of each powers of $\zeta$, we obtain the asymptotic behavior of $M_{n}(\zeta)$ as $\zeta\rightarrow\infty$:
\begin{equation}\label{2.42}
M_{n}(\zeta)=\left(
\begin{array}{cc}
q_{n-1}+\mathcal{O}(\zeta^{-1})\\
\zeta+\mathcal{O}(1)\\
\end{array}
\right)\,,\zeta\rightarrow\infty\,.
\end{equation}

In the same way, we get the asymptotic behavior of $N_{n}(\zeta)$ as $\zeta\rightarrow\infty$:
\begin{equation}\label{2.43}
N_{n}(\zeta)=\frac{1}{\Theta_{n}}\,\left(
\begin{array}{cc}
r+\mathcal{O}(\zeta^{-1})\\
-r_{n}+\mathcal{O}(\zeta^{-1})\\
\end{array}
\right)\,,\,\zeta\rightarrow\infty\,,
\end{equation}
the asymptotic behavior of
$M_{n}(\zeta)$, $N_{n}(\zeta)$ as $\zeta\rightarrow \frac{1}{r}$:
\begin{equation}\label{2.44}
M_{n}(\zeta)=q_{-}\left(
\begin{array}{cc}
1+\mathcal{O}(\zeta-\frac{1}{r})\\
\frac{r_{n-1}}{r}+\mathcal{O}(\zeta-\frac{1}{r})\\
\end{array}
\right)\,,\,N_{n}(\zeta)=-\frac{r_{+}}{\Theta_{n}}\left(
\begin{array}{cc}
\mathcal{O}(\zeta-\frac{1}{r})\\
1+\mathcal{O}(\zeta-\frac{1}{r})\\
\end{array}
\right)\,,\,\zeta\rightarrow \frac{1}{r}\,,
\end{equation}
and the asymptotic behavior of $\bar{M}_{n}(\zeta)$, $\bar{N}_{n}(\zeta)$ as $\zeta\rightarrow 0$, $\zeta\rightarrow r$, respectively:
\begin{subequations}\label{2.45}
\begin{align}
\bar{M}_{n}(\zeta)&=\left(
\begin{array}{cc}
-\frac{1}{\zeta}+\mathcal{O}(1)\\
-r_{n-1}+\mathcal{O}(\zeta)\\
\end{array}
\right)\,,\,\bar{N}_{n}(\zeta)=\frac{1}{\Theta_{n}}\,\left(
\begin{array}{cc}
q_{n}+\mathcal{O}(\zeta)\\
-r+\mathcal{O}(\zeta)\\
\end{array}
\right)\,,\,\zeta\rightarrow 0\,,\\
\bar{M}_{n}(\zeta)&=-r_{-}\left(
\begin{array}{cc}
\frac{q_{n-1}}{r}+\mathcal{O}(\zeta-r)\\
1+\mathcal{O}(\zeta-r)\\
\end{array}
\right)\,,\,\bar{N}_{n}(\zeta)=\frac{q_{+}}{\Theta_{n}}\,\left(
\begin{array}{cc}
1+\mathcal{O}(\zeta-r)\\
\mathcal{O}(\zeta-r)\\
\end{array}
\right)\,,\,\zeta\rightarrow r.
\end{align}
\end{subequations}

According to Eq.~(\ref{2.23}a), we can also get the asymptotic behavior of scattering coefficients $t_{11}(\zeta)$ and $t_{22}(\zeta)$:
\begin{subequations}\label{2.46}
\begin{align}
t_{11}(\zeta)&=1+\mathcal{O}(\frac{1}{\zeta})\,,\,\zeta\rightarrow \infty\,,\ \ \ \ \
t_{11}(\zeta)=1+\mathcal{O}(\zeta-\frac{1}{r})\,,\,\zeta\rightarrow \frac{1}{r}\,,\\
t_{22}(\zeta)&=1+\mathcal{O}(\zeta)\,,\,\zeta\rightarrow 0\,,\ \ \ \ \ \ \
t_{22}(\zeta)=1+\mathcal{O}(\zeta-r)\,,\,\zeta\rightarrow r\,.
\end{align}
\end{subequations}

\vspace{5mm} \noindent\textbf{2.1.4 Symmetries of scattering coefficients}
\\\hspace*{\parindent}

Based on the fact that the scattering problem~(\ref{1.5}a) is independent with $\lambda$, we know that $\phi_{n}(z,\frac{1}{\lambda})$ is the solution to Eq.~(\ref{1.5}a) as well. It is easy to prove that $\frac{r/\lambda-z}{- r_{-}}\bar{\phi}_{n}(z,\lambda)$ is also the solution to Eq.~(\ref{1.5}a) and has the same boundary condition with $\phi_{n}(z,\frac{1}{\lambda})$ as $n\rightarrow-\infty$. Therefore, we can obtain the following symmetries of eigenfunctions:
\begin{equation}\label{2.47}
\phi_{n}(z,\frac{1}{\lambda})=\frac{r/\lambda-z}{- r_{-}}\bar{\phi}_{n}(z,\lambda)\,.
\end{equation}
Under the same consideration, we obtain the another symmetries of eigenfunctions:
\begin{equation}\label{2.48}
\psi_{n}(z,\frac{1}{\lambda})=\frac{r/\lambda-z}{q_{+}}\bar{\psi}_{n}(z,\lambda)\,.
\end{equation}

According to Eqs.~(\ref{2.47}), (\ref{2.48}) and~(\ref{2.10}) and the fact when $z\mapsto z$ and $\lambda\mapsto\frac{1}{\lambda}$, $\zeta\mapsto\bar{\zeta}=\frac{r\zeta-1}{\zeta-r}$, we have the following symmetries of scattering coefficients:
\begin{equation}\label{2.49}
t_{11}(\zeta)=t_{22}(\bar{\zeta})\,,\,t_{21}(\zeta)=-\frac{q_{+}}{r_{-}}\,t_{12}(\bar{\zeta}).
\end{equation}
In addition, according to Eq.~(\ref{2.49}), we can obtain
\begin{equation}\label{2.50}
t_{11}'(\zeta)=\frac{q_{0}^{2}}{(\zeta-r)^{2}}t_{22}'(\bar{\zeta})\,.
\end{equation}
From Eq.~(\ref{2.49}), we also obtain that $\zeta_{j}$ is a simple zero of $t_{11}(\zeta)$ in $D_{-}$ if and only if $\bar{\zeta}_{j}=\frac{r\zeta_{j}-1}{\zeta_{j}-r}$ is a simple zero of $t_{22}(\zeta)$ in $D_{+}$, which indicates that discrete eigenvalues always appear in pairs. By the defination of complex parameters $b_{k}$, $\bar{b}_{k}$ before, the following symmetry of these two parameters is
\begin{equation}\label{2.51}
b_{k}=-\frac{r_{-}}{q_{+}}\bar{b}_{k}\,,
\end{equation}
this, as well as Eq.~(\ref{2.50}), will be used to calculate symmetry for norming constants that will be defined after.

In fact, the second symmetry can be analyzed by the modified scattering problems of modified eigenfunctions, which corresponds to that when $z\mapsto z^{*}$ and $\lambda\mapsto\frac{1}{\lambda^{*}}$, $\zeta\mapsto\bar{\zeta}^{*}=\frac{r\zeta^{*}-1}{\zeta^{*}-r}$.

Due to nonlocal in space, it is necessary to consider the ``backward" modified scattering problem of Eq.~(\ref{2.16}) as follows:
\begin{equation}\label{2.52}
\mathcal{V}_{n-1}(z)=\left(
\begin{array}{cc}
\lambda/(rz) &-q_{n}/r\\
-\lambda^{2}r_{n}/r &z\lambda/r\\
\end{array}
\right)\mathcal{V}_{n}(z)\,,
\end{equation}
where $\mathcal{V}_{n}(z)=v_{n}\mathcal{W}_{n+1}(z)$, $v_{n}=\prod_{k=-\infty}^{n}\frac{1+q^{2}_{0}}{1-q_{k}r_{k}}$.
According to $M_{n}(z)$ meeting difference equation~(\ref{2.16}), we can obtain that $M_{n}'(z)=(M_{-n}^{(2)*}(z^{*}),-M_{-n}^{(1)*}(z^{*}))^{T}$ is a solution of Eq.~(\ref{2.52}). In addition, $\bar{N}_{n}(z)$ is a solution of Eq.~(\ref{2.16}) so that $C_{0}\, v_{n}\bar{N}_{n+1}(z)$ is a solution of Eq.~(\ref{2.52}) with $n$-independent unknown variable $C_{0}$. By checking the boundary conditions of $M_{n}'(z)$ and $C_{0}\, v_{n}\bar{N}_{n+1}(z)$ as $n\to +\infty$, we have
\begin{equation}\label{2.53}
{\lim_{n\to +\infty}}M_{n}'^{(1)}(z)=-\frac{q_{0}^{2}}{\lambda/z-r}={\lim_{n\to +\infty}}C_{0}\, v_{n}\bar{N}_{n+1}^{(1)}(z)=\frac{C_{0}\,q_{+}}{\Theta_{-\infty}}\,.
\end{equation}
If Eq.~(\ref{2.53}) holds, we have $C_{0}=-\frac{r_{+}}{\lambda/z-r}\Theta_{-\infty}$, which can also be obtained by comparing the boundary conditions of $M_{n}'^{(2)}(z)$ and $C_{0}\, v_{n}\bar{N}_{n+1}^{(2)}(z)$. Thus, we have the following symmetry between the modified eigenfunctions of $\zeta$:
\begin{equation}\label{2.54}
\left(
\begin{array}{cc}
\bar{N}_{n+1}^{(1)}(\zeta)\\
\bar{N}_{n+1}^{(2)}(\zeta)\\
\end{array}
\right)=
-\frac{\zeta-r}{r_{+}\Theta_{-\infty}v_{n}}\left(
\begin{array}{cc}
M_{-n}^{(2)*}(\bar{\zeta}^{*})\\
-M_{-n}^{(1)*}(\bar{\zeta}^{*})\\
\end{array}
\right)\,.
\end{equation}

By the similar process as above, we can get that
$\bar{M}_{n}'(z)=(\bar{M}_{-n}^{(2)*}(z^{*}),-\bar{M}_{-n}^{(1)*}(z^{*}))^{T}$ is a solution of
\begin{equation}\label{2.55}
\mathcal{\tilde{V}}_{n-1}(z)=\left(
\begin{array}{cc}
(r\lambda z)^{-1} &-q_{n}/(r\lambda^{2})\\
-r_{n}/r &z/(\lambda r)\\
\end{array}
\right)\mathcal{\tilde{V}}_{n}(z)\,,
\end{equation}
where $\mathcal{\tilde{V}}_{n}(z)=v_{n}\mathcal{\tilde{W}}_{n+1}(z)$, $\mathcal{\tilde{W}}_{n}(z)$ meets difference equation~(\ref{2.17}). In addition, it is easy know that $\bar{C}_{0}\, v_{n}N_{n+1}(z)$ is a solution of Eq.~(\ref{2.55}) with $n$-independent unknown variable $\bar{C}_{0}$. By checking the boundary conditions of $\bar{M}_{n}'(z)$ and $\bar{C}_{0}\, v_{n}N_{n+1}(z)$ as $n\to +\infty$, we have
\begin{equation}\label{2.56}
{\lim_{n\to +\infty}}\bar{M}_{n}'^{(1)}(z)=-q_{+}={\lim_{n\to +\infty}}\bar{C}_{0}\, v_{n}N_{n+1}^{(1)}(z)=\frac{C_{0}\,(r-z/\lambda)}{\Theta_{-\infty}}\,,
\end{equation}
in which  $\bar{C}_{0}=\frac{q_{+}}{z/\lambda-r}\Theta_{-\infty}$ that can also be obtained by comparing the boundary conditions of $\bar{M}_{n}'^{(2)}(z)$ and $\bar{C}_{0}\, v_{n}N_{n+1}^{(2)}(z)$. Thus, we have the following symmetry be expressed as $\zeta$:
\begin{equation}\label{2.57}
\left(
\begin{array}{cc}
N_{n+1}^{(1)}(\zeta)\\
N_{n+1}^{(2)}(\zeta)\\
\end{array}
\right)=
\frac{1/\zeta-r}{q_{+}\Theta_{-\infty}v_{n}}\left(
\begin{array}{cc}
\bar{M}_{-n}^{(2)*}(\bar{\zeta}^{*})\\
-\bar{M}_{-n}^{(1)*}(\bar{\zeta}^{*})\\
\end{array}
\right)\,.
\end{equation}

According to Eqs.~(\ref{2.54}), (\ref{2.57}), (\ref{2.23}) and the fact when $z\mapsto z^{*}$ and $\lambda\mapsto\frac{1}{\lambda^{*}}$, $\zeta\mapsto\bar{\zeta}^{*}=\frac{r\zeta^{*}-1}{\zeta^{*}-r}$, we have the following symmetries of scattering coefficients:
\begin{equation}\label{2.58}
t_{11}(\zeta)=t_{22}^{*}\left(\bar{\zeta}^{*}\right)\,.
\end{equation}
From Eq.~(\ref{2.58}), we also obtain that $\zeta_{j}$ is a simple zero of $t_{11}(\zeta)$ in $D_{-}$ if and only if $\bar{\zeta}^{*}_{j}=\frac{r\zeta^{*}_{j}-1}{\zeta^{*}_{j}-r}$ is a simple zero of $t_{22}(\zeta)$ in $D_{+}$, which indicates that discrete eigenvalues always appear in pairs.

\vspace{5mm} \noindent\textbf{2.2 The inverse scattering problem}
\\\hspace*{\parindent}

We will accomplish this progress by looking for the RH problem and solve it to construct the reconstructed potentials.

\vspace{5mm}\noindent\textbf{2.2.1 The RH problem and reconstructed potential}
\\\hspace*{\parindent}

We rewrite Eqs.~(\ref{2.21}) as the following jump condition on $\left|\zeta\right|=1$
\begin{subequations}\label{2.59}
\begin{align}
\mu(\zeta)-\bar{N}_{n}(\zeta)&=\lambda(\zeta)^{-2n}\rho(\zeta)N_{n}(\zeta)\,,\\
\bar{\mu}(\zeta)-N_{n}(\zeta)&=\lambda(\zeta)^{2n}\bar{\rho}(\zeta)\bar{N}_{n}(\zeta)\,,
\end{align}
\end{subequations}
where the new functions are defined as $\mu(\zeta)=\frac{M_{n}(\zeta)}{t_{11}(\zeta)}$, $\bar{\mu}(\zeta)=\frac{\bar{M}_{n}(\zeta)}{t_{22}(\zeta)}$ and the reflection coefficients are $\rho(\zeta)=\frac{\tilde{t}_{21}(\zeta)}{t_{11}(\zeta)}$ and $\bar{\rho}(\zeta)=\frac{\tilde{t}_{12}(\zeta)}{t_{22}(\zeta)}$. In the respective analytic regions, since the functions $\mu(\zeta)$ and $\bar{\mu}(\zeta)$ are meromorphic at the zeros of $t_{11}(\zeta)$, $t_{22}(\zeta)$, respectively, besides that, $\mu(\zeta)$, $N_{n}(\zeta)$ and $\bar{\mu}(\zeta)$, $\bar{N}_{n}(\zeta)$ have the poles at $\zeta=\infty$, $\zeta=0$, respectively, we will discuss the asymptotic behavior and residue conditions of these functions to detail and solve the RH problem.

In Section 2.1, the asymptotic behavior of modified eigenfunctions and scattering coefficients is all analyzed so that we have
\begin{subequations}\label{2.60}
\begin{align}
&\mu_{n}(\zeta)=\left(
\begin{array}{cc}
q_{n-1}+\mathcal{O}(\zeta^{-1})\\
\zeta+\mathcal{O}(1)\\
\end{array}
\right)\,,\ \ \ \ \ \ \ \ \ \ \ \ \ \ \ N_{n}(\zeta)=\frac{1}{\Theta_{n}}\,\left(
\begin{array}{cc}
r+\mathcal{O}(\zeta^{-1})\\
-r_{n}+\mathcal{O}(\zeta^{-1})\\
\end{array}
\right)\,,\,\zeta\rightarrow\infty\,,\\
&\bar{\mu}_{n}(\zeta)=\left(
\begin{array}{cc}
-\frac{1}{\zeta}+\mathcal{O}(1)\\
-r_{n-1}+\mathcal{O}(\zeta)\\
\end{array}
\right)\,,\ \ \ \ \ \ \ \ \ \ \ \ \ \ \ \ \bar{N}_{n}(\zeta)=\frac{1}{\Theta_{n}}\,\left(
\begin{array}{cc}
q_{n}+\mathcal{O}(\zeta)\\
-r+\mathcal{O}(\zeta)\\
\end{array}
\right)\,,\,\zeta\rightarrow 0\,,\\
&\mu_{n}(\zeta)=q_{-}\left(
\begin{array}{cc}
1+\mathcal{O}(\zeta-\frac{1}{r})\\
\frac{r_{n-1}}{r}+\mathcal{O}(\zeta-\frac{1}{r})\\
\end{array}
\right)\,,\ \ \ \ \ \ \ \ \ N_{n}(\zeta)=-\frac{r_{+}}{\Theta_{n}}\left(
\begin{array}{cc}
\mathcal{O}(\zeta-\frac{1}{r})\\
1+\mathcal{O}(\zeta-\frac{1}{r})\\
\end{array}
\right)\,,\,\zeta\rightarrow \frac{1}{r}\,,\,\\
&\bar{\mu}_{n}(\zeta)=-r_{-}\left(
\begin{array}{cc}
\frac{q_{n-1}}{r}+\mathcal{O}(\zeta-r)\\
1+\mathcal{O}(\zeta-r)\\
\end{array}
\right)\,,\ \ \ \ \ \ \ \bar{N}_{n}(\zeta)=\frac{q_{+}}{\Theta_{n}}\,\left(
\begin{array}{cc}
1+\mathcal{O}(\zeta-r)\\
\mathcal{O}(\zeta-r)\\
\end{array}
\right)\,,\,\zeta\rightarrow r.
\end{align}
\end{subequations}

In addition, we can easily obtain the following residue conditions
\begin{subequations}\label{2.61}
\begin{align}
\underset{\zeta=\zeta_{j}}{\mathrm{Res}}\mu_{n}(\zeta)&=C_{j}\,\lambda(\zeta_{j})^{-2n}\,N_{n}(\zeta_{j})\,,\,C_{j}=\frac{\tilde{t}_{21}(\zeta_{j})}{t_{11}'(\zeta_{j})},\\
\underset{\zeta=\bar{\zeta}_{j}}{\mathrm{Res}}\bar{\mu}_{n}(\zeta)&=\bar{C}_{j}\,\lambda(\bar{\zeta}_{j})^{2n}\,\bar{N}_{n}(\bar{\zeta}_{j})\,,\,\bar{C}_{j}=\frac{ \tilde{t}_{12}(\bar{\zeta}_{j})}{t_{22}'(\bar{\zeta}_{j})}\,,
\end{align}
\end{subequations}
where we assume that $t_{11}(\zeta)$ has $J$ zeros, $t_{22}(\zeta)$ has $\bar{J}$ zeros, $\frac{1}{t_{11}^{'}(\zeta_{j})}$, $\frac{1}{t_{22}^{'}(\bar{\zeta}_{j})}$ are the first derivatives of $t_{11}(\zeta)$ and $t_{22}(\zeta)$ at $\zeta_{j}$ and $\bar{\zeta}_{j}$, respectively and $C_{j}$, $\bar{C}_{j}$ are the norming constants. In fact, we have $\tilde{t}_{21}(\zeta_{j})=\lambda(\zeta_{j})b_{k}$, $\tilde{t}_{12}(\bar{\zeta}_{j})=\lambda(\bar{\zeta}_{j})\bar{b}_{k}$. We also can get $C_{j}=-\frac{q_{+}^{2}}{(\bar{\zeta}_{j}-r)^{2}}\bar{C}_{j}$ according to Eqs.~(\ref{2.50}) and~(\ref{2.51}).

Dividing both sides of Eq.~(\ref{2.59}a) by $\zeta-r$, then subtracting the asymptotic behavior of $\frac{\mu(\zeta)}{\zeta-r}$ as $\zeta\rightarrow\infty$ and the effect of pole $\zeta_{j}$, we have
\begin{equation}\label{2.62}
\frac{\mu(\zeta)}{\zeta-r}-\left(
\begin{array}{cc}
0\\
1\\
\end{array}
\right)-\sum\limits _{j=1}^{J}\frac{\underset{\zeta=\zeta_{j}}{\mathrm{Res}}\mu_{n}(\zeta)}{(\zeta_{j}-r)(\zeta-\zeta_{j})}=\frac{\bar{N}_{n}(\zeta)}{\zeta-r}-\left(
\begin{array}{cc}
0\\
1\\
\end{array}
\right)-\sum\limits _{j=1}^{J}\frac{\underset{\zeta=\zeta_{j}}{\mathrm{Res}}\mu_{n}(\zeta)}{(\zeta_{j}-r)(\zeta-\zeta_{j})}+\frac{\lambda(\zeta)^{-2n}}{\zeta-r}\rho(\zeta)N_{n}(\zeta)\,.
\end{equation}
Considering the asymptotic behavior of the above left side as $\zeta\rightarrow\infty$ and combining with Eqs.~(\ref{2.60}d), (\ref{2.61}a), we can get
\begin{equation}\label{2.63}
\bar{N}_{n}(\zeta)=\left(
\begin{array}{cc}
\frac{q_{+}}{\Theta_{n}}\\
\zeta-r\\
\end{array}
\right)+(\zeta-r)\sum\limits _{j=1}^{J}\frac{C_{j}\,\lambda(\zeta_{j})^{-2n}}{(\zeta_{j}-r)(\zeta-\zeta_{j})}\,N_{n}(\zeta_{j})-
\frac{\zeta-r}{2\pi i}\oint_{|\omega|=1}\frac{\lambda(\omega)^{-2n}}{(\omega-\zeta)(\omega-r)}\rho(\omega)N_{n}(\omega)
d\omega\,,\zeta\in D_{+}\,.
\end{equation}

Dividing both sides of Eq.~(\ref{2.59}b) by $\zeta-\frac{1}{r}$ and dealing with the formula with the same technique as above, we obtain
\begin{equation}\label{2.64}
N_{n}(\zeta)=\left(
\begin{array}{cc}
r-\frac{1}{\zeta}\\
-\frac{r_{+}}{\Theta_{n}}\\
\end{array}
\right)+(\zeta-\frac{1}{r})\sum\limits _{j=1}^{\bar{J}}\frac{\bar{C}_{j}\,\lambda(\bar{\zeta}_{j})^{2n}}{(\bar{\zeta}_{j}-\frac{1}{r})(\zeta-\bar{\zeta}_{j})}\,\bar{N}_{n}(\bar{\zeta}_{j})+
\frac{\zeta-\frac{1}{r}}{2\pi i}\oint_{|\omega|=1}\frac{\lambda(\omega)^{2n}}{(\omega-\zeta)(\omega-\frac{1}{r})}\bar{\rho}(\omega)\bar{N}_{n}(\omega)
d\omega\,,\zeta\in D_{-}\,.
\end{equation}

Comparing the second component of Eq.~(\ref{2.63}) with $\zeta=0$ and asymptotic behavior of $\bar{N}_{n}(\zeta)$ as $\zeta\rightarrow0$ in Eq.~(\ref{2.60}b), we can obtain
\begin{equation}\label{2.65}
\frac{1}{\Theta_{n}}=1-\sum\limits _{j=1}^{J}\frac{C_{j}\,\lambda(\zeta_{j})^{-2n}}{\zeta_{j}(\zeta_{j}-r)}\,N^{(2)}_{n}(\zeta_{j})-
\frac{1}{2\pi i}\oint_{|\omega|=1}\frac{\lambda(\omega)^{-2n}}{\omega(\omega-r)}\rho(\omega)N^{(2)}_{n}(\omega)
d\omega\,,
\end{equation}
which is also an important equation to solve the reconstructed potential. Next, by comparing the first component of Eq.~(\ref{2.63}) with $\zeta=0$ and the asymptotic behavior of $\bar{N}_{n}(\zeta)$ as $\zeta\rightarrow0$ in Eq.~(\ref{2.60}b), the reconstructed potential will be constructed as
\begin{equation}\label{2.66}
q_{n}=q_{+}+\Theta_{n}\sum\limits _{j=1}^{J}\frac{r C_{j}\,\lambda(\zeta_{j})^{-2n}}{\zeta_{j}(\zeta_{j}-r)}\,N^{(1)}_{n}(\zeta_{j})+\frac{\Theta_{n}}{2\pi i}\oint_{|\omega|=1}\frac{r\lambda(\omega)^{-2n}}{\omega(\omega-r)}\rho(\omega)N^{(1)}_{n}(\omega)
d\omega
\end{equation}
with the items $\Theta_{n}$, $N^{(1)}_{n}(\zeta_{j})$ that will be solved later.

\vspace{5mm} \noindent\textbf{2.2.2 The reflectionless condition}
\\\hspace*{\parindent}

We can obtain the soliton solutions of Eq.~(\ref{1.3}) with the reflectionless condition $\rho(\zeta)=0$, $\bar{\rho}(\zeta)=0$ for $|\zeta|=1$, under which Eqs.~(\ref{2.63}),~(\ref{2.64}),~(\ref{2.65}),~(\ref{2.66}) reduce to the solvable algebraic set of equations:
\begin{subequations}\label{2.67}
\begin{align}
&\bar{N}^{(1)}_{n}(\zeta)-(\zeta-r)\sum\limits _{j=1}^{J}\frac{C_{j}\,\lambda(\zeta_{j})^{-2n}}{(\zeta_{j}-r)(\zeta-\zeta_{j})}\,N^{(1)}_{n}(\zeta_{j})-\frac{q_{+}}{\Theta_{n}}=0\,,\\
&\bar{N}^{(2)}_{n}(\zeta)-(\zeta-r)\sum\limits _{j=1}^{J}\frac{C_{j}\,\lambda(\zeta_{j})^{-2n}}{(\zeta_{j}-r)(\zeta-\zeta_{j})}\,N^{(2)}_{n}(\zeta_{j})=\zeta-r\,,\\
&N^{(1)}_{n}(\zeta)-(\zeta-\frac{1}{r})\sum\limits _{j=1}^{\bar{J}}\frac{\bar{C}_{j}\,\lambda(\bar{\zeta}_{j})^{2n}}{(\bar{\zeta}_{j}-\frac{1}{r})(\zeta-\bar{\zeta}_{j})}\,\bar{N}^{(1)}_{n}(\bar{\zeta}_{j})=r-\frac{1}{\zeta}\,,\\
&N^{(2)}_{n}(\zeta)+\frac{r_{+}}{\Theta_{n}}-(\zeta-\frac{1}{r})\sum\limits _{j=1}^{\bar{J}}\frac{\bar{C}_{j}\,\lambda(\bar{\zeta}_{j})^{2n}}{(\bar{\zeta}_{j}-\frac{1}{r})(\zeta-\bar{\zeta}_{j})}\,\bar{N}^{(2)}_{n}(\bar{\zeta}_{j})=0\,,\\
&\frac{1}{\Theta_{n}}=1-\sum\limits _{j=1}^{J}\frac{C_{j}\,\lambda(\zeta_{j})^{-2n}}{\zeta_{j}(\zeta_{j}-r)}\,N^{(2)}_{n}(\zeta_{j})
\end{align}
\end{subequations}
and the reconstructed potential
\begin{equation}\label{2.68}
q_{n}=q_{+}+\Theta_{n}\sum\limits _{j=1}^{J}\frac{r\,C_{j}\,\lambda(\zeta_{j})^{-2n}}{\zeta_{j}(\zeta_{j}-r)}\,N^{(1)}_{n}(\zeta_{j})\,.
\end{equation}
Substituting the scattering coefficients $\{\zeta_{j},\bar{\zeta}_{j},C_{j},\bar{C}_{j},j=1,2,\cdot\cdot\cdot,J\}$ into Eqs.~(\ref{2.67}), we will get the solvable algebraic set of equations and then get the potential~(\ref{2.68}).

\vspace{5mm} \noindent\textbf{2.2.3 The trace formula and discrete eigenvalues}
\\\hspace*{\parindent}

From Eqs.~(\ref{2.49}) and~(\ref{2.58}), we know that $\zeta_{j}$, $\zeta^{*}_{j}$ are simple zeros of $t_{11}(\zeta)$ in $D_{-}$ if and only if $\bar{\zeta}_{j}=\frac{r\zeta_{j}-1}{\zeta_{j}-r}$ and $\bar{\zeta}^{*}_{j}=\frac{r\zeta^{*}_{j}-1}{\zeta^{*}_{j}-r}$ are simple zeros of $t_{22}(\zeta)$ in $D_{+}$ so that the numbers of simple zeros for $t_{11}(\zeta)$ and $t_{22}(\zeta)$ are equal, namely $J=\bar{J}$ and the discrete eigenvalues are in quartets: $\{\zeta_{j}, \zeta^{*}_{j}, \bar{\zeta}_{j}, \bar{\zeta}^{*}_{j}\}$, where the first pair is in $D_{-}$ and the second pair is in $D_{+}$. In addition, the all $J$ simple zeros of $t_{11}(\zeta)$ can be divided into two classes denoted by $\zeta_{j}$, $j=1,2,\cdot\cdot\cdot,J_{1}$ and $\hat{\zeta}_{k}$, $k=1,2,\cdot\cdot\cdot,J_{2}$, where $2J_{1}+J_{2}=J$, Im$(\zeta_{j})\neq0$, Im$(\hat{\zeta}_{k})=0$, then we obtain that $t_{22}(\zeta)$ has $J$ simple zeros $\bar{\zeta}_{j}$, $\bar{\zeta}^{*}_{j}$, $\bar{\hat{\zeta}}_{k}=\frac{r\hat{\zeta}_{k}-1}{\hat{\zeta}_{k}-r}$.

From the front discussion, we know that $t_{11}(\zeta)$ and $t_{22}(\zeta)$ are analytic in $D_{-}$, $D_{+}$, respectively and therefore the new functions
\begin{equation}\label{2.69}
\begin{aligned}
A(\zeta)=\prod_{j=1}^{J_{1}}\frac{\zeta-\bar{\zeta}_{j}}{\zeta-\zeta_{j}}\frac{\zeta-\bar{\zeta}^{*}_{j}}{\zeta-\zeta^{*}_{j}}\prod_{k=1}^{J_{2}}\frac{\zeta-\bar{\hat{\zeta}}_{k}}{\zeta-\hat{\zeta}_{k}}t_{11}(\zeta)\,,\,
\bar{A}(\zeta)=\prod_{j=1}^{J_{1}}\frac{\zeta-\zeta_{j}}{\zeta-\bar{\zeta}_{j}}\frac{\zeta-\zeta^{*}_{j}}{\zeta-\bar{\zeta}^{*}_{j}}\prod_{k=1}^{J_{2}}\frac{\zeta-\hat{\zeta}_{k}}{\zeta-\bar{\hat{\zeta}}_{k}}t_{22}(\zeta)
\end{aligned}
\end{equation}
are analytic in $D_{-}$ and $D_{+}$, respectively but both have no zeros and $A(\zeta)\bar{A}(\zeta)=t_{11}(\zeta) t_{22}(\zeta)$.

According to the asymptotic behavior of $t_{11}(\zeta)$ as $\zeta\rightarrow\infty$, the following equation holds for $\zeta\in D_{+}$
\begin{equation}\label{2.70}
\frac{1}{2\pi i}\oint_{|\omega|=1}\frac{\ln (A(\omega))}{\omega-\zeta}d\omega=0.
\end{equation}
In addition, we have
\begin{equation}\label{2.71}
\ln (\bar{A}(\zeta))=\frac{1}{2\pi i}\oint_{|\omega|=1}\frac{\ln (\bar{A}(\omega))}{\omega-\zeta}d\omega\,,\,\zeta\in D_{+}
\end{equation}
so that
\begin{equation}\label{2.72}
\ln (\bar{A}(\zeta))=\frac{1}{2\pi i}\oint_{|\omega|=1}\frac{\ln (t_{11}(\omega)t_{22}(\omega))}{\omega-\zeta}d\omega\,,\,\zeta\in D_{+}.
\end{equation}

Combining with Eq.~(\ref{2.8}) and the definition of reflection coefficients, we obtain the trace formula
\begin{equation}\label{2.73}
t_{22}(\zeta)=\Theta_{-\infty}\prod_{j=1}^{J_{1}}\frac{\zeta-\bar{\zeta}_{j}}{\zeta-\zeta_{j}}\frac{\zeta-\bar{\zeta}^{*}_{j}}{\zeta-\zeta^{*}_{j}}\prod_{k=1}^{J_{2}}\frac{\zeta-\bar{\hat{\zeta}}_{k}}{\zeta-\hat{\zeta}_{k}} e^{-\frac{1}{2\pi i}\oint_{|\omega|=1}\frac{\ln \left(1-\rho(\omega)\bar{\rho}(\omega)\right)}{\omega-\zeta}d\omega}\,,\zeta\in D_{+}\,.
\end{equation}
In the same way, we can get another trace formula
\begin{equation}\label{2.74}
t_{11}(\zeta)=\prod_{j=1}^{J_{1}}\frac{\zeta-\zeta_{j}}{\zeta-\bar{\zeta}_{j}}\frac{\zeta-\zeta^{*}_{j}}{\zeta-\bar{\zeta}^{*}_{j}}\prod_{k=1}^{J_{2}}\frac{\zeta-\hat{\zeta}_{k}}{\zeta-\bar{\hat{\zeta}}_{k}} e^{\frac{1}{2\pi i}\oint_{|\omega|=1}\frac{\ln \left(1-\rho(\omega)\bar{\rho}(\omega)\right)}{\omega-\zeta}d\omega}\,,\zeta\in D_{-}\,.
\end{equation}
Under the reflectionless condition and taking the derivative of above trace formulae with respect to $\zeta$ gives
\begin{subequations}\label{2.75}
\begin{align}
&t_{22}^{'}(\zeta)=\Theta_{-\infty}\prod_{j=1}^{J_{1}}\frac{\zeta-\bar{\zeta}_{j}}{\zeta-\zeta_{j}}\frac{\zeta-\bar{\zeta}^{*}_{j}}{\zeta-\zeta^{*}_{j}}\prod_{k=1}^{J_{2}}\frac{\zeta-\bar{\hat{\zeta}}_{k}}{\zeta-\hat{\zeta}_{k}}\left[
\sum_{j=1}^{J_{1}}\left(\frac{1}{\zeta-\bar{\zeta}_{j}}+\frac{1}{\zeta-\bar{\zeta}^{*}_{j}}-\frac{1}{\zeta-\zeta_{j}}-\frac{1}{\zeta-\zeta^{*}_{j}}\right)+\sum_{k=1}^{J_{2}}\left(\frac{1}{\zeta-\bar{\hat{\zeta}}_{j}}-\frac{1}{\zeta-\hat{\zeta}_{k}}\right)\right] \,,\\
&t_{11}^{'}(\zeta)=\prod_{j=1}^{J_{1}}\frac{\zeta-\zeta_{j}}{\zeta-\bar{\zeta}_{j}}\frac{\zeta-\zeta^{*}_{j}}{\zeta-\bar{\zeta}^{*}_{j}}\prod_{k=1}^{J_{2}}\frac{\zeta-\hat{\zeta}_{k}}{\zeta-\bar{\hat{\zeta}}_{k}} \left[\sum_{j=1}^{J_{1}}\left(\frac{1}{\zeta-\zeta_{j}}+\frac{1}{\zeta-\zeta^{*}_{j}}-\frac{1}{\zeta-\bar{\zeta}_{j}}-\frac{1}{\zeta-\bar{\zeta}^{*}_{j}}\right)+\sum_{k=1}^{J_{2}}\left(\frac{1}{\zeta-\hat{\zeta}_{j}}-\frac{1}{\zeta-\bar{\hat{\zeta}}_{k}}\right)\right]\,.
\end{align}
\end{subequations}

We consider the discrete eigenvalues under the asymptotic behavior~(\ref{2.46}) of scattering coefficients $t_{11}(\zeta)$ and $t_{22}(\zeta)$. The trace formula of $t_{11}(\zeta)$ and $t_{22}(\zeta)$ meeting the following asymptotic behavior under the reflectionless condition:
\begin{align}
&{\lim_{\zeta\to 1/r}}t_{11}(\zeta)=\prod_{j=1}^{J_{1}}\frac{1/r-\zeta_{j}}{1/r-\bar{\zeta}_{j}}\frac{1/r-\zeta^{*}_{j}}{1/r-\bar{\zeta}^{*}_{j}}\prod_{k=1}^{J_{2}}\frac{1/r-\hat{\zeta}_{k}}{1/r-\bar{\hat{\zeta}}_{k}}=\prod_{j=1}^{J_{1}}\frac{|1/r-\zeta_{j}|^{2}}{|1/r-\bar{\zeta}_{j}|^{2}}\prod_{k=1}^{J_{2}}\frac{1/r-\hat{\zeta}_{k}}{1/r-\bar{\hat{\zeta}}_{k}}=1\,,\label{2.76}\\
&{\lim_{\zeta\to 0}}t_{22}(\zeta)=\Theta_{-\infty}\prod_{j=1}^{J_{1}}\frac{\bar{\zeta}_{j}}{\zeta_{j}}\frac{\bar{\zeta}^{*}_{j}}{\zeta^{*}_{j}}\prod_{k=1}^{J_{2}}\frac{\bar{\hat{\zeta}}_{k}}{\hat{\zeta}_{k}}=\Theta_{-\infty}\prod_{j=1}^{J_{1}}\frac{|\bar{\zeta}_{j}|^{2}}{|\zeta_{j}|^{2}}\prod_{k=1}^{J_{2}}\frac{\bar{\hat{\zeta}}_{k}}{\hat{\zeta}_{k}}=1\,,\label{2.77}\\
&{\lim_{\zeta\to r}}t_{22}(\zeta)=\Theta_{-\infty}\prod_{j=1}^{J_{1}}\frac{r-\bar{\zeta}_{j}}{r-\zeta_{j}}\frac{r-\bar{\zeta}^{*}_{j}}{r-\zeta^{*}_{j}}\prod_{k=1}^{J_{2}}\frac{r-\bar{\hat{\zeta}}_{k}}{r-\hat{\zeta}_{k}}=\Theta_{-\infty}\prod_{j=1}^{J_{1}}\frac{|r-\bar{\zeta}_{j}|^{2}}{|r-\zeta_{j}|^{2}}\prod_{k=1}^{J_{2}}\frac{r-\bar{\hat{\zeta}}_{k}}{r-\hat{\zeta}_{k}}=1\,.\label{2.78}
\end{align}

Setting $J=1$, then $J_{1}=0$, $J_{2}=1$, we get $\hat{\zeta}_{1}=r\pm iq_{0}$, which are complex leading to a contradiction. When $J=2$, there are two scenarios. The first scenario is $J_{1}=1$, $J_{2}=0$, we can get that $\bar{\zeta}_{1}, \bar{\zeta}^{*}_{1}\in D_{+}$ and $\zeta_{1}, \zeta^{*}_{1}\in D_{-}$ are on the circle $|\zeta-1/r|=q_{0}/r$, which can be used to construct the 2-eigenvalue soliton solution. The second scenario is $J_{1}=0$, $J_{2}=2$, then we can get $\bar{\hat{\zeta}}_{1}, \bar{\hat{\zeta}}_{2}=\frac{1}{\hat{\zeta}_{1}}\in D_{+}$, $\hat{\zeta}_{1}, \hat{\zeta}_{2}=\frac{1}{\bar{\hat{\zeta}}_{1}}\in D_{-}$. It should be emphasized that the complexity of obtaining discrete eigenvalues by the asymptotic behavior of $t_{22}(\zeta)$ lies in finding $\Theta_{-\infty}$. By using the fact that $|\lambda(\zeta_{1})|>1$ and $|\lambda(\bar{\zeta}_{1})|<1$, we can take the limit $n\rightarrow-\infty$ of $\Theta_{n}$ obtained by solving the linear system~(\ref{2.67}) with discrete eigenvalues under the reflectionless condition at $J=1$ and $J=2$, respectively, to obtain $\Theta_{-\infty}$.

When $J_{1}=1$, $J_{2}=0$, $\bar{\zeta}_{1}=\frac{1+q_{0}e^{i\eta_{1}}}{r}\in D_{+}$, $\eta_{1}\in\mathbb{R}$, $|\pi-\eta_{1}|<$ arctan$(r/q_{0})$, and the corresponding $\zeta_{1}=\frac{r}{1+q_{0}e^{-i\eta_{1}}}\in D_{-}$, $\bar{\zeta}_{2}=\bar{\zeta}^{*}_{1}$. According to the symmetry about eigenfunctions, we have $|\bar{b}_{1}\bar{b}_{2}|=\lambda(\bar{\zeta}_{1})^{-8}$ so that we can write $\bar{b}_{1}=\kappa_{1}e^{i\bar{\theta}_{1}}$, $\bar{b}_{2}=\frac{\lambda(\bar{\zeta}_{1})^{-4}}{\kappa_{1}}e^{i\bar{\theta}_{2}}$, where $\kappa_{1}$, $\bar{\theta}_{1}$, $\bar{\theta}_{2}$ are real parameters. When $J_{1}=0$, $J_{2}=2$, according to the symmetry about eigenfunctions and direct calculation, we have $|\bar{b}_{1}|^{2}<0$ leading to a contradiction so that we don't consider this scenario any more.

According to Eq.~(\ref{2.75}a), we also have $t_{22}^{'}(\bar{\zeta}_{1})=\frac{\bar{\zeta}_{1}-\bar{\zeta}^{*}_{1}}{(\bar{\zeta}_{1}-\zeta_{1})(\bar{\zeta}_{1}-\zeta^{*}_{1})}$, $t_{22}^{'}(\bar{\zeta}^{*}_{1})=\frac{\bar{\zeta}^{*}_{1}-\bar{\zeta}_{1}}{(\bar{\zeta}^{*}_{1}-\zeta_{1})(\bar{\zeta}^{*}_{1}-\zeta^{*}_{1})}$, therefore, we get
\begin{subequations}\label{2.79}
\begin{align}
&\bar{C}_{1}=\frac{\kappa_{1}(\lambda(\bar{\zeta}_{1}))^{3}e^{i\bar{\theta}_{1}}}{\bar{\zeta}_{1}-\bar{\zeta}^{*}_{1}}(\bar{\zeta}_{1}-\zeta_{1})(\bar{\zeta}_{1}-\zeta^{*}_{1})\,,\\ &\bar{C}_{2}=\frac{\kappa_{1}^{-1}(\lambda(\bar{\zeta}_{1}))^{-1}e^{i\bar{\theta}_{2}}}{\bar{\zeta}^{*}_{1}-\bar{\zeta}_{1}}(\bar{\zeta}^{*}_{1}-\zeta_{1})(\bar{\zeta}^{*}_{1}-\zeta^{*}_{1})\,.
\end{align}
\end{subequations}

\vspace{5mm} \noindent\textbf{2.3 The time evolution}
\\\hspace*{\parindent}

In this section, we will determine the time evolution of eigenfunctions and scattering coefficients. We denote the potentials as $q_{n}(t)$, $r_{n}(t)$ whose boundary conditions are given before in the forms
\begin{equation}
{\lim_{n\to \pm \infty}}q_{n}(t)=q_{\pm}(t)=q_{0}e^{i (\theta+2q_{0}^{2}t)}\,,\,{\lim_{n\to \pm\infty}}r_{n}(t)=r_{\pm}(t)=q_{\mp}^{*}(t)=q_{0}e^{-i (\theta+2q_{0}^{2}t)}\,.
\end{equation}

Next, we will discuss the time evolution of the eigenfunctions. According to the above nonzero boundary conditions, Eq.~(\ref{1.5}b) has the asymptotic form
\begin{equation}
\varphi_{n,t}=i\left (
\begin{array}{cc}
q_{\pm}(t)r_{\pm}(t)-\frac{1}{2}(z-z^{-1})^{2} &(z^{-1}-z)\,q_{\pm}(t)\\
(z^{-1}-z)\,r_{\pm}(t) &-q_{\pm}(t)r_{\pm}(t)+\frac{1}{2}(z-z^{-1})^{2}\\
\end{array}
\right)\varphi_{n}\,,\,n\rightarrow\pm\infty.
\end{equation}
Combining with the asymptotic form as $n\rightarrow\pm\infty$ of Eq.~(\ref{1.5}a), we get the differential-difference equations of every components as $n\rightarrow\pm\infty$
\begin{subequations}\label{2.82}
\begin{align}
&\varphi^{(1)}_{n,t}=i\left(q_{0}^{2}+\frac{z^{2}-z^{-2}}{2}\right)\varphi^{(1)}_{n}+i\left(z^{-1}-z\right)\varphi^{(1)}_{n+1}\,,\\
&\varphi^{(2)}_{n,t}=i\left(-q_{0}^{2}+\frac{z^{2}-z^{-2}}{2}\right)\varphi^{(2)}_{n}+i\left(z^{-1}-z\right)\varphi^{(2)}_{n+1}\,.
\end{align}
\end{subequations}
Introducing the solutions of Eqs.~(\ref{1.5}) in the following forms
\begin{subequations}\label{2.83}
\begin{align}
&\dot{\phi}_{n}(t)=e^{\eta_{1}t}\phi_{n}(t)\,,\,\dot{\bar{\phi}}_{n}(t)=e^{\eta_{2}t}\bar{\phi}_{n}(t)\,,\\
&\dot{\psi}_{n}(t)=e^{\eta_{2}t}\psi_{n}(t)\,,\,\dot{\bar{\psi}}_{n}(t)=e^{\eta_{1}t}\bar{\psi}_{n}(t)\,,
\end{align}
\end{subequations}
which meet Eqs.~(\ref{2.82}) and $\eta_{1}$, $\eta_{2}$ are unknown parameters, according to the boundary conditions~(\ref{2.2}) and Eqs.~(\ref{2.82}), we can obtain
\begin{equation}\label{2.84}
\eta_{1}=-q_{0}^{2}+\frac{z^{2}-z^{-2}}{2}-(z-z^{-1})r\lambda\,,\,
\eta_{2}=q_{0}^{2}+\frac{z^{2}-z^{-2}}{2}-(z-z^{-1})r\lambda^{-1}\,.
\end{equation}

Since Eqs.~(\ref{2.83}) are solutions of Eqs.~(\ref{1.5}), we can get the time evolution of the eigenfunctions as
\begin{subequations}\label{2.85}
\begin{align}
&\phi_{n,t}(t)=\left(\mathbf{V}_{n}-i\,\eta_{1}\mathbf{I_{2}}\right)\phi_{n}(t)\,,\,\bar{\phi}_{n,t}(t)=\left(\mathbf{V}_{n}-i\,\eta_{2}\mathbf{I_{2}}\right)\bar{\phi}_{n}(t)\,,\\
&\psi_{n,t}(t)=\left(\mathbf{V}_{n}-i\,\eta_{2}\mathbf{I_{2}}\right)\psi_{n}(t)\,,\,\bar{\psi}_{n,t}(t)=\left(\mathbf{V}_{n}-i\,\eta_{1}\mathbf{I_{2}}\right)\bar{\psi}_{n}(t)\,.
\end{align}
\end{subequations}
Eqs.~(\ref{2.85}) can also be written as
\begin{equation}\label{2.86}
\mathbf{\Phi}_{n,t}(z,t)=\mathbf{V}_{n}(z,t)\mathbf{\Phi}_{n}(z,t)-\mathbf{\Phi}_{n}(z,t)\mathcal{Y}(z)\,,\,
\mathbf{\Psi}_{n,t}(z,t)=\mathbf{V}_{n}(z,t)\mathbf{\Psi}_{n}(z,t)-\mathbf{\Psi}_{n}(z,t)\mathcal{Y}(z)
\end{equation}
with the new matrix $\mathcal{Y}(z)=diag\left(i\,\eta_{1}(z),i\,\eta_{2}(z)\right)$,
In addition, taking the partial derivative of $t$ on Eq.~(\ref{2.7}) and comparing it with Eq.~(\ref{2.86}), we know that $t_{11}(z,t)$ and $t_{22}(z,t)$ are independent of $t$ but
\begin{equation}
t_{21}(z,t)=t_{21}(z,0)e^{2iq_{0}^{2}t+i\,\gamma(z)t}\,,\,t_{12}(z,t)=t_{12}(z,0)e^{-2iq_{0}^{2}t-i\,\gamma(z)t}
\end{equation}
with $\gamma(z)=r(\lambda-\lambda^{-1})(z-z^{-1})$.
Furthermore, we can also obtain the time evolution of norming constants
\begin{equation}
C_{j}(t)=C_{j}(0)e^{2iq_{0}^{2}t+i\,\gamma(z_{j})t}\,,\,\bar{C}_{j}(t)=\bar{C}_{j}(0)e^{-2iq_{0}^{2}t-i\,\gamma(\bar{z}_{j})t}.
\end{equation}
Combining with Eqs.~(\ref{2.79}), we obtain the determinant formulae of norming constants as follows:
\begin{subequations}
\begin{align}
&\bar{C}_{1}=\frac{\kappa_{1}(\lambda(\bar{\zeta}_{1}))^{3}(\bar{\zeta}_{1}-\zeta_{1})(\bar{\zeta}_{1}-\zeta^{*}_{1})}{\bar{\zeta}_{1}-\bar{\zeta}^{*}_{1}}e^{i(\bar{\theta}_{1}-2q_{0}^{2}t-\gamma(\bar{\zeta}_{1})t)}\,,\\ &\bar{C}_{2}=\frac{\kappa_{1}^{-1}(\lambda(\bar{\zeta}_{1}))^{-1}(\bar{\zeta}^{*}_{1}-\zeta_{1})(\bar{\zeta}^{*}_{1}-\zeta^{*}_{1})}{\bar{\zeta}^{*}_{1}-\bar{\zeta}_{1}}e^{i(\bar{\theta}_{2}-2q_{0}^{2}t-\gamma(\bar{\zeta}^{*}_{1})t)}
\end{align}
\end{subequations}
with
\begin{equation}
\gamma(\zeta)=r^{2}\frac{(\zeta-\frac{2}{r}+\frac{1}{\zeta})(\zeta-2r+\frac{1}{\zeta})}{(\zeta-r)(\frac{1}{\zeta}-r)}.\nonumber
\end{equation}

\vspace{5mm} \noindent\textbf{2.4 The  soliton solutions}
\\\hspace*{\parindent}

The soliton solutions can be obtained by solving algebraic system~(\ref{2.67}) with discrete eigenvalues and corresponding norming constants $\{\zeta_{j},\bar{\zeta}_{j},C_{j}(t),\bar{C}_{j}(t),j=1,2,\cdot\cdot\cdot,J\}$ under the reflectionless condition. We have obtained the suitable discrete eigenvalues under the asymptotic behavior of scattering coefficients and the constraint of $\bar{b}_{j}$ above so that in this part, we construct the 2-eigenvalue soliton solution by setting $J=2$ and the discrete eigenvalues $\{\zeta_{1}, \zeta_{2}, \bar{\zeta}_{1}, \bar{\zeta}_{2}\}$ with $\bar{\zeta}_{1}=\frac{1+q_{0}e^{i\eta_{1}}}{r}$, $\eta_{1}\in\mathbb{R}$, $|\pi-\eta_{1}|<$ arctan$(r/q_{0})$, $\bar{\zeta}_{2}=\bar{\zeta}^{*}_{1}$, $\zeta_{2}=\zeta^{*}_{1}$.

According to Eqs.~(\ref{2.67}) and~(\ref{2.68}), we introduce $\mathbf{X}=(X_{1},...,X_{9})^{T}$, $\mathbf{Y}=(r-\frac{1}{\zeta_{1}},r-\frac{1}{\zeta_{2}},0,0,0,0,\bar{\zeta}_{1}-r,\bar{\zeta}_{2}-r,1)^{T}$ and matrix $\mathbf{B}=(\mathbf{B}_{1},...,\mathbf{B}_{9})$, where
\begin{align}
&X_{1}=N^{(1)}_{n}(\zeta_{1})\,,\,X_{2}=N^{(1)}_{n}(\zeta_{2})\,,\,X_{3}=N^{(2)}_{n}(\zeta_{1})\,,\,X_{4}=N^{(2)}_{n}(\zeta_{2})\,,\,\nonumber\\
&X_{5}=\bar{N}^{(1)}_{n}(\bar{\zeta}_{1})\,,\,X_{6}=\bar{N}^{(1)}_{n}(\bar{\zeta}_{2})\,,\,X_{7}=\bar{N}^{(2)}_{n}(\bar{\zeta}_{1})\,,\,
X_{8}=\bar{N}^{(2)}_{n}(\bar{\zeta}_{2})\,,\,X_{9}=\frac{1}{\Theta_{n}}\,,\,\nonumber
\end{align}
\begin{equation}
\mathbf{B}=\left (
\begin{array}{ccccccccc}
1 &0 &0 &0 &R_{1} &R_{2} &0 &0 &0\\
0 &1 &0 &0 &R_{3} &R_{4} &0 &0 &0\\
0 &0 &1 &0 &0 &0 &R_{1} &R_{2} &r_{+}\\
0 &0 &0 &1 &0 &0 &R_{3} &R_{4} &r_{+}\\
R_{5} &R_{6} &0 &0 &1 &0 &0 &0 &-q_{+}\\
R_{7} &R_{8} &0 &0 &0 &1 &0 &0 &-q_{+}\\
0 &0 &R_{5} &R_{6} &0 &0 &1 &0 &0\\
0 &0 &R_{7} &R_{8} &0 &0 &0 &1 &0\\
0 &0 &R_{9} &R_{10} &0 &0 &0 &0 &1
\end{array}
\right)\nonumber
\end{equation}
with
\begin{align}
&R_{1}=-\frac{(\zeta_{1}-\frac{1}{r})\bar{C}_{1}(t)\lambda(\bar{\zeta}_{1})^{2n}}{(\bar{\zeta}_{1}-\frac{1}{r})(\zeta_{1}-\bar{\zeta}_{1})}\,,\,
R_{2}=-\frac{(\zeta_{1}-\frac{1}{r})\bar{C}_{2}(t)\lambda(\bar{\zeta}_{2})^{2n}}{(\bar{\zeta}_{2}-\frac{1}{r})(\zeta_{1}-\bar{\zeta}_{2})}\,,\,
R_{3}=-\frac{(\zeta_{2}-\frac{1}{r})\bar{C}_{1}(t)\lambda(\bar{\zeta}_{1})^{2n}}{(\bar{\zeta}_{1}-\frac{1}{r})(\zeta_{2}-\bar{\zeta}_{1})}\,,\nonumber\\
&R_{4}=-\frac{(\zeta_{2}-\frac{1}{r})\bar{C}_{2}(t)\lambda(\bar{\zeta}_{2})^{2n}}{(\bar{\zeta}_{2}-\frac{1}{r})(\zeta_{2}-\bar{\zeta}_{2})}\,,\,R_{5}=-\frac{(\bar{\zeta}_{1}-r)C_{1}(t)\,\lambda(\zeta_{1})^{-2n}}{(\zeta_{1}-r)(\bar{\zeta}_{1}-\zeta_{1})}\,,\,R_{6}=-\frac{(\bar{\zeta}_{1}-r)C_{2}(t)\,\lambda(\zeta_{2})^{-2n}}{(\zeta_{2}-r)(\bar{\zeta}_{1}-\zeta_{2})}\,,\nonumber\\
&R_{7}=-\frac{(\bar{\zeta}_{2}-r)C_{1}(t)\,\lambda(\zeta_{1})^{-2n}}{(\zeta_{1}-r)(\bar{\zeta}_{2}-\zeta_{1})}\,,\,R_{8}=-\frac{(\bar{\zeta}_{2}-r)C_{2}(t)\,\lambda(\zeta_{2})^{-2n}}{(\zeta_{2}-r)(\bar{\zeta}_{2}-\zeta_{2})}\,,\,R_{9}=\frac{C_{1}(t)\,\lambda(\zeta_{1})^{-2n}}{\zeta_{1}(\zeta_{1}-r)}\,,\,R_{10}=\frac{C_{2}(t)\,\lambda(\zeta_{2})^{-2n}}{\zeta_{2}(\zeta_{2}-r)}\nonumber
\end{align}
and $\mathbf{B}_{j},\,j=1,...,9$ is the $j$-th column of matrix $\mathbf{B}$. Then we solve the linear system $\mathbf{B}\mathbf{X}=\mathbf{Y}$ and obtain that
\begin{equation}
X_{1}=\frac{Det(\mathbf{B}_{1}^{c})}{Det(\mathbf{B})}\,,\,X_{2}=\frac{Det(\mathbf{B}_{2}^{c})}{Det(\mathbf{B})}\,,\,X_{9}=\frac{Det(\mathbf{B}_{9}^{c})}{Det(\mathbf{B})}\,,
\end{equation}
where $\mathbf{B}_{1}^{c}=(\mathbf{Y},\mathbf{B}_{2},...,\mathbf{B}_{9})$, $\mathbf{B}_{2}^{c}=(\mathbf{B}_{1},\mathbf{Y},\mathbf{B}_{3},...,\mathbf{B}_{9})$, $\mathbf{B}_{9}^{c}=(\mathbf{B}_{1},...,\mathbf{B}_{8},\mathbf{Y})$.
Then the second-order soliton solution can be written as
\begin{equation}
q_{n}^{[2]}(t)=q_{+}(t)+r\,R_{9}\frac{X_{1}}{X_{9}}+r\,R_{10}\frac{X_{2}}{X_{9}}\,,
\end{equation}
where the superscript $[2]$ denotes the second-order soliton solution.

Some typical second-order soliton solutions with the nonzero boundary coefficients $q_{0}=\frac{2}{3}$ are shown in Fig.~$2$. Fig.~$2$(a) shows the second-order dark-dark soliton solution that is an interaction of two waves having the same amplitude with the nonzero boundary condition $q_{0}=\frac{2}{3}, \theta=0$, scattering coefficients $\bar{\zeta}_{1}=\frac{1-q_{0}e^{i\pi/7}}{r}$, $\kappa_{1}=1$, $\bar{\theta}_{1}=0$, $\bar{\theta}_{2}=0$. Fig.~$2$(b) shows the second-order bright-dark soliton solution with the nonzero boundary condition $q_{0}=\frac{2}{3}, \theta=\frac{2\pi}{5}$, scattering coefficients $\bar{\zeta}_{1}=\frac{1-q_{0}e^{i\pi/7}}{r}$, $\kappa_{1}=1$, $\bar{\theta}_{1}=0$, $\bar{\theta}_{2}=0$. Fig.~$2$(c) shows the second-order bright-bright soliton solution with the nonzero boundary condition $q_{0}=\frac{2}{3}, \theta=\pi$, scattering coefficients $\bar{\zeta}_{1}=\frac{1-q_{0}e^{i\pi/7}}{r}$, $\kappa_{1}=1$, $\bar{\theta}_{1}=0$, $\bar{\theta}_{2}=0$.

We found that under the nonzero boundary condition $q_{0}=\frac{2}{3}$ and scattering coefficient $\bar{\zeta}_{1}=\frac{1-q_{0}e^{i\pi/7}}{r}$ (similar conclusions have been obtained for other parameters as well), as $\theta$ increases from $0$ to $2\pi$, the second-order soliton transforms from dark-dark soliton $\rightarrow$ bright-dark soliton $\rightarrow$ bright-bright soliton $\rightarrow$ bright-dark soliton $\rightarrow$ dark-dark soliton. Some cross-sectional views of second-order solitons under the continuous limit of $n$ (to make a more intuitive comparison) are given in Fig.~$1$(d)-(f) with $t=10$. Fig.~$1$(d) shows three dark-dark solitons with $\theta=0$, $\frac{\pi}{4}$, $\frac{\pi}{3}$ in the interval $[0, \frac{\pi}{2}]$, respectively. Fig.~$1$(e) shows three bright-dark solitons with $\theta=\frac{11\pi}{30}$, $\frac{2\pi}{5}$, $\frac{5\pi}{11}$ in the interval $[0, \frac{\pi}{2}]$, respectively. Fig.~$1$(f) shows three bright-bright solitons with $\theta=\frac{3\pi}{4}$, $\frac{9\pi}{10}$, $\pi$ in the interval $[\frac{\pi}{2}, \pi]$, respectively. What we also have to emphasize is that when the phase $\theta=\pi/2$ or $3\pi/2$, the bright component of the bright-dark soliton will exhibit singularity, that is, the amplitude will be infinite, which can be deduced in Fig.~$2$(e). The transformation of the second-order soliton in the interval $[0, 2\pi]$ is shown in Fig.~$3$.

\begin{center}
\includegraphics[scale=0.21]{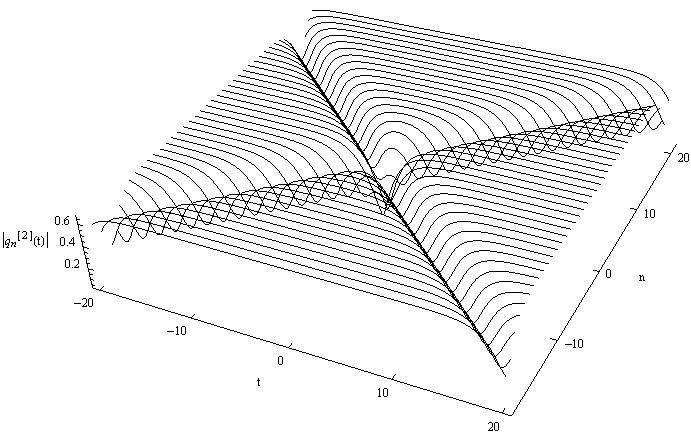}\hfill
\includegraphics[scale=0.2]{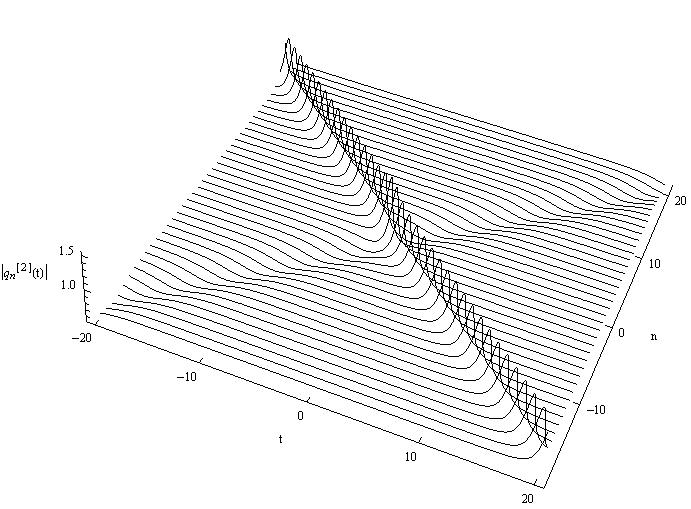}\hfill
\includegraphics[scale=0.21]{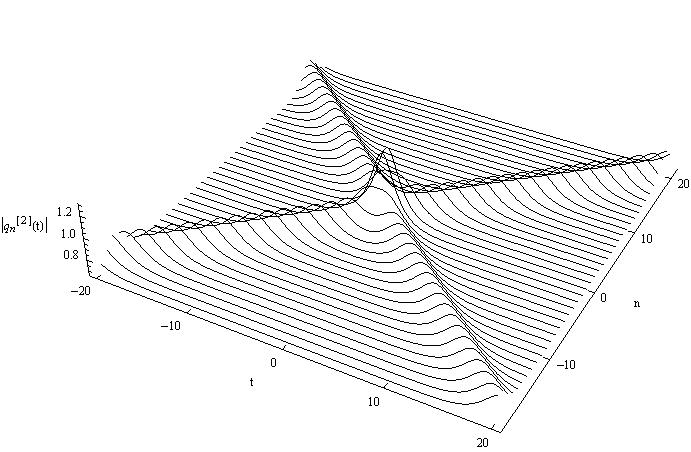}\hfill
{\footnotesize\hspace{6cm}(a)\hspace{6cm}(b)\hspace{6cm}(c)}\\
\includegraphics[scale=0.25]{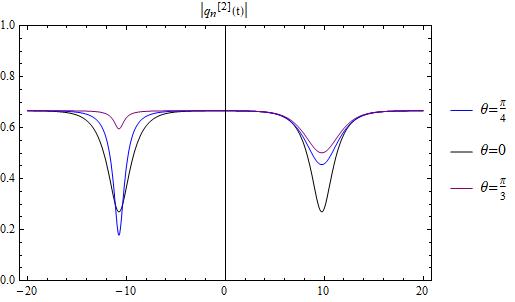}\hfill
\includegraphics[scale=0.25]{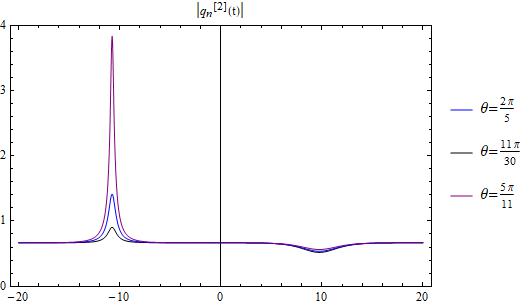}\hfill
\includegraphics[scale=0.25]{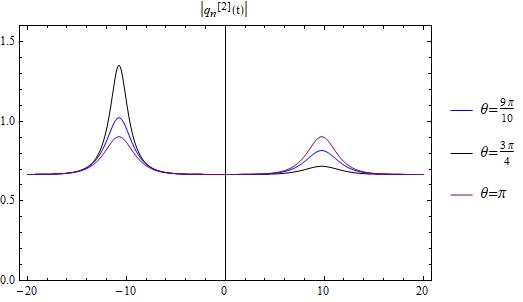}\hfill
{\footnotesize\hspace{5cm}(d)\hspace{5.5cm}(e)\hspace{5cm}(f)}\\
\flushleft{\footnotesize
\textbf{Fig.~$2$.} The second-order soliton solutions with the nonzero boundary condition $q_{0}=\frac{2}{3}$ and scattering coefficients $\kappa_{1}=1$, $\bar{\theta}_{1}=0$, $\bar{\theta}_{2}=0$, $\bar{\zeta}_{1}=\frac{1-q_{0}e^{i\pi/7}}{r}$. (a) The dark-dark soliton solution $q_{n}^{[2]}(t)$ with $\theta=0$. (b) The bright-dark soliton solution $q_{n}^{[2]}(t)$ with $\theta=\frac{2\pi}{5}$. (c) The bright-bright soliton solution $q_{n}^{[2]}(t)$ with $\theta=\pi$. (d) The dark-dark soliton solution $q_{n}^{[2]}(t)$ with $\theta=0$, $\frac{\pi}{4}$, $\frac{\pi}{3}$, respectively. (e) The bright-dark soliton solution $\theta=\frac{11\pi}{30}$, $\frac{2\pi}{5}$, $\frac{5\pi}{11}$, respectively. (f) The bright-bright soliton solution $\theta=\frac{3\pi}{4}$, $\frac{9\pi}{10}$, $\pi$, respectively.}
\end{center}

\begin{center}
\includegraphics[scale=0.15]{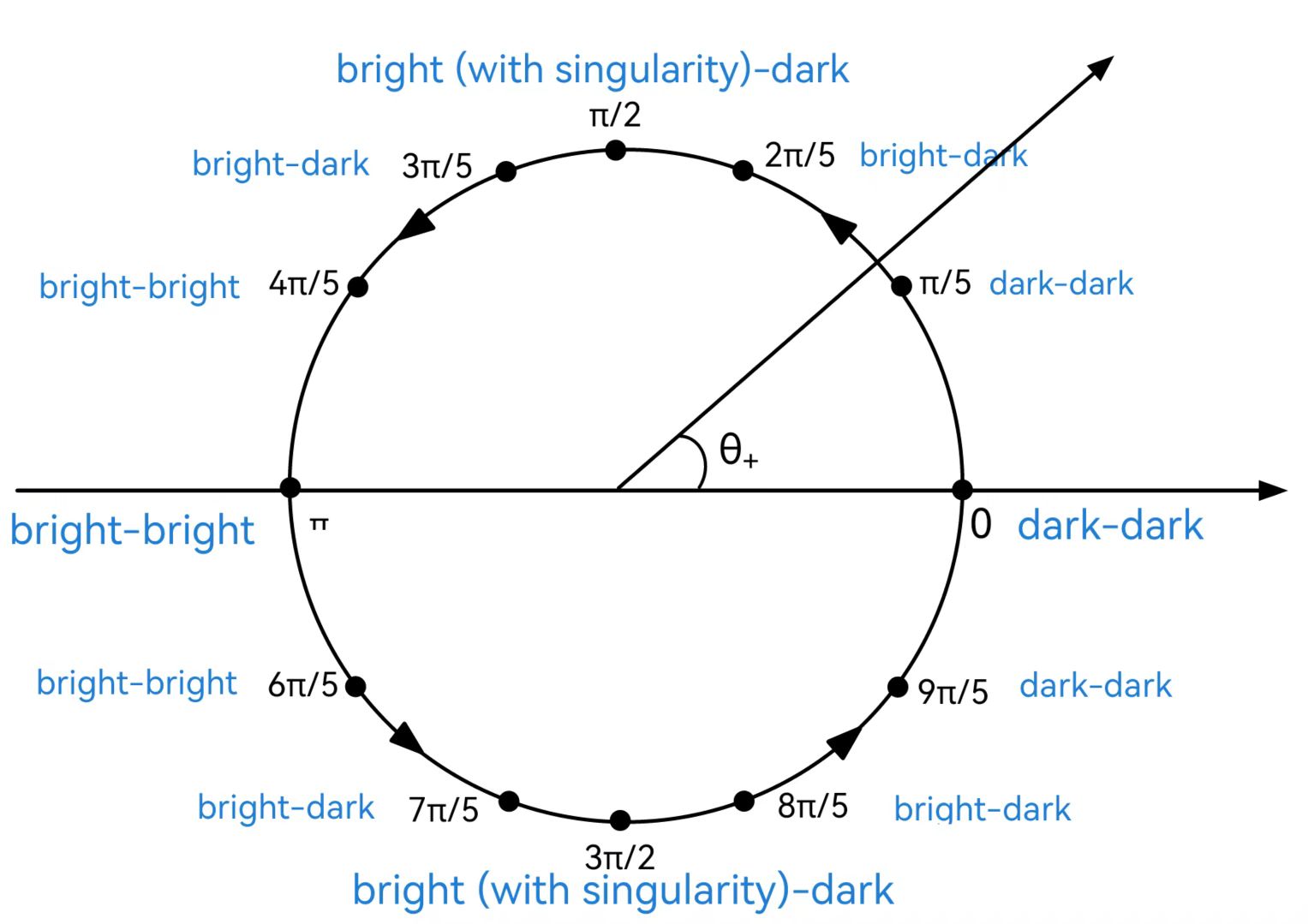}\hfill
\flushleft{\footnotesize
\textbf{Fig.~$3$.} The transformation of second-order soliton solutions under the nonzero boundary condition $q_{0}=\frac{2}{3}$ and scattering coefficients $\kappa_{1}=1$, $\bar{\theta}_{1}=0$, $\bar{\theta}_{2}=0$, $\bar{\zeta}_{1}=\frac{1-q_{0}e^{i\pi/7}}{r}$.}
\end{center}

\vspace{7mm}\noindent\textbf{3 Case (\ref{c2}): $\sigma=1$, $\Delta\theta=\pi$}
\hspace*{\parindent}
\renewcommand{\theequation}{3.\arabic{equation}}\setcounter{equation}{0}\\

In this section, we will investigate the IST for Eq.~(\ref{1.3}) with NZBCs~(\ref{1.4}) meeting the case~(\ref{c2}).

\vspace{5mm} \noindent\textbf{3.1 The direct scattering problem}
\\\hspace*{\parindent}

We also rewrite the NZBCs~(\ref{1.4}) and ${\lim_{n\to \pm \infty}}r_{n}(t)$ without the time variable $t$ as
\begin{equation}\label{3.1}
{\lim_{n\to \pm \infty}}q_{n}=q_{\pm}=q_{0}e^{i \vartheta_{\pm}}\,,{\lim_{n\to \pm \infty}}r_{n}=r_{\pm}=q_{0}e^{-i \vartheta_{\mp}}\,,
\end{equation}
where $\vartheta_{\pm}=\theta_{\pm}-2q_{0}^{2}t$ is also real, $\theta_{+}=\pi+\theta_{-}$.

\vspace{5mm}\noindent\textbf{3.1.1 Eigenfunctions, scattering matrix, uniformization variable}
\\\hspace*{\parindent}

Under the boundary conditions~(\ref{3.1}) and symmetry reduction condition $r_{n}(t)=q_{-n}^{*}(t)$, the solution matrices $\mathbf{\Phi}_{n}(z)$ and $\mathbf{\Psi}_{n}(z)$ of Eq.~(\ref{1.5}a)  satisfy the following boundary conditions:
\begin{subequations}\label{3.2}
\begin{align}
\mathbf{\Phi}_{n}(z)&=\left(\phi_{n}(z), \bar{\phi}_{n}(z)\right)\sim r^{n}\left (
\begin{array}{cc}
q_{-} &\lambda r-z\\
\lambda r-z &-r_{-}\\
\end{array}
\right)\mathbf{\Lambda}^{n},\ n\to -\infty\,,\\
\mathbf{\Psi}_{n}(z)&=\left(\bar{\psi}_{n}(z), \psi_{n}(z)\right)\sim r^{n}\left (
\begin{array}{cc}
q_{+} &\lambda r-z\\
\lambda r-z &-r_{+}\\
\end{array}
\right)\mathbf{\Lambda}^{n}\,,\,n\rightarrow+\infty\,,
\end{align}
\end{subequations}
where $\phi_{n}(z)$, $\bar{\phi}_{n}(z)$, $\bar{\psi}_{n}(z)$, $\psi_{n}(z)$ are also eigenfunctions but $r=\sqrt{1+q_{0}^{2}}$ with $q_{0}>0$, which is dirrerent from the Case (\ref{c1}). $\lambda(z)$ also can be written as $\xi\pm\sqrt{\xi^{2}-1}$ by introducing $\xi(z)=\frac{z+\frac{1}{z}}{2r}$, $\mathbf{\Lambda}=diag(\lambda,\frac{1}{\lambda})$. $\lambda$ and $z$ also satisfy $r(\lambda+\frac{1}{\lambda})=z+\frac{1}{z}$ but $\lambda(z)$ is a double-valued function of $z$ with four branch points $\pm z_{1}$ and $\pm z_{2}$ meeting $\xi(z)^{2}=1$, where
\begin{equation}\label{3.3}
z_{1}=r+q_{0}>1\,,z_{2}=r-q_{0}<1\,.
\end{equation}
Furthermore, we can also prove that the continuous spectrum of the scattering problem is any $z$ that meet $|\lambda(z)|=1$ in the set:
\begin{equation}\label{3.4}
\begin{aligned}
\Sigma=[-r-q_{0},-r+q_{0}]\cup[r-q_{0},r+q_{0}]\cup\{z\in\mathbb{C}:|z|=1\}\,.
\end{aligned}
\end{equation}

Two solutions $\varphi_{n}$, $\bar{\varphi}_{n}$ of scattering problem obey the
recursion relation:
\begin{equation}\label{3.5}
\begin{aligned}
Det\left(\varphi_{n+1}, \bar{\varphi}_{n+1}\right)=(1-q_{n}r_{n})Det\left(\varphi_{n}, \bar{\varphi}_{n}\right)\,,
\end{aligned}
\end{equation}
and according to Eqs.~(\ref{3.2}), we can obtain
\begin{subequations}\label{3.6}
\begin{align}
Det\left(\mathbf{\Phi}_{n}(z)\right)&=-\left((\lambda r-z)^{2}-q^{2}_{0}\right)r^{2n}\prod_{k=-\infty}^{n-1}\frac{1-q_{k}r_{k}}{1+q^{2}_{0}}\,,\\
Det\left(\mathbf{\Psi}_{n}(z)\right)&=-\left((\lambda r-z)^{2}-q^{2}_{0}\right)r^{2n}\prod_{k=n}^{+\infty}\frac{1+q^{2}_{0}}{1-q_{k}r_{k}}\,.
\end{align}
\end{subequations}
The factor $(\lambda r-z)^{2}+q^{2}_{0}=0$ if and only if $z$ are also the four branch points $\pm z_{1}$, $\pm z_{2}$ of $\lambda(z)$. In addition, we assume that potential functions $q_{n}$, $r_{n}$ meet the condition $1-q_{n}r_{n}\neq0$ for any $n\in\mathbb{Z}$, then Eqs.~(\ref{3.6}) don't equal zero so that both $\mathbf{\Phi}_{n}(z)$ and $\mathbf{\Psi}_{n}(z)$ are the fundamental solutions of scattering problem for any $z\in\mathbb{C}$ except $\pm z_{1}$, $\pm z_{2}$. Thus, scattering matrix $T(z)$ is exist to relate these two solution matrices
\begin{equation}\label{3.7}
\begin{aligned}
\mathbf{\Phi}_{n}(z)=\mathbf{\Psi}_{n}(z)\mathbf{T}(z)\,,\,\mathbf{T}(z)=\left(
\begin{array}{cc}
t_{11}(z) &t_{12}(z)\\
t_{21}(z) &t_{22}(z)\\
\end{array}
\right)
\end{aligned}
\end{equation}
and it yields
\begin{equation}\label{3.8}
\begin{aligned}
t_{11}(z)t_{22}(z)-t_{21}(z)t_{12}(z)=\Theta_{-\infty}\,,\Theta_{-\infty}=\lim_{n\to -\infty}\Theta_{n}\,,\Theta_{n}=\prod_{k=n}^{+\infty}\frac{1-q_{k}r_{k}}{1+q^{2}_{0}}\,.
\end{aligned}
\end{equation}
The scattering coefficients $t_{ij}(z)$, $i,j=1,2$ can be represented as Eqs.~(\ref{2.10}). We also define the uniformization variable as $\zeta(z)=\lambda(z)/z$, then we have the same mappings as Eq.~(\ref{2.12}) but $r=\sqrt{1+q_{0}^{2}}$. In addition, the continuous spectrum in this case is mapped onto $\{\zeta\in\mathbb{C}:|\zeta|=1\}\cup\{\zeta\in\mathbb{C}:|\zeta-r|=q_{0}\}$ and four branch points $\pm z_{1}$, $\pm z_{2}$ are mapped onto two points $\zeta_{1}=\zeta(\pm z_{1})=z_{2}$, $\zeta_{2}=\zeta(\pm z_{2})=z_{1}$ in the $\zeta-$plane. Furthermore, we can show that
\begin{equation}\label{3.9}
|\lambda|\leq 1 \Leftrightarrow(|\zeta|-1)(|\zeta-r|-q_{0})\leq0\,.
\end{equation}
Defining $\Sigma^{'}=\{\zeta\in\mathbb{C}:|\zeta|=1\}\cup\{\zeta\in\mathbb{C}:|\zeta-r|=q_{0}\}$ with the orientation as shown in Fig.~$4$, then the regions
\begin{equation}\label{3.10}
D_{+}=\{\zeta\in\mathbb{C}\mid (|\zeta|-1)(|\zeta-r|-q_{0})<0\}\,,\,D_{-}=\{\zeta\in\mathbb{C}\mid(|\zeta|-1)(|\zeta-r|-q_{0})>0\}
\end{equation}
are the corresponding positive region for $|\lambda|< 1$ and negative region for $|\lambda|> 1$.

The modified eigenfunctions
\begin{subequations}\label{3.11}
\begin{align}
&(M_{n}(z),\bar{M}_{n}(z))=r^{-n}\mathbf{\Omega}\,\mathbf{\Phi}_{n}(z)\mathbf{\Omega}^{-1}\mathbf{\Lambda}^{-n}\,,\\
&(\bar{N}_{n}(z),N_{n}(z))=r^{-n}\mathbf{\Omega}\,\mathbf{\Psi}_{n}(z)\mathbf{\Omega}^{-1}\mathbf{\Lambda}^{-n}
\end{align}
\end{subequations}
with $\mathbf{\Omega}=diag(1,\lambda)$ also have the same forms of difference equations as Eqs.~(\ref{2.16}), (\ref{2.17}) except for $r$ and the same boundary conditions except for $q_{\pm}$, $r_{\pm}$ and $r$ at $n\rightarrow\pm\infty$. They meet
\begin{subequations}\label{3.12}
\begin{align}
M_{n}(\zeta)&=\tilde{t}_{11}(\zeta)\bar{N}_{n}(\zeta)+\lambda^{-2n}\tilde{t}_{21}(\zeta)N_{n}(\zeta)\,,\\
\bar{M}_{n}(\zeta)&=\tilde{t}_{22}(\zeta)N_{n}(\zeta)+\lambda^{2n}\tilde{t}_{12}(\zeta)\bar{N}_{n}(\zeta).
\end{align}
\end{subequations}
with modified scattering matrix
\begin{equation}\label{3.13}
\tilde{\mathbf{T}}(\zeta)=\mathbf{\Omega}\,\mathbf{T}(z)\mathbf{\Omega}^{-1}=\left (
\begin{array}{cc}
\tilde{t}_{11}(\zeta) &\tilde{t}_{12}(\zeta)\\
\tilde{t}_{21}(\zeta) &\tilde{t}_{22}(\zeta)\\
\end{array}
\right)\,,
\end{equation}
where $\tilde{t}_{11}(\zeta)=t_{11}(\zeta)\,,\,\tilde{t}_{22}(\zeta)=t_{22}(\zeta)\,,\,\tilde{t}_{12}(\zeta)=\frac{t_{12}(\zeta)}{\lambda}\,,\,
\tilde{t}_{21}(\zeta)=\lambda t_{21}(\zeta)$ are in the forms as Eqs.~(\ref{2.23}).

\vspace{5mm} \noindent\textbf{3.1.2 Analyticity of modified eigenfunctions and scattering coefficients}
\\\hspace*{\parindent}

To analyze the analyticity of modified eigenfunctions, we introduce three modifications similar to Eqs.~(\ref{2.24})-(\ref{2.31}) in the Case (\ref{c1}), but the difference is that the modified potentials $\dot{q}_{n}$, $\dot{r}_{n}$ in $\hat{\mathbf{E}}_{n}$ satisfy the constraint $\dot{q}_{n}\dot{r}_{n}=-q_{0}^{2}\,,\,n\in\mathbb{Z}$ and $\dot{\mathbf{U}}'_{n}$ becomes

\begin{center}
\includegraphics[scale=0.6]{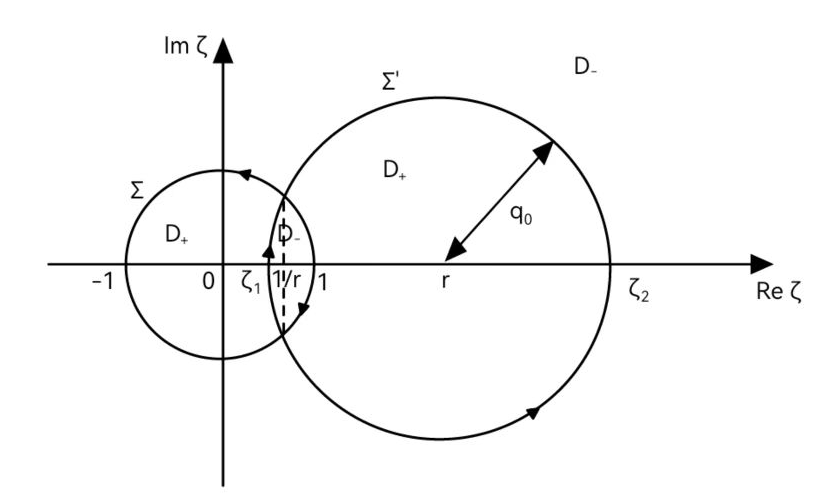}\hfill
\flushleft{\footnotesize
\textbf{Fig.~$4$.} The $\zeta$-plane:  $\Sigma^{'}=\{\zeta\in\mathbb{C}:|\zeta|=1\}\cup\{\zeta\in\mathbb{C}:|\zeta-r|=q_{0}\}$ is the continuous spectrum and it prescribed orientation identifies the regions $D_{+}$ and $D_{-}$ as in Eq.~(\ref{3.10}).}
\end{center}

\begin{align}\label{3.14}
&\dot{\mathbf{U}}'_{n}=\frac{1}{-q_{0}^{2}+(r\lambda-z)^{2}}\nonumber\\
&\left(
\begin{array}{cc}
(r\lambda-z)(\eta_{n}^{[1]}+\eta_{-n}^{[1]*})-(r\lambda q_{n}-z \eta_{n}^{[2]})\eta_{-(n+1)}^{[3]*} &\frac{(r\lambda-z)^{2}\eta_{-n}^{[2]*}-\dot{r}_{n}\eta_{-n}^{[1]*}-(z(r\lambda-z)-q_{n}\dot{r}_{n})\eta_{-(n+1)}^{[3]*}}{\lambda}\\
\lambda\left((r\lambda-z)^{2}\eta_{n}^{[2]}-\dot{q}_{n}\eta_{n}^{[1]}-(z^{-1}(r\lambda-z)+\dot{q}_{n}r_{n})\eta_{n}^{[3]}\right) &-(r\lambda-z)(\eta_{n}^{[1]}+\eta_{-n}^{[1]*})+(r\lambda^{-1} r_{n}-z^{-1}\eta_{-n}^{[2]*})\eta_{n}^{[3]}\\
\end{array}
\right)
\end{align}
in which $\eta_{n}^{[1]}=\dot{q}_{n}r_{n}+q_{0}^{2}$, $\eta_{n}^{[2]}=q_{n}-\dot{q}_{n}$, $\eta_{n}^{[3]}=\dot{q}_{n+1}-\dot{q}_{n}$ are all summable and decaying as $n\rightarrow\pm\infty$ as well so that $\dot{\mathbf{U}}'_{n}$ decays as $n\rightarrow\pm\infty$. Then we can obtain following $\mathbf{Theorem~\ref{T2}}$ by the same analysis as Case (\ref{c1}):
\begin{theorem}\label{T2}
Under the condition of the potentials $q_{n}, r_{n}, \dot{q}_{n}, \dot{r}_{n}\in \left\{ f_{n}: \sum\limits _{l=\pm\infty}^{n}|f_{l}-\lim_{k\to \pm \infty}f_{k}|<\infty, \forall n\in\mathbb{Z}\right\}$, $M_{n}$, $N_{n}$, $t_{11}(\zeta)=\tilde{t}_{11}(\zeta)$ are analytic in $D_{-}$ defined in Eq.~(\ref{3.10}) (except for the only possible pole $\zeta=\infty$) and $\bar{M}_{n}$, $\bar{N}_{n}$, $t_{22}(\zeta)=\tilde{t}_{22}(\zeta)$ are analytic in $D_{+}$ defined in Eq.~(\ref{3.10}) (except for the only possible pole $\zeta=0$).
\end{theorem}

\vspace{5mm} \noindent\textbf{3.1.3 Asymptotic behavior of modified eigenfunctions and scattering coefficients}
\\\hspace*{\parindent}

According the definition of uniformization variable $\zeta$ and by direct calculations, the asymptotic behavior of $z^{2}$, $\lambda^{2}$ and $z\lambda$ can also be obtained as Table~\ref{tab:1} and the asymptotic behavior of modified eigenfunctions is the same as Eqs.~(\ref{2.42})-(\ref{2.45}) except that $q_{\pm}$ and $r_{\pm}$ corresponds to Eq.~(\ref{3.1}), but the one of scattering coefficients $t_{11}(\zeta)$ and $t_{22}(\zeta)$ is as follows that is different from Eqs.~(\ref{2.46}):
\begin{subequations}\label{3.15}
\begin{align}
t_{11}(\zeta)&=1+\mathcal{O}(\frac{1}{\zeta})\,,\,\zeta\rightarrow \infty\,,\ \ \ \ \
t_{11}(\zeta)=-1+\mathcal{O}(\zeta-\frac{1}{r})\,,\,\zeta\rightarrow \frac{1}{r}\,,\\
t_{22}(\zeta)&=1+\mathcal{O}(\zeta)\,,\,\zeta\rightarrow 0\,,\ \ \ \ \ \ \
t_{22}(\zeta)=-1+\mathcal{O}(\zeta-r)\,,\,\zeta\rightarrow r\,.
\end{align}
\end{subequations}

\vspace{5mm} \noindent\textbf{3.1.4 Symmetries of scattering coefficients}
\\\hspace*{\parindent}

By the same consideration as Case (\ref{c1}), we have the following first symmetry about eigenfunctions and scattering coefficients under the fact when $z\mapsto z$ and $\lambda\mapsto\frac{1}{\lambda}$, $\zeta\mapsto\bar{\zeta}=\frac{r\zeta-1}{\zeta-r}$:
\begin{equation}\label{3.16}
\phi_{n}(z,\frac{1}{\lambda})=\frac{r/\lambda-z}{- r_{-}}\bar{\phi}_{n}(z,\lambda)\,,
\end{equation}
\begin{equation}\label{3.17}
\psi_{n}(z,\frac{1}{\lambda})=\frac{r/\lambda-z}{q_{+}}\bar{\psi}_{n}(z,\lambda)\,,
\end{equation}
\begin{equation}\label{3.18}
t_{11}(\zeta)=t_{22}(\bar{\zeta})\,,\,t_{21}(\zeta)=-\frac{q_{+}}{r_{-}}\,t_{12}(\bar{\zeta}).
\end{equation}

The second symmetry can also be analyzed by the modified scattering problems corresponding to that when $z\mapsto z^{*}$ and $\lambda\mapsto\frac{1}{\lambda^{*}}$, $\zeta\mapsto\bar{\zeta}^{*}=\frac{r\zeta^{*}-1}{\zeta^{*}-r}$:
\begin{align}
&\left(
\begin{array}{cc}
\bar{N}_{n+1}^{(1)}(\zeta)\\
\bar{N}_{n+1}^{(2)}(\zeta)\\
\end{array}
\right)=
-\frac{\zeta-r}{r_{+}\Theta_{-\infty}v_{n}}\left(
\begin{array}{cc}
M_{-n}^{(2)*}(\bar{\zeta}^{*})\\
-M_{-n}^{(1)*}(\bar{\zeta}^{*})\\
\end{array}
\right)\,,\label{3.19}\\
&\left(
\begin{array}{cc}
N_{n+1}^{(1)}(\zeta)\\
N_{n+1}^{(2)}(\zeta)\\
\end{array}
\right)=
\frac{1/\zeta-r}{q_{+}\Theta_{-\infty}v_{n}}\left(
\begin{array}{cc}
\bar{M}_{-n}^{(2)*}(\bar{\zeta}^{*})\\
-\bar{M}_{-n}^{(1)*}(\bar{\zeta}^{*})\\
\end{array}
\right)\,.\label{3.20}
\end{align}

According to Eqs.~(\ref{3.19}), (\ref{3.20}), (\ref{2.23}) and the fact when $z\mapsto z^{*}$ and $\lambda\mapsto\frac{1}{\lambda^{*}}$, $\zeta\mapsto\bar{\zeta}^{*}=\frac{r\zeta^{*}-1}{\zeta^{*}-r}$, we have the following symmetries of scattering coefficients:
\begin{equation}\label{3.21}
t_{11}(\zeta)=-t_{22}^{*}\left(\bar{\zeta}^{*}\right)\,.
\end{equation}

\vspace{5mm} \noindent\textbf{3.2 The inverse scattering problem}
\\\hspace*{\parindent}

We will also construct and solve a RH problem to get the reconstructed potential by considering the asymptotic behavior of the modified eigenfunctions.

\vspace{5mm}\noindent\textbf{3.2.1 The RH problem and reconstructed potential}
\\\hspace*{\parindent}

We rewrite Eqs.~(\ref{3.12}) as the following jump condition on $\Sigma^{'}$
\begin{subequations}\label{3.22}
\begin{align}
\mu(\zeta)-\bar{N}_{n}(\zeta)&=\lambda(\zeta)^{-2n}\rho(\zeta)N_{n}(\zeta)\,,\\
\bar{\mu}(\zeta)-N_{n}(\zeta)&=\lambda(\zeta)^{2n}\bar{\rho}(\zeta)\bar{N}_{n}(\zeta)\,,
\end{align}
\end{subequations}
where $\mu(\zeta)$, $\bar{\mu}(\zeta)$, $\rho(\zeta)$, $\bar{\rho}(\zeta)$ are defined as before and all functions have similar poles. In addition, the asymptotic behavior of $\mu(\zeta)$ as $\zeta\rightarrow\infty$, $\bar{\mu}(\zeta)$ as $\zeta\rightarrow 0$, $\bar{N}(\zeta)$ as $\zeta\rightarrow r$, $N(\zeta)$ as $\zeta\rightarrow \frac{1}{r}$ in Eqs.~(\ref{3.22}) is the same as Eqs.~(\ref{2.60}) and the reduce conditions have the similar forms as Eqs.~(\ref{2.61}) except the different region of poles not repeated here.

By direct calculation, we finally obtain
\begin{equation}\label{3.23}
\bar{N}_{n}(\zeta)=\left(
\begin{array}{cc}
\frac{q_{+}}{\Theta_{n}}\\
\zeta-r\\
\end{array}
\right)+(\zeta-r)\sum\limits _{j=1}^{J}\frac{C_{j}\,\lambda(\zeta_{j})^{-2n}}{(\zeta_{j}-r)(\zeta-\zeta_{j})}\,N_{n}(\zeta_{j})-
\frac{\zeta-r}{2\pi i}\oint_{\Sigma^{'}}\frac{\lambda(\omega)^{-2n}}{(\omega-\zeta)(\omega-r)}\rho(\omega)N_{n}(\omega)
d\omega\,,\zeta\in D_{+}\,,
\end{equation}
\begin{equation}\label{3.24}
N_{n}(\zeta)=\left(
\begin{array}{cc}
r-\frac{1}{\zeta}\\
-\frac{r_{+}}{\Theta_{n}}\\
\end{array}
\right)+(\zeta-\frac{1}{r})\sum\limits _{j=1}^{\bar{J}}\frac{\bar{C}_{j}\,\lambda(\bar{\zeta}_{j})^{2n}}{(\bar{\zeta}_{j}-\frac{1}{r})(\zeta-\bar{\zeta}_{j})}\,\bar{N}_{n}(\bar{\zeta}_{j})+
\frac{\zeta-\frac{1}{r}}{2\pi i}\oint_{\Sigma^{'}}\frac{\lambda(\omega)^{2n}}{(\omega-\zeta)(\omega-\frac{1}{r})}\bar{\rho}(\omega)\bar{N}_{n}(\omega)
d\omega\,,\zeta\in D_{-}\,,
\end{equation}
where the norming constants also meet the  symmetry $C_{j}=-\frac{q_{+}^{2}}{(\bar{\zeta}_{j}-r)^{2}}\bar{C}_{j}$.

Comparing the second and first component of Eq.~(\ref{3.23}) with $\zeta=0$ and asymptotic behavior of $\bar{N}_{n}(\zeta)$ as $\zeta\rightarrow0$, respectively, we obtain
\begin{equation}\label{3.25}
\frac{1}{\Theta_{n}}=1-\sum\limits _{j=1}^{J}\frac{C_{j}\,\lambda(\zeta_{j})^{-2n}}{\zeta_{j}(\zeta_{j}-r)}\,N^{(2)}_{n}(\zeta_{j})-
\frac{1}{2\pi i}\oint_{\Sigma^{'}}\frac{\lambda(\omega)^{-2n}}{\omega(\omega-r)}\rho(\omega)N^{(2)}_{n}(\omega)
d\omega\,,
\end{equation}
\begin{equation}\label{3.26}
q_{n}=q_{+}+\Theta_{n}\sum\limits _{j=1}^{J}\frac{r C_{j}\,\lambda(\zeta_{j})^{-2n}}{\zeta_{j}(\zeta_{j}-r)}\,N^{(1)}_{n}(\zeta_{j})+\frac{\Theta_{n}}{2\pi i}\oint_{\Sigma^{'}}\frac{r\lambda(\omega)^{-2n}}{\omega(\omega-r)}\rho(\omega)N^{(1)}_{n}(\omega)
d\omega
\end{equation}
with the items $\Theta_{n}$, $N^{(1)}_{n}(\zeta_{j})$ that can be solved by Cramer's Rule.

\vspace{5mm} \noindent\textbf{3.2.2 The reflectionless condition}
\\\hspace*{\parindent}

We can obtain the soliton solutions of Eq.~(\ref{1.3}) with the reflectionless condition $\rho(\zeta)=0$, $\bar{\rho}(\zeta)=0$ for $\Sigma^{'}$, under which Eqs.~(\ref{3.23}),~(\ref{3.24}),~(\ref{3.25}),~(\ref{3.26}) reduce to the solvable algebraic set of equations:
\begin{subequations}\label{3.27}
\begin{align}
&\bar{N}^{(1)}_{n}(\zeta)-(\zeta-r)\sum\limits _{j=1}^{J}\frac{C_{j}\,\lambda(\zeta_{j})^{-2n}}{(\zeta_{j}-r)(\zeta-\zeta_{j})}\,N^{(1)}_{n}(\zeta_{j})-\frac{q_{+}}{\Theta_{n}}=0\,,\\
&\bar{N}^{(2)}_{n}(\zeta)-(\zeta-r)\sum\limits _{j=1}^{J}\frac{C_{j}\,\lambda(\zeta_{j})^{-2n}}{(\zeta_{j}-r)(\zeta-\zeta_{j})}\,N^{(2)}_{n}(\zeta_{j})=\zeta-r\,,\\
&N^{(1)}_{n}(\zeta)-(\zeta-\frac{1}{r})\sum\limits _{j=1}^{\bar{J}}\frac{\bar{C}_{j}\,\lambda(\bar{\zeta}_{j})^{2n}}{(\bar{\zeta}_{j}-\frac{1}{r})(\zeta-\bar{\zeta}_{j})}\,\bar{N}^{(1)}_{n}(\bar{\zeta}_{j})=r-\frac{1}{\zeta}\,,\\
&N^{(2)}_{n}(\zeta)+\frac{r_{+}}{\Theta_{n}}-(\zeta-\frac{1}{r})\sum\limits _{j=1}^{\bar{J}}\frac{\bar{C}_{j}\,\lambda(\bar{\zeta}_{j})^{2n}}{(\bar{\zeta}_{j}-\frac{1}{r})(\zeta-\bar{\zeta}_{j})}\,\bar{N}^{(2)}_{n}(\bar{\zeta}_{j})=0\,,\\
&\frac{1}{\Theta_{n}}=1-\sum\limits _{j=1}^{J}\frac{C_{j}\,\lambda(\zeta_{j})^{-2n}}{\zeta_{j}(\zeta_{j}-r)}\,N^{(2)}_{n}(\zeta_{j})
\end{align}
\end{subequations}
and the reconstructed potential
\begin{equation}\label{3.28}
q_{n}=q_{+}+\Theta_{n}\sum\limits _{j=1}^{J}\frac{r C_{j}\,\lambda(\zeta_{j})^{-2n}}{\zeta_{j}(\zeta_{j}-r)}\,N^{(1)}_{n}(\zeta_{j})
\end{equation}
according which we can obtain the soliton solutions of Eq.~(\ref{1.3}) if the scattering coefficients $\{\zeta_{j},\bar{\zeta}_{k},C_{j},\bar{C}_{k},j=1,2,\cdot\cdot\cdot,J,,k=1,2,\cdot\cdot\cdot,\bar{J}\}$ exist.

\vspace{5mm} \noindent\textbf{3.2.3 The trace formula and discrete eigenvalues}
\\\hspace*{\parindent}

As before, $\zeta_{j}$, $\zeta^{*}_{j}$ are simple zeros of $t_{11}(\zeta)$ in $D_{-}$ if and only if $\bar{\zeta}_{j}=\frac{r\zeta_{j}-1}{\zeta_{j}-r}$ and $\bar{\zeta}^{*}_{j}=\frac{r\zeta^{*}_{j}-1}{\zeta^{*}_{j}-r}$ are simple zeros of $t_{22}(\zeta)$ in $D_{+}$ so that the numbers of simple zeros for $t_{11}(\zeta)$ and $t_{22}(\zeta)$ are equal, namely $J=\bar{J}$ and the discrete eigenvalues are in quartets: $\{\zeta_{j}, \zeta^{*}_{j}, \bar{\zeta}_{j}, \bar{\zeta}^{*}_{j}\}$, where the first pair is in $D_{-}$ and the second pair is in $D_{+}$. The all $J$ simple zeros of $t_{11}(\zeta)$ can also be divided into two classes denoted by $\zeta_{j}$, $j=1,2,\cdot\cdot\cdot,J_{1}$ and $\hat{\zeta}_{k}$, $k=1,2,\cdot\cdot\cdot,J_{2}$, where $2J_{1}+J_{2}=J$, Im$(\zeta_{j})\neq0$, Im$(\hat{\zeta}_{k})=0$ and $t_{22}(\zeta)$ has $J$ simple zeros $\bar{\zeta}_{j}$, $\bar{\zeta}^{*}_{j}$, $\bar{\hat{\zeta}}_{k}=\frac{r\hat{\zeta}_{k}-1}{\hat{\zeta}_{k}-r}$.

Defining the new functions
\begin{equation}\label{3.29}
\begin{aligned}
A(\zeta)=\prod_{j=1}^{J_{1}}\frac{\zeta-\bar{\zeta}_{j}}{\zeta-\zeta_{j}}\frac{\zeta-\bar{\zeta}^{*}_{j}}{\zeta-\zeta^{*}_{j}}\prod_{k=1}^{J_{2}}\frac{\zeta-\bar{\hat{\zeta}}_{k}}{\zeta-\hat{\zeta}_{k}}t_{11}(\zeta)\,,\,
\bar{A}(\zeta)=\prod_{j=1}^{J_{1}}\frac{\zeta-\zeta_{j}}{\zeta-\bar{\zeta}_{j}}\frac{\zeta-\zeta^{*}_{j}}{\zeta-\bar{\zeta}^{*}_{j}}\prod_{k=1}^{J_{2}}\frac{\zeta-\hat{\zeta}_{k}}{\zeta-\bar{\hat{\zeta}}_{k}}t_{22}(\zeta)\,,
\end{aligned}
\end{equation}
they are analytic in $D_{-}$ and $D_{+}$, respectively but both have no zeros and $A(\zeta)\bar{A}(\zeta)=t_{11}(\zeta) t_{22}(\zeta)$. According to the asymptotic behavior of $t_{11}(\zeta)$ as $\zeta\rightarrow\infty$ and combining with Eq.~(\ref{3.8}) as well as the definition of reflection coefficients, we obtain the trace formula
\begin{equation}\label{3.30}
t_{22}(\zeta)=\Theta_{-\infty}\prod_{j=1}^{J_{1}}\frac{\zeta-\bar{\zeta}_{j}}{\zeta-\zeta_{j}}\frac{\zeta-\bar{\zeta}^{*}_{j}}{\zeta-\zeta^{*}_{j}}\prod_{k=1}^{J_{2}}\frac{\zeta-\bar{\hat{\zeta}}_{k}}{\zeta-\hat{\zeta}_{k}} e^{-\frac{1}{2\pi i}\oint_{\Sigma^{'}}\frac{\ln \left(1-\rho(\omega)\bar{\rho}(\omega)\right)}{\omega-\zeta}d\omega}\,,\zeta\in D_{+}\,.
\end{equation}
In the same way, we can get another trace formula
\begin{equation}\label{3.31}
t_{11}(\zeta)=\prod_{j=1}^{J_{1}}\frac{\zeta-\zeta_{j}}{\zeta-\bar{\zeta}_{j}}\frac{\zeta-\zeta^{*}_{j}}{\zeta-\bar{\zeta}^{*}_{j}}\prod_{k=1}^{J_{2}}\frac{\zeta-\hat{\zeta}_{k}}{\zeta-\bar{\hat{\zeta}}_{k}} e^{\frac{1}{2\pi i}\oint_{\Sigma^{'}}\frac{\ln \left(1-\rho(\omega)\bar{\rho}(\omega)\right)}{\omega-\zeta}d\omega}\,,\zeta\in D_{-}\,.
\end{equation}

We consider the discrete eigenvalues under the asymptotic behavior~(\ref{3.15}) of scattering coefficients $t_{11}(\zeta)$ and $t_{22}(\zeta)$. The trace formulae meet the following asymptotic behavior under the reflectionless condition:
\begin{subequations}\label{3.32}
\begin{align}
&\prod_{j=1}^{J_{1}}\frac{|1/r-\zeta_{j}|^{2}}{|1/r-\bar{\zeta}_{j}|^{2}}\prod_{k=1}^{J_{2}}\frac{1/r-\hat{\zeta}_{k}}{1/r-\bar{\hat{\zeta}}_{k}}=-1\,,\\
&\Theta_{-\infty}\prod_{j=1}^{J_{1}}\frac{|\bar{\zeta}_{j}|^{2}}{|\zeta_{j}|^{2}}\prod_{k=1}^{J_{2}}\frac{\bar{\hat{\zeta}}_{k}}{\hat{\zeta}_{k}}=1\,,\\
&\Theta_{-\infty}\prod_{j=1}^{J_{1}}\frac{|r-\bar{\zeta}_{j}|^{2}}{|r-\zeta_{j}|^{2}}\prod_{k=1}^{J_{2}}\frac{r-\bar{\hat{\zeta}}_{k}}{r-\hat{\zeta}_{k}}=-1\,.
\end{align}
\end{subequations}
By direct calculation, there are no discrete eigenvalues $\zeta_{j}$, $\hat{\zeta}_{k}$ in $D_{-}$ that satisfy Eqs.~(\ref{3.32}), namely, this case exists no soliton solutions.

\vspace{5mm} \noindent\textbf{3.3 The time evolution}
\\\hspace*{\parindent}

We denote the potentials as $q_{n}(t)$, $r_{n}(t)$ with the boundary conditions
\begin{equation}\label{3.33}
{\lim_{n\to \pm \infty}}q_{n}(t)=q_{\pm}(t)=q_{0}e^{i (\theta_{\pm}-2q_{0}^{2}t)}\,, \,{\lim_{n\to \pm\infty}}r_{n}(t)=q_{\mp}^{*}(t)=q_{0}e^{-i (\theta_{\mp}-2q_{0}^{2}t)}.
\end{equation}
Through the same analysis as Case~(\ref{c1}), we get that $t_{11}(z,t)$ and $t_{22}(z,t)$ are also independent of $t$ but
\begin{equation}\label{3.34}
t_{21}(z,t)=t_{21}(z,0)e^{-2iq_{0}^{2}t+i\,\gamma(z)t}\,,\,t_{12}(z,t)=t_{12}(z,0)e^{2iq_{0}^{2}t-i\,\gamma(z)t}
\end{equation}
with $\gamma(z)=r(\lambda-\lambda^{-1})(z-z^{-1})$.
Furthermore, we can also obtain the time evolution of norming constants
\begin{equation}\label{3.35}
C_{j}(t)=C_{j}(0)e^{-2iq_{0}^{2}t+i\,\gamma(z_{j})t}\,,\,\bar{C}_{j}(t)=\bar{C}_{j}(0)e^{2iq_{0}^{2}t-i\,\gamma(\bar{z}_{j})t}.
\end{equation}

\vspace{7mm}\noindent\textbf{4 Case (\ref{c3}): $\sigma=-1$, $\Delta\theta=0$}
\hspace*{\parindent}
\renewcommand{\theequation}{4.\arabic{equation}}\setcounter{equation}{0}\\

In this section, we will investigate the IST for Eq.~(\ref{1.3}) with NZBCs~(\ref{1.4}) meeting the case~(\ref{c3}).

\vspace{5mm} \noindent\textbf{4.1 The direct scattering problem}
\\\hspace*{\parindent}

We also rewrite the NZBCs~(\ref{1.4}) and ${\lim_{n\to \pm \infty}}r_{n}(t)$ without the time variable $t$ as
\begin{equation}\label{4.1}
{\lim_{n\to \pm \infty}}q_{n}=q_{\pm}=q_{0}e^{i \vartheta_{\pm}}\,,{\lim_{n\to \pm \infty}}r_{n}=r_{\pm}=-q_{0}e^{-i \vartheta_{\mp}}\,,
\end{equation}
where $\vartheta_{\pm}=\theta-2q_{0}^{2}t$ is also real.

\vspace{5mm}\noindent\textbf{4.1.1 Eigenfunctions, scattering matrix, uniformization variable}
\\\hspace*{\parindent}

Under the boundary conditions~(\ref{4.1}) and symmetry reduction condition $r_{n}(t)=-q_{-n}^{*}(t)$, the solution matrices $\mathbf{\Phi}_{n}(z)$ and $\mathbf{\Psi}_{n}(z)$ of Eq.~(\ref{1.5}a)  satisfy the similar boundary conditions with Eqs.~(\ref{3.2}) and determinants~(\ref{3.6}), except that $q_{\pm}$ and $r_{\pm}$ corresponds to Eq.~(\ref{4.1}).

In fact, in this case, lots of analyses are similar to case~(\ref{c2}), including the defination of continuous spectrum~(\ref{3.4}), scattering matrices~(\ref{3.7}), uniformization variable $\zeta$, modified eigenfunctions~(\ref{3.11}) and modified scattering matrices~(\ref{3.13}).

\vspace{5mm}\noindent\textbf{4.1.2 Analyticity of modified eigenfunctions and scattering coefficients}
\\\hspace*{\parindent}

By the similar analysis except $\dot{\mathbf{U}}'_{n}$ becoming
\begin{align}\label{4.2}
&\dot{\mathbf{U}}'_{n}=\frac{1}{-q_{0}^{2}+(r\lambda-z)^{2}}\nonumber\\
&\left(
\begin{array}{cc}
(r\lambda-z)(\eta_{n}^{[1]}+\eta_{-n}^{[1]*})+(r\lambda q_{n}-z \eta_{n}^{[2]})\eta_{-(n+1)}^{[3]*} &-\frac{(r\lambda-z)^{2}\eta_{-n}^{[2]*}+\dot{r}_{n}\eta_{-n}^{[1]*}-(z(r\lambda-z)-q_{n}\dot{r}_{n})\eta_{-(n+1)}^{[3]*}}{\lambda}\\
\lambda\left((r\lambda-z)^{2}\eta_{n}^{[2]}-\dot{q}_{n}\eta_{n}^{[1]}-(z^{-1}(r\lambda-z)+\dot{q}_{n}r_{n})\eta_{n}^{[3]}\right) &-(r\lambda-z)(\eta_{n}^{[1]}+\eta_{-n}^{[1]*})+(r\lambda^{-1} r_{n}+z^{-1}\eta_{-n}^{[2]*})\eta_{n}^{[3]}\\
\end{array}
\right)
\end{align}
with $\eta_{n}^{[j]}$, $j=1,2,3$ same as Section 3, we can also obtain $\mathbf{Theorem~\ref{T3}}$:
\begin{theorem}\label{T3}
Under the condition of the potentials $q_{n}, r_{n}, \dot{q}_{n}, \dot{r}_{n}\in \left\{ f_{n}: \sum\limits _{l=\pm\infty}^{n}|f_{l}-\lim_{k\to \pm \infty}f_{k}|<\infty, \forall n\in\mathbb{Z}\right\}$, $M_{n}$, $N_{n}$, $t_{11}(\zeta)=\tilde{t}_{11}(\zeta)$ are analytic in $D_{-}$ defined as Eq.~(\ref{3.10}) (except for the only possible pole $\zeta=\infty$) and $\bar{M}_{n}$, $\bar{N}_{n}$, $t_{22}(\zeta)=\tilde{t}_{22}(\zeta)$ are analytic in $D_{+}$ defined as Eq.~(\ref{3.10}) (except for the only possible pole $\zeta=0$).
\end{theorem}

\vspace{5mm}\noindent\textbf{4.1.3 Asymptotic behavior of modified eigenfunctions and scattering coefficients}
\\\hspace*{\parindent}

The asymptotic behavior of $z^{2}$, $\lambda^{2}$, $z\lambda$ is also same as Table~\ref{tab:1} and the asymptotics of modified eigenfunctions and scattering coefficients are the same as Eqs.~(\ref{2.42})-(\ref{2.46}) except $q_{\pm}$ and $r_{\pm}$ as well.

\vspace{5mm} \noindent\textbf{4.1.4 Symmetries of scattering coefficients}
\\\hspace*{\parindent}

By the same consideration as before, the first symmetry about eigenfunctions and scattering coefficients under the fact when $z\mapsto z$ and $\lambda\mapsto\frac{1}{\lambda}$, $\zeta\mapsto\bar{\zeta}=\frac{r\zeta-1}{\zeta-r}$ is:
\begin{equation}\label{4.3}
\phi_{n}(z,\frac{1}{\lambda})=\frac{r/\lambda-z}{- r_{-}}\bar{\phi}_{n}(z,\lambda)\,,
\end{equation}
\begin{equation}\label{4.4}
\psi_{n}(z,\frac{1}{\lambda})=\frac{r/\lambda-z}{q_{+}}\bar{\psi}_{n}(z,\lambda)\,,
\end{equation}
\begin{equation}\label{4.5}
t_{11}(\zeta)=t_{22}(\bar{\zeta})\,,\,t_{21}(\zeta)=-\frac{q_{+}}{r_{-}}\,t_{12}(\bar{\zeta}).
\end{equation}

The second symmetry when $z\mapsto z^{*}$ and $\lambda\mapsto\frac{1}{\lambda^{*}}$, $\zeta\mapsto\bar{\zeta}^{*}=\frac{r\zeta^{*}-1}{\zeta^{*}-r}$ is:
\begin{equation}\label{4.6}
\left(
\begin{array}{cc}
\bar{N}_{n+1}^{(1)}(\zeta)\\
\bar{N}_{n+1}^{(2)}(\zeta)\\
\end{array}
\right)=
-\frac{\zeta-r}{r_{+}\Theta_{-\infty}v_{n}}\left(
\begin{array}{cc}
M_{-n}^{(2)*}(\bar{\zeta}^{*})\\
M_{-n}^{(1)*}(\bar{\zeta}^{*})\\
\end{array}
\right)\,,
\end{equation}
\begin{equation}\label{4.7}
\left(
\begin{array}{cc}
N_{n+1}^{(1)}(\zeta)\\
N_{n+1}^{(2)}(\zeta)\\
\end{array}
\right)=-
\frac{1/\zeta-r}{q_{+}\Theta_{-\infty}v_{n}}\left(
\begin{array}{cc}
\bar{M}_{-n}^{(2)*}(\bar{\zeta}^{*})\\
\bar{M}_{-n}^{(1)*}(\bar{\zeta}^{*})\\
\end{array}
\right)\,,
\end{equation}
\begin{equation}\label{4.8}
t_{11}(\zeta)=t_{22}^{*}(\bar{\zeta}^{*})\,.
\end{equation}

\vspace{5mm} \noindent\textbf{4.2 The inverse scattering problem}
\\\hspace*{\parindent}

In this section, we simply provide a series of expressions for modified eigenfunctions, $\frac{1}{\Theta_{n}}$, reconstructed potential obtained from analyzing the RH problem.

\vspace{5mm}\noindent\textbf{4.2.1 The RH problem and reconstructed potential}
\\\hspace*{\parindent}

By direct calculation, we finally obtain
\begin{align}\label{4.9}
&\bar{N}_{n}(\zeta)=\left(
\begin{array}{cc}
\frac{q_{+}}{\Theta_{n}}\\
\zeta-r\\
\end{array}
\right)+(\zeta-r)\sum\limits _{j=1}^{J}\frac{C_{j}\,\lambda(\zeta_{j})^{-2n}}{(\zeta_{j}-r)(\zeta-\zeta_{j})}\,N_{n}(\zeta_{j})-
\frac{\zeta-r}{2\pi i}\oint_{\Sigma^{'}}\frac{\lambda(\omega)^{-2n}}{(\omega-\zeta)(\omega-r)}\rho(\omega)N_{n}(\omega)
d\omega\,,\zeta\in D_{+}\,,\\
&N_{n}(\zeta)=\left(
\begin{array}{cc}
r-\frac{1}{\zeta}\\
-\frac{r_{+}}{\Theta_{n}}\\
\end{array}
\right)+(\zeta-\frac{1}{r})\sum\limits _{j=1}^{\bar{J}}\frac{\bar{C}_{j}\,\lambda(\bar{\zeta}_{j})^{2n}}{(\bar{\zeta}_{j}-\frac{1}{r})(\zeta-\bar{\zeta}_{j})}\,\bar{N}_{n}(\bar{\zeta}_{j})+
\frac{\zeta-\frac{1}{r}}{2\pi i}\oint_{\Sigma^{'}}\frac{\lambda(\omega)^{2n}}{(\omega-\zeta)(\omega-\frac{1}{r})}\bar{\rho}(\omega)\bar{N}_{n}(\omega)
d\omega\,,\zeta\in D_{-}\,,\\
&\frac{1}{\Theta_{n}}=1-\sum\limits _{j=1}^{J}\frac{C_{j}\,\lambda(\zeta_{j})^{-2n}}{\zeta_{j}(\zeta_{j}-r)}\,N^{(2)}_{n}(\zeta_{j})-
\frac{1}{2\pi i}\oint_{\Sigma^{'}}\frac{\lambda(\omega)^{-2n}}{\omega(\omega-r)}\rho(\omega)N^{(2)}_{n}(\omega)
d\omega\,,\\
&q_{n}=q_{+}+\Theta_{n}\sum\limits _{j=1}^{J}\frac{r C_{j}\,\lambda(\zeta_{j})^{-2n}}{\zeta_{j}(\zeta_{j}-r)}\,N^{(1)}_{n}(\zeta_{j})+\frac{\Theta_{n}}{2\pi i}\oint_{\Sigma^{'}}\frac{r\lambda(\omega)^{-2n}}{\omega(\omega-r)}\rho(\omega)N^{(1)}_{n}(\omega)
d\omega
\end{align}
with the symmetry of norming constants $C_{j}=-\frac{q_{+}^{2}}{(\bar{\zeta}_{j}-r)^{2}}\bar{C}_{j}$.

\vspace{5mm} \noindent\textbf{4.2.2 The reflectionless condition}
\\\hspace*{\parindent}

Under the reflectionless condition $\rho(\zeta)=0$, $\bar{\rho}(\zeta)=0$ for $\Sigma^{'}$, we can also obtain:
\begin{subequations}\label{4.13}
\begin{align}
&\bar{N}^{(1)}_{n}(\zeta)-(\zeta-r)\sum\limits _{j=1}^{J}\frac{C_{j}\,\lambda(\zeta_{j})^{-2n}}{(\zeta_{j}-r)(\zeta-\zeta_{j})}\,N^{(1)}_{n}(\zeta_{j})-\frac{q_{+}}{\Theta_{n}}=0\,,\\
&\bar{N}^{(2)}_{n}(\zeta)-(\zeta-r)\sum\limits _{j=1}^{J}\frac{C_{j}\,\lambda(\zeta_{j})^{-2n}}{(\zeta_{j}-r)(\zeta-\zeta_{j})}\,N^{(2)}_{n}(\zeta_{j})=\zeta-r\,,\\
&N^{(1)}_{n}(\zeta)-(\zeta-\frac{1}{r})\sum\limits _{j=1}^{\bar{J}}\frac{\bar{C}_{j}\,\lambda(\bar{\zeta}_{j})^{2n}}{(\bar{\zeta}_{j}-\frac{1}{r})(\zeta-\bar{\zeta}_{j})}\,\bar{N}^{(1)}_{n}(\bar{\zeta}_{j})=r-\frac{1}{\zeta}\,,\\
&N^{(2)}_{n}(\zeta)+\frac{r_{+}}{\Theta_{n}}-(\zeta-\frac{1}{r})\sum\limits _{j=1}^{\bar{J}}\frac{\bar{C}_{j}\,\lambda(\bar{\zeta}_{j})^{2n}}{(\bar{\zeta}_{j}-\frac{1}{r})(\zeta-\bar{\zeta}_{j})}\,\bar{N}^{(2)}_{n}(\bar{\zeta}_{j})=0\,,\\
&\frac{1}{\Theta_{n}}=1-\sum\limits _{j=1}^{J}\frac{C_{j}\,\lambda(\zeta_{j})^{-2n}}{\zeta_{j}(\zeta_{j}-r)}\,N^{(2)}_{n}(\zeta_{j})
\end{align}
\end{subequations}
and the reconstructed potential
\begin{equation}\label{4.14}
q_{n}=q_{+}+\Theta_{n}\sum\limits _{j=1}^{J}\frac{r C_{j}\,\lambda(\zeta_{j})^{-2n}}{\zeta_{j}(\zeta_{j}-r)}\,N^{(1)}_{n}(\zeta_{j})
\end{equation}
according which we can obtain the soliton solutions of Eq.~(\ref{1.3}) after substituting the scattering coefficients $\{\zeta_{j},\bar{\zeta}_{k},C_{j},\bar{C}_{k},j=1,2,\cdot\cdot\cdot,J,,k=1,2,\cdot\cdot\cdot,\bar{J}\}$ that we will analyze next.

\vspace{5mm} \noindent\textbf{4.2.3 The trace formula and discrete eigenvalues}
\\\hspace*{\parindent}

From Eqs.~(\ref{4.5}) and~(\ref{4.8}), we know that the zeros of scattering coefficients also have the symmetry so that the discrete eigenvalues are in quartets: $\{\zeta_{j}, \zeta^{*}_{j}, \bar{\zeta}_{j}, \bar{\zeta}^{*}_{j}\}$, $j=1,2,\cdot\cdot\cdot,J$, in each set the first pair is the simple zeros of $t_{11}(\zeta)$ in $D_{-}$ and the second pair is the simple zeros of $t_{22}(\zeta)$ in $D_{+}$. In addition, the all $J$ simple zeros of $t_{11}(\zeta)$ can also be divided into two classes denoted by $\zeta_{j}$, $j=1,2,\cdot\cdot\cdot,J_{1}$ and $\hat{\zeta}_{k}$, $k=1,2,\cdot\cdot\cdot,J_{2}$, where $2J_{1}+J_{2}=J$, Im$(\zeta_{j})\neq0$, Im$(\hat{\zeta}_{k})=0$, then we obtain that $t_{22}(\zeta)$ has $J$ simple zeros $\bar{\zeta}_{j}$, $\bar{\zeta}^{*}_{j}$, $\bar{\hat{\zeta}}_{k}=\frac{r\hat{\zeta}_{k}-1}{\hat{\zeta}_{k}-r}$ as well. Using similar calculation as before, we can obtain the trace formula
\begin{subequations}\label{4.15}
\begin{align}
&t_{22}(\zeta)=\Theta_{-\infty}\prod_{j=1}^{J_{1}}\frac{\zeta-\bar{\zeta}_{j}}{\zeta-\zeta_{j}}\frac{\zeta-\bar{\zeta}^{*}_{j}}{\zeta-\zeta^{*}_{j}}\prod_{k=1}^{J_{2}}\frac{\zeta-\bar{\hat{\zeta}}_{k}}{\zeta-\hat{\zeta}_{k}} e^{-\frac{1}{2\pi i}\oint_{\Sigma^{'}}\frac{\ln \left(1-\rho(\omega)\bar{\rho}(\omega)\right)}{\omega-\zeta}d\omega}\,,\zeta\in D_{+}\,,\\
&t_{11}(\zeta)=\prod_{j=1}^{J_{1}}\frac{\zeta-\zeta_{j}}{\zeta-\bar{\zeta}_{j}}\frac{\zeta-\zeta^{*}_{j}}{\zeta-\bar{\zeta}^{*}_{j}}\prod_{k=1}^{J_{2}}\frac{\zeta-\hat{\zeta}_{k}}{\zeta-\bar{\hat{\zeta}}_{k}} e^{\frac{1}{2\pi i}\oint_{\Sigma^{'}}\frac{\ln \left(1-\rho(\omega)\bar{\rho}(\omega)\right)}{\omega-\zeta}d\omega}\,,\zeta\in D_{-}\,.
\end{align}
\end{subequations}
and consider the asymptotic behavior of the scattering coefficients, under the reflectionless condition, we also have the following constraint:
\begin{subequations}\label{4.16}
\begin{align}
&\prod_{j=1}^{J_{1}}\frac{|1/r-\zeta_{j}|^{2}}{|1/r-\bar{\zeta}_{j}|^{2}}\prod_{k=1}^{J_{2}}\frac{1/r-\hat{\zeta}_{k}}{1/r-\bar{\hat{\zeta}}_{k}}=1\,,\\
&\Theta_{-\infty}\prod_{j=1}^{J_{1}}\frac{|\bar{\zeta}_{j}|^{2}}{|\zeta_{j}|^{2}}\prod_{k=1}^{J_{2}}\frac{\bar{\hat{\zeta}}_{k}}{\hat{\zeta}_{k}}=1\,,\\
&\Theta_{-\infty}\prod_{j=1}^{J_{1}}\frac{|r-\bar{\zeta}_{j}|^{2}}{|r-\zeta_{j}|^{2}}\prod_{k=1}^{J_{2}}\frac{r-\bar{\hat{\zeta}}_{k}}{r-\hat{\zeta}_{k}}=1\,.
\end{align}
\end{subequations}

Setting $J=1$, then $J_{1}=0$, $J_{2}=1$, which means that $\hat{\zeta}_{1}=r\pm q_{0}$ are branch points we won't consider. When $J=2$, there are two scenarios. The first scenario is $J_{1}=1$, $J_{2}=0$, we can get that $\zeta_{1}$ is on the circle $|\zeta-r|=q_{0}$ that is a part of continuous spectrum we won't deal with either. The second scenario is $J_{1}=0$, $J_{2}=2$, then we can get $\hat{\zeta}_{2}=\frac{\hat{\zeta}_{1}-r}{r\hat{\zeta}_{1}-1}=\frac{1}{\bar{\hat{\zeta}}_{1}}$ and $\{\hat{\zeta}_{1}, \bar{\hat{\zeta}}_{1}\}$, $\{\hat{\zeta}_{2}, \bar{\hat{\zeta}}_{2}\}$ are two sets of discrete eigenvalues used to construct the 2-eigenvalue soliton solution.

\vspace{5mm} \noindent\textbf{4.3 The time evolution}
\\\hspace*{\parindent}

We denote the potentials as $q_{n}(t)$, $r_{n}(t)$ with the boundary conditions
\begin{equation}\label{4.17}
{\lim_{n\to \pm \infty}}q_{n}(t)=q_{\pm}(t)=q_{0}e^{i (\theta-2q_{0}^{2}t)}\,, \,{\lim_{n\to \pm\infty}}r_{n}(t)=-q_{\mp}^{*}(t)=-q_{0}e^{-i (\theta-2q_{0}^{2}t)}.
\end{equation}
and we can get that $t_{11}(z,t)$ and $t_{22}(z,t)$ are still independent of $t$ but
\begin{equation}\label{4.18}
t_{21}(z,t)=t_{21}(z,0)e^{-2iq_{0}^{2}t+i\,\gamma(z)t}\,,\,t_{12}(z,t)=t_{12}(z,0)e^{2iq_{0}^{2}t-i\,\gamma(z)t}
\end{equation}
with $\gamma(z)=r(\lambda-\lambda^{-1})(z-z^{-1})$.
Furthermore, the time evolution of norming constants are:
\begin{equation}\label{4.19}
C_{j}(t)=C_{j}(0)e^{-2iq_{0}^{2}t+i\,\gamma(z_{j})t}\,,\,\bar{C}_{j}(t)=\bar{C}_{j}(0)e^{2iq_{0}^{2}t-i\,\gamma(\bar{z}_{j})t}.
\end{equation}

\vspace{5mm} \noindent\textbf{4.4 The soliton solutions}
\\\hspace*{\parindent}

In this part, we construct the 2-eigenvalue soliton solution by setting $J=2$ and getting the discrete eigenvalues $\zeta_{1}$, $\bar{\zeta}_{1}=\frac{r\zeta_{1}-1}{\zeta_{1}-r}$, $\zeta_{2}=\frac{1}{\bar{\zeta}_{1}}$, $\bar{\zeta}_{2}=\frac{r\zeta_{2}-1}{\zeta_{2}-r}\in\mathbb{R}$ (For convenience, we also omit the `` $\hat{}$ " above the variables). By the similar integration and introducing notations as Case (\ref{c1}), we obtain the second-order soliton solution:
\begin{equation}
q_{n}^{[2]}(t)=q_{+}(t)+r\,R_{9}\frac{X_{1}}{X_{9}}+r\,R_{10}\frac{X_{2}}{X_{9}}\,,
\end{equation}
in which the parameters are given in Section $2.4$.

\vspace{7mm}\noindent\textbf{5 Case (\ref{c4}): $\sigma=-1$, $\Delta\theta=\pi$}
\hspace*{\parindent}
\renewcommand{\theequation}{5.\arabic{equation}}\setcounter{equation}{0}\\

The last case is that Eq.~(\ref{1.3}) with the NZBCs~(\ref{1.4}) and ${\lim_{n\to \pm \infty}}r_{n}(t)$, which can also be temporarily written as
\begin{equation}\label{5.1}
{\lim_{n\to \pm \infty}}q_{n}=q_{\pm}=q_{0}e^{i \vartheta_{\pm}}\,,{\lim_{n\to \pm \infty}}r_{n}=r_{\pm}=-q_{0}e^{-i \vartheta_{\mp}}\,,
\end{equation}
where $\vartheta_{\pm}=\theta_{\pm}+2q_{0}^{2}t$ is real.

\vspace{5mm} \noindent\textbf{5.1 The direct scattering problem}
\\\hspace*{\parindent}

Under the boundary conditions~(\ref{5.1}) and symmetry reduction condition $r_{n}(t)=-q_{-n}^{*}(t)$, the solution matrices $\mathbf{\Phi}_{n}(z)$ and $\mathbf{\Psi}_{n}(z)$ of Eq.~(\ref{1.5}a)  satisfy the similar boundary conditions with Eqs.~(\ref{2.2}) and determinants~(\ref{2.6}), except that $q_{\pm}$ and $r_{\pm}$ corresponds to Eq.~(\ref{5.1}).

In fact, in this case, lots of analyses are also similar to case~(\ref{c1}) such as the defination of continuous spectrum~(\ref{2.4}), scattering matrices~(\ref{2.7}), uniformization variable $\zeta$, modified eigenfunctions~(\ref{2.15}), modified scattering matrices~(\ref{2.20}).

By the similar analysis except different $\dot{\mathbf{U}}'_{n}$, we can also prove that under the condition of the potentials $q_{n}, r_{n}, \dot{q}_{n}, \dot{r}_{n}\in \left\{ f_{n}: \sum\limits _{l=\pm\infty}^{n}|f_{j}-\lim_{k\to \pm \infty}f_{k}|<\infty, \forall n\in\mathbb{Z}\right\}$, $M_{n}$, $N_{n}$, $t_{11}(\zeta)=\tilde{t}_{11}(\zeta)$ are analytic in $D_{-}$ (except for the only possible pole $\zeta=\infty$) and $\bar{M}_{n}$, $\bar{N}_{n}$, $t_{22}(\zeta)=\tilde{t}_{22}(\zeta)$ are analytic in $D_{+}$ (except for the only possible pole $\zeta=0$), where $D_{\pm}$ are defined as Eq.~(\ref{2.14}).

The asymptotic behavior of $z^{2}$, $\lambda^{2}$, $z\lambda$ is also as before, the asymptotic behavior of modified eigenfunctions is the same as Eqs.~(\ref{2.42})-(\ref{2.45}) except $q_{\pm}$ and $r_{\pm}$ as well and the one of scattering coefficients is similar to Eqs.~(\ref{3.15}).

By the same consideration as before, the symmetry about eigenfunctions and scattering coefficients is
\begin{align}\label{5.2}
\phi_{n}(z,\frac{1}{\lambda})=\frac{r/\lambda-z}{- r_{-}}\bar{\phi}_{n}(z,\lambda)\,,\,
\psi_{n}(z,\frac{1}{\lambda})=\frac{r/\lambda-z}{q_{+}}\bar{\psi}_{n}(z,\lambda)\,,\,t_{11}(\zeta)=t_{22}(\bar{\zeta})\,,\,t_{21}(\zeta)=-\frac{q_{+}}{r_{-}}\,t_{12}(\bar{\zeta})\,,
\end{align}
\begin{align}\label{5.3}
\bar{N}_{n+1}(\zeta)=
-\frac{\zeta-r}{r_{+}\Theta_{-\infty}v_{n}}\left(
\begin{array}{cc}
M_{-n}^{(2)*}(\bar{\zeta}^{*})\\
M_{-n}^{(1)*}(\bar{\zeta}^{*})\\
\end{array}
\right)\,,\,
N_{n+1}(\zeta)=-
\frac{1/\zeta-r}{q_{+}\Theta_{-\infty}v_{n}}\left(
\begin{array}{cc}
\bar{M}_{-n}^{(2)*}(\bar{\zeta}^{*})\\
\bar{M}_{-n}^{(1)*}(\bar{\zeta}^{*})\\
\end{array}
\right)\,,\,
t_{11}(\zeta)=-t_{22}^{*}(\bar{\zeta}^{*})\,,
\end{align}
and the discrete eigenvalues are also in quartets: $\{\zeta_{j}, \zeta^{*}_{j}, \bar{\zeta}_{j}, \bar{\zeta}^{*}_{j}\}$, $j=1,2,\cdot\cdot\cdot,J$, where $J=2J_{1}+J_{2}$, $J_{1}$, $J_{2}$ are the quantities of $\zeta_{j}$ meeting Im$(\zeta_{j})\neq0$ and $\hat{\zeta}_{k}$ meeting Im$(\hat{\zeta}_{k})=0$, respectively.

\vspace{5mm} \noindent\textbf{5.2 The inverse scattering problem}
\\\hspace*{\parindent}

In the inverse scattering problem, we still define and solve a RH problem to reconstruct the potential so that under the reflectionless condition, we obtain the linear system:
\begin{subequations}\label{5.4}
\begin{align}
&\bar{N}^{(1)}_{n}(\zeta)-(\zeta-r)\sum\limits _{j=1}^{J}\frac{C_{j}\,\lambda(\zeta_{j})^{-2n}}{(\zeta_{j}-r)(\zeta-\zeta_{j})}\,N^{(1)}_{n}(\zeta_{j})-\frac{q_{+}}{\Theta_{n}}=0\,,\\
&\bar{N}^{(2)}_{n}(\zeta)-(\zeta-r)\sum\limits _{j=1}^{J}\frac{C_{j}\,\lambda(\zeta_{j})^{-2n}}{(\zeta_{j}-r)(\zeta-\zeta_{j})}\,N^{(2)}_{n}(\zeta_{j})=\zeta-r\,,\\
&N^{(1)}_{n}(\zeta)-(\zeta-\frac{1}{r})\sum\limits _{j=1}^{\bar{J}}\frac{\bar{C}_{j}\,\lambda(\bar{\zeta}_{j})^{2n}}{(\bar{\zeta}_{j}-\frac{1}{r})(\zeta-\bar{\zeta}_{j})}\,\bar{N}^{(1)}_{n}(\bar{\zeta}_{j})=r-\frac{1}{\zeta}\,,\\
&N^{(2)}_{n}(\zeta)+\frac{r_{+}}{\Theta_{n}}-(\zeta-\frac{1}{r})\sum\limits _{j=1}^{\bar{J}}\frac{\bar{C}_{j}\,\lambda(\bar{\zeta}_{j})^{2n}}{(\bar{\zeta}_{j}-\frac{1}{r})(\zeta-\bar{\zeta}_{j})}\,\bar{N}^{(2)}_{n}(\bar{\zeta}_{j})=0\,,\\
&\frac{1}{\Theta_{n}}=1-\sum\limits _{j=1}^{J}\frac{C_{j}\,\lambda(\zeta_{j})^{-2n}}{\zeta_{j}(\zeta_{j}-r)}\,N^{(2)}_{n}(\zeta_{j})
\end{align}
\end{subequations}
and the reconstructed potential
\begin{equation}\label{5.5}
q_{n}=q_{+}+\Theta_{n}\sum\limits _{j=1}^{J}\frac{r C_{j}\,\lambda(\zeta_{j})^{-2n}}{\zeta_{j}(\zeta_{j}-r)}\,N^{(1)}_{n}(\zeta_{j})\,.
\end{equation}

By the same way as before, the trace formulae of $t_{11}(\zeta)$ and $t_{22}(\zeta)$ are
\begin{subequations}\label{5.6}
\begin{align}
&t_{22}(\zeta)=\Theta_{-\infty}\prod_{j=1}^{J_{1}}\frac{\zeta-\bar{\zeta}_{j}}{\zeta-\zeta_{j}}\frac{\zeta-\bar{\zeta}^{*}_{j}}{\zeta-\zeta^{*}_{j}}\prod_{k=1}^{J_{2}}\frac{\zeta-\bar{\hat{\zeta}}_{k}}{\zeta-\hat{\zeta}_{k}} e^{-\frac{1}{2\pi i}\oint_{|\omega|=1}\frac{\ln \left(1-\rho(\omega)\bar{\rho}(\omega)\right)}{\omega-\zeta}d\omega}\,,\zeta\in D_{+}\,,\\
&t_{11}(\zeta)=\prod_{j=1}^{J_{1}}\frac{\zeta-\zeta_{j}}{\zeta-\bar{\zeta}_{j}}\frac{\zeta-\zeta^{*}_{j}}{\zeta-\bar{\zeta}^{*}_{j}}\prod_{k=1}^{J_{2}}\frac{\zeta-\hat{\zeta}_{k}}{\zeta-\bar{\hat{\zeta}}_{k}} e^{\frac{1}{2\pi i}\oint_{|\omega|=1}\frac{\ln \left(1-\rho(\omega)\bar{\rho}(\omega)\right)}{\omega-\zeta}d\omega}\,,\zeta\in D_{-}\,.
\end{align}
\end{subequations}
Under the reflectionless condition and taking the derivative of above trace formulae with respect to $\zeta$ gives
\begin{subequations}\label{5.7}
\begin{align}
&t_{22}^{'}(\zeta)=\Theta_{-\infty}\prod_{j=1}^{J_{1}}\frac{\zeta-\bar{\zeta}_{j}}{\zeta-\zeta_{j}}\frac{\zeta-\bar{\zeta}^{*}_{j}}{\zeta-\zeta^{*}_{j}}\prod_{k=1}^{J_{2}}\frac{\zeta-\bar{\hat{\zeta}}_{k}}{\zeta-\hat{\zeta}_{k}}\left[
\sum_{j=1}^{J_{1}}\left(\frac{1}{\zeta-\bar{\zeta}_{j}}+\frac{1}{\zeta-\bar{\zeta}^{*}_{j}}-\frac{1}{\zeta-\zeta_{j}}-\frac{1}{\zeta-\zeta^{*}_{j}}\right)+\sum_{k=1}^{J_{2}}\left(\frac{1}{\zeta-\bar{\hat{\zeta}}_{j}}-\frac{1}{\zeta-\hat{\zeta}_{k}}\right)\right] \,,\\
&t_{11}^{'}(\zeta)=\prod_{j=1}^{J_{1}}\frac{\zeta-\zeta_{j}}{\zeta-\bar{\zeta}_{j}}\frac{\zeta-\zeta^{*}_{j}}{\zeta-\bar{\zeta}^{*}_{j}}\prod_{k=1}^{J_{2}}\frac{\zeta-\hat{\zeta}_{k}}{\zeta-\bar{\hat{\zeta}}_{k}} \left[\sum_{j=1}^{J_{1}}\left(\frac{1}{\zeta-\zeta_{j}}+\frac{1}{\zeta-\zeta^{*}_{j}}-\frac{1}{\zeta-\bar{\zeta}_{j}}-\frac{1}{\zeta-\bar{\zeta}^{*}_{j}}\right)+\sum_{k=1}^{J_{2}}\left(\frac{1}{\zeta-\hat{\zeta}_{j}}-\frac{1}{\zeta-\bar{\hat{\zeta}}_{k}}\right)\right]\,.
\end{align}
\end{subequations}

Considering the asymptotic behavior of the scattering coefficient $t_{11}(\zeta)$ as $\zeta\rightarrow\frac{1}{r}$, $t_{22}(\zeta)$ as $\zeta\rightarrow 0$, $\zeta\rightarrow r$ under the reflectionless condition, we also have the constraint of discrete eigenvalues similar to Eqs.~(\ref{3.32}). Setting $J=1$, which only means $J_{1}=0$, $J_{2}=1$, we get $\bar{\hat{\zeta}}_{1}=\frac{1-q_{0}}{r}\in D_{+}$ on the circle $|\zeta-1/r|=q_{0}/r$, which can be used to construct the first-order soliton solution later. When $J=2$, there are no discrete eigenvalues that simultaneously meet the above three conditions.

When $J_{1}=0$, $J_{2}=1$, $\bar{\hat{\zeta}}_{1}=\frac{1-q_{0}}{r}\in D_{+}$ and the corresponding $\hat{\zeta}_{1}=\frac{r}{1-q_{0}}\in D_{-}$, according to the symmetry about eigenfunctions and scattering coefficients, we have $\bar{b}_{1}\bar{b}_{1}^{*}=\lambda(\bar{\hat{\zeta}}_{1})^{-4}$ so that we write $\bar{b}_{1}=\lambda(\bar{\hat{\zeta}}_{1})^{-2}e^{i\bar{\theta}_{1}}$, where $\bar{\theta}_{1}$ is a real parameter. According to Eqs.~(\ref{5.7}), we also have $t_{22}^{'}(\bar{\hat{\zeta}}_{1})=\frac{\lambda(\zeta_{1})^{2}}{\bar{\hat{\zeta}}_{1}-\hat{\zeta}_{1}}$, therefore, we get
\begin{align}\label{5.8}
\bar{C}_{1}=(\lambda(\bar{\hat{\zeta}}_{1}))^{-5}(\bar{\hat{\zeta}}_{1}-\hat{\zeta}_{1})e^{i\bar{\theta}_{1}}.
\end{align}

\vspace{5mm} \noindent\textbf{5.3 The time evolution}
\\\hspace*{\parindent}

The time evolution of scattering coefficients and norming constants are identical with case~(\ref{c1}), namely, $t_{11}(z,t)$ and $t_{22}(z,t)$ are independent of $t$ but $t_{21}(z,t)$, $t_{12}(z,t)$, $C_{j}(t)$ , $\bar{C}_{j}(t)$ are related to $t$:
\begin{align}
&t_{21}(z,t)=t_{21}(z,0)e^{2iq_{0}^{2}t+i\,\gamma(z)t}\,,\,t_{12}(z,t)=t_{12}(z,0)e^{-2iq_{0}^{2}t-i\,\gamma(z)t}\,,\\
&C_{j}(t)=C_{j}(0)e^{2iq_{0}^{2}t+i\,\gamma(z_{j})t}\,,\,\bar{C}_{j}(t)=\bar{C}_{j}(0)e^{-2iq_{0}^{2}t-i\,\gamma(\bar{z}_{j})t}.
\end{align}
Combining with Eq.~(\ref{5.8}), we have
\begin{equation}
\bar{C}_{1}(t)=(\lambda(\bar{\hat{\zeta}}_{1}))^{-5}(\bar{\hat{\zeta}}_{1}-\hat{\zeta}_{1})e^{i(\bar{\theta}_{1}-2q_{0}^{2}t-\gamma(\bar{\hat{\zeta}}_{1})t)}
\end{equation}
with
\begin{equation}
\gamma(\bar{\hat{\zeta}}_{1})=r^{2}\frac{(\bar{\hat{\zeta}}_{1}-\frac{2}{r}+\frac{1}{\bar{\hat{\zeta}}_{1}})(\bar{\hat{\zeta}}_{1}-2r+\frac{1}{\bar{\hat{\zeta}}_{1}})}{(\bar{\hat{\zeta}}_{1}-r)(\frac{1}{\bar{\hat{\zeta}}_{1}}-r)}.\nonumber
\end{equation}

\vspace{5mm} \noindent\textbf{5.4 The soliton solutions}
\\\hspace*{\parindent}

The soliton solutions can be obtained by solving algebraic system~(\ref{5.4}) with discrete eigenvalues and corresponding norming constants under the reflectionless condition. For $J=1$, considering discrete eigenvalues $\bar{\zeta}_{1}=\frac{1-q_{0}}{r}$, $\zeta_{1}=\frac{r\bar{\zeta}_{1}-1}{\bar{\zeta}_{1}-r}$ (We also omit the `` $\hat{}$ " of the real variables in discrete eigenvalues for brevity), norming constants $\bar{C}_{1}(t)$, $C_{1}(t)=-\frac{q_{+}^{2}(t)}{(\bar{\zeta}_{1}-r)^{2}}\bar{C}_{1}(t)$ and introducing $\mathbf{X}=(X_{1},...,X_{5})^{T}$, $\mathbf{Y}=(r-\frac{1}{\zeta_{1}},0,0,\bar{\zeta}_{1}-r,1)^{T}$, matrix $\mathbf{B}=(\mathbf{B}_{1},...,\mathbf{B}_{5})$, where
\begin{equation}
X_{1}=N^{(1)}_{n}(\zeta_{1})\,,\,X_{2}=N^{(2)}_{n}(\zeta_{1})\,,\,
X_{3}=\bar{N}^{(1)}_{n}(\bar{\zeta}_{1})\,,\,X_{4}=\bar{N}^{(2)}_{n}(\bar{\zeta}_{1})\,,\,
X_{5}=\frac{1}{\Theta_{n}}\,,\,\nonumber\\
\end{equation}
\begin{equation}
\mathbf{B}=\left (
\begin{array}{ccccc}
1 &0 &R_{1} &0 &0\\
0 &1 &0 &R_{1} &r_{+}\\
R_{2} &0 &1 &0 &-q_{+}\\
0 &R_{2} &0 &1 &0\\
0 &R_{3} &0 &0 &1\\
\end{array}
\right)\nonumber
\end{equation}
with
\begin{align}
R_{1}=-\frac{(\zeta_{1}-\frac{1}{r})\bar{C}_{1}(t)\lambda(\bar{\zeta}_{1})^{2n}}{(\bar{\zeta}_{1}-\frac{1}{r})(\zeta_{1}-\bar{\zeta}_{1})}\,,\,
R_{2}=-\frac{(\bar{\zeta}_{1}-r)C_{1}(t)\,\lambda(\zeta_{1})^{-2n}}{(\zeta_{1}-r)(\bar{\zeta}_{1}-\zeta_{1})}\,,\,
R_{3}=\frac{C_{1}(t)\,\lambda(\zeta_{1})^{-2n}}{\zeta_{1}(\zeta_{1}-r)}\,\nonumber
\end{align}
and $\mathbf{B}_{j},\,j=1,...,5$ is the $j$-th column of matrix $\mathbf{B}$, by Cramer's Rule, we solve the linear system $\mathbf{B}\mathbf{X}=\mathbf{Y}$ and obtain
\begin{equation}
X_{1}=\frac{Det(\mathbf{B}_{1}^{c})}{Det(\mathbf{B})}\,,\,X_{5}=\frac{Det(\mathbf{B}_{5}^{c})}{Det(\mathbf{B})}\,,
\end{equation}
where $\mathbf{B}_{1}^{c}=(\mathbf{Y},\mathbf{B}_{2},\mathbf{B}_{3},\mathbf{B}_{4},\mathbf{B}_{5})$, $\mathbf{B}_{5}^{c}=(\mathbf{B}_{1},\mathbf{B}_{2},\mathbf{B}_{3},\mathbf{B}_{4},\mathbf{Y})$.
Then the first-order soliton solution can be written as
\begin{equation}\label{5.11}
q_{n}^{[1]}(t)=q_{+}(t)+r\,R_{3}\frac{X_{1}}{X_{5}}\,,
\end{equation}
where the superscript $[1]$ denotes the first-order soliton solution.

We can further simplify the solution~(\ref{5.11}). Setting $R_{1}R_{2}=v_{n}^{2}$, where $v_{n}=\frac{q_{+}(t)\bar{C}_{1}(t)}{\bar{\zeta}_{1}^{2}-2r\,\bar{\zeta}_{1}+1}\lambda(\bar{\zeta}_{1})^{2n+1}$ and symmetry $C_{j}=-\frac{q_{+}^{2}}{(\bar{\zeta}_{j}-r)^{2}}\bar{C}_{j}$ is used, we can get
\begin{equation}
X_{5}=\frac{v_{n}^{2}\bar{\zeta}_{1}/\zeta_{1}-1}{v_{n}^{2}+r_{+}(t)R_{3}-1}\,,\,X_{1}=\frac{(q_{0}^{2}\bar{\zeta}_{1}/(r\,\bar{\zeta}_{1}-1))+q_{+}(t)R_{1}X_{5}}{v_{n}^{2}-1}\,.
\end{equation}
Then first-order soliton solution has the following form:
\begin{align}
q_{n}^{[1]}(t)=q_{+}(t)\left[1+\frac{1}{v_{n}^{2}-1}\left(\frac{r\,\bar{\zeta}_{1}(v_{n}^{2}\lambda(\bar{\zeta}_{1})^{2}-1)\lambda(\bar{\zeta}_{1})^{2n}}{(r\bar{\zeta}_{1}-1)^{2}(v_{n}^{2}+R_{3}\,r_{+}(t)-1)}q_{+}(t)\bar{C}_{1}(t)-\frac{\lambda(\bar{\zeta}_{1})^{2}-1}{\bar{\zeta}_{1}/r-1}v_{n}^{2}\right)\right]\,.
\end{align}

The first-order soliton solutions with the nonzero boundary coefficients $q_{0}=\frac{2}{3}$ and scattering coefficients $\bar{\zeta}_{1}=\frac{1-q_{0}}{r}$, $\bar{\theta}_{1}=\frac{\pi}{3}$ are shown in Fig.~$5$. Fig.~$5$(a), (b) show the first-order bright soliton solution with the nonzero boundary condition $q_{0}=\frac{2}{3}$, $\theta_{+}=0$. Fig.~$5$(c), (d) show the first-order dark soliton solution with the nonzero boundary condition $q_{0}=\frac{2}{3}$, $\theta_{+}=\frac{\pi}{3}$.

\begin{center}
\includegraphics[scale=0.28]{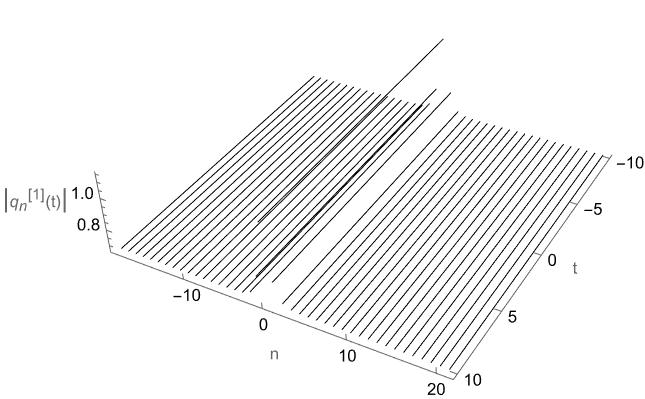}\hfill
\includegraphics[scale=0.24]{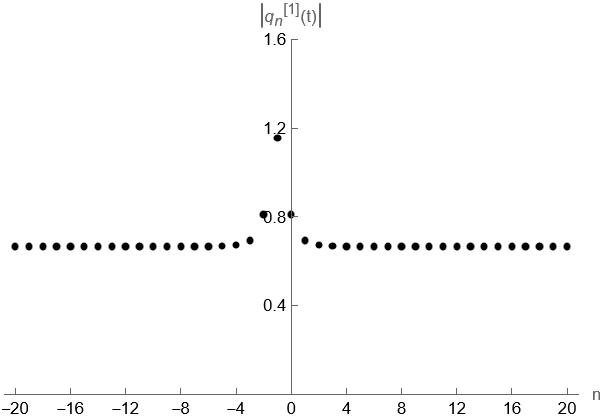}\hfill
\includegraphics[scale=0.24]{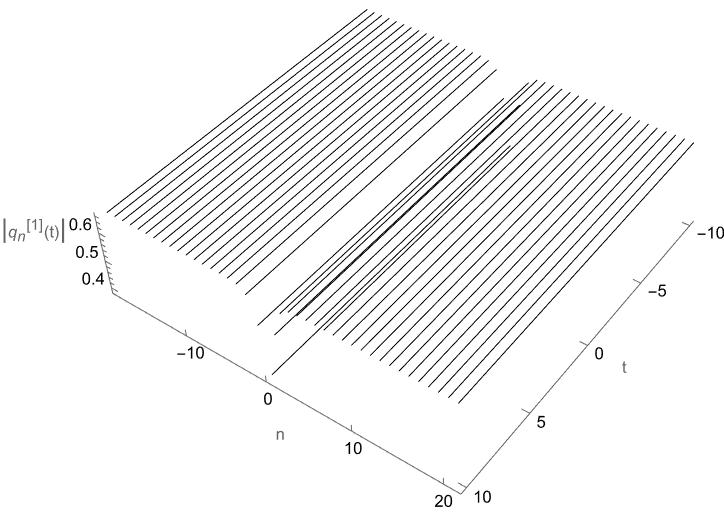}\hfill
\includegraphics[scale=0.24]{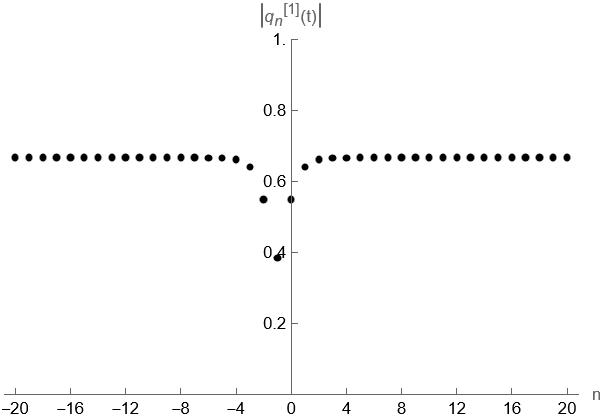}\hfill
{\footnotesize\hspace{4cm}(a)\hspace{4cm}(b)\hspace{4cm}(c)\hspace{4cm}(d)}\\
\flushleft{\footnotesize
\textbf{Fig.~$5$.} The first-order soliton solutions with the nonzero boundary coefficients $q_{0}=\frac{2}{3}$ and scattering coefficients $\bar{\zeta}_{1}=\frac{1-q_{0}}{r}$, $\bar{\theta}_{1}=\frac{\pi}{3}$. (a) The bright soliton solution $q_{n}^{[1]}(t)$ with $\theta_{+}=0$. (b) The cross-sectional view of (a) with $t=0$. (c) The dark soliton solution $q_{n}^{[1]}(t)$ with $\theta_{+}=\frac{\pi}{3}$. (d) The cross-sectional view of (c) with $t=0$.}
\end{center}

\vspace{7mm}\noindent\textbf{6 Conclusions}
\hspace*{\parindent}
\renewcommand{\theequation}{6.\arabic{equation}}\setcounter{equation}{0}\\

In this paper, we have thoroughly developed the IST for the integrable discrete nonlocal $PT$ symmetric NLS equation with nonzero boundary conditions. According to the two different signs $\sigma=\pm1$ and two different values of the phase difference $\Delta\theta=0$ or $\Delta\theta=\pi$, we have discussed four cases with significant differences about analytical regions, symmetry, asymptotic behavior and the presence or absence of discrete eigenvalues, namely, the existence or absence of soliton solutions. In the direct scattering problem, we have all studied the analyticity, asymptotic behavior and symmetries of the modified eigenfunctions and scattering coefficients. In the inverse scattering problem, the RH problem has been constructed and solved to get the reconstructed potential. Finally, combined with the time evolution, the dark, bright, dark-bright soliton solutions in existence have been obtained under the reflectionless condition.

In this work, we have pointed out some key points and solved a number of difficulties. Firstly, the second symmetry about $\zeta\mapsto\bar{\zeta}^{*}=\frac{r\zeta^{*}-1}{\zeta^{*}-r}$ has been deduced by introducing the ``backward" modified scattering problem and analysing modified eigenfunctions. Secondly, which is unlike some local cases that the values of discrete eigenvalues are relatively free except restricted within the domains of scattering coefficients, the discrete eigenvalues here are only obtained by comparing the trace formula with asymptotic behavior of scattering coefficients. Thirdly, due to the strong limitations on discrete eigenvalues mentioned above, under the $PT$ symmetry reduction condition, not all cases exist soliton solutions. For example, in the Case (\ref{c1}): $\sigma=1$, $\Delta\theta=0$, the single eigenvalue is not exist so that there is no first-order soliton solution. The simplest reflectionless potential generates second-order soliton solutions, which include dark-dark, dark-bright, bright-bright ones. In the Case (\ref{c2}): $\sigma=1$, $\Delta\theta=\pi$, we have showed there are no discrete eigenvalues so that the soliton solutions are not exist.

The research processes and results in this work also provide instructive technology to investigate other types of integrable nonlocal system, as well as to further investigate the long-time asymptotic behavior of solutions for nonlocal systems.

\vspace{5mm} \noindent\textbf{Acknowledgments}\\
\hspace*{\parindent}

We express our sincere thanks to each member of our discussion group for their suggestions. This work has been supported by the Fund Program for the Scientific Activities of Selected Returned Overseas Scholars in Shanxi Province under Grant No. 20220008 and the Graduate Education Innovation Project in Shanxi Province of China under Grant No. 2023KY189.

\end{document}